\documentclass[11pt]{article}
\pdfoutput=1
\usepackage{jheppub}
\usepackage{tabu}
\usepackage[vcentermath]{youngtab}
\usepackage[usenames,dvipsnames,table]{xcolor}
\usepackage{graphicx,amsmath,amssymb,amsthm,multirow,array,bm,bbm,esint}
\usepackage[mathscr]{eucal}
\usepackage[bbgreekl]{mathbbol}
\usepackage{epsf,amsfonts}
\usepackage{slashed}
\usepackage[numbers,sort&compress]{natbib}
\usepackage{minibox}
\usepackage{rotating}
\usepackage{pdflscape}
\usepackage{array,tikz-cd}

 \def\shadeB{\cellcolor{blue!5}}
\def\shadeR{\cellcolor{red!5}}

\definecolor{rust}{rgb}{0.8,0.2,0.2}

\usepackage{slashed}

\newcommand{\cov}{{{\tiny cov.}}}

 \def\shadeB{\cellcolor{blue!5}}
\def\shadeR{\cellcolor{red!5}}

\definecolor{rust}{rgb}{0.8,0.2,0.2}

\newcommand{\prn}[1]{\left ( #1 \right )}
\newcommand{\brk}[1]{\left [ #1 \right ]}
\newcommand{\bigbr}[1]{\left\{ #1 \right\} }

\newcommand{\half}{\frac{1}{2}}

\newcommand{\Tr}[1]{\hbox{Tr}\left(#1\right)}

\newcommand{\group}{\mathcal{G}}

\newcommand{\vev}[1]{\langle #1 \rangle}

\newcommand{\rhoi}{\hat{\rho}_{\text{initial}}}
\newcommand{\rhoT}{\hat{\rho}_{_T}}

\newcommand{\QSK}{\mathcal{Q}_{_{SK}}}
\newcommand{\QSKb}{\overline{\mathcal{Q}}_{_{SK}}}
\newcommand{\QKMS}{\mathcal{Q}_{_{KMS}}}
\newcommand{\QKMSb}{\overline{\mathcal{Q}}_{_{KMS}}}

\newcommand{\dSK}{\mathbb{d}_{_\text{SK}}}
\newcommand{\dSKb}{\bar{\mathbb{d}}_{_\text{SK}}}

\newcommand{\Qzero}{\mathcal{Q}^0_{_{KMS}}}
\newcommand{\Qbeta}{\mathscr{L}_{_{KMS}}}

\newcommand{\IKMS}{\SF{\cal I}^{\text{\tiny{KMS}}}}
\newcommand{\LKMS}{\SF{\cal L}^{\text{\tiny{KMS}}}}
\newcommand{\IKMSb}{\SF{\overline{\cal I}}{}^{^{\text{\tiny{KMS}}}}}
\newcommand{\IKMSzero}{\SF{\cal I}^{\text{\tiny{KMS}}}_0\,}
\newcommand{\iKMS}{{\iota}^{\text{\tiny{KMS}}}}
\newcommand{\iKMSb}{{\overline{\iota}^{\text{\tiny{KMS}}}}}
\newcommand{\iKMSzero}{{\iota}^{\text{\tiny{KMS}}}_0\,}
\newcommand{\lKMS}{{\ell}^{\text{\tiny{KMS}}}}
\newcommand{\imap}{\iota}
\newcommand{\ibmap}{\overline{\iota}}
\newcommand{\izmap}{\iota^0}
\newcommand{\lmap}{\ell}

\newcommand{\Q}{\mathcal{Q}}
\newcommand{\Qb}{\overline{\mathcal{Q}}}

\newcommand{\QC}{\mathbb{d}_{_{\text{\tiny C}}}}
\newcommand{\QCb}{\overline{\mathbb{d}}_{_{\text{\tiny C}}}}
\newcommand{\QW}{\mathbb{d}_{_{\text{\tiny W}}}}

\newcommand{\QWb}{\overline{\mathbb{d}}_{_{\text{\tiny W}}}}
\newcommand{\QWEG}{\mathbb{d}_{_{\text{\tiny W}}}^{\text{\tiny{E}}}}
\newcommand{\QBRST}{\mathbb{d}_{_{\text{\tiny BRST}}}}

\newcommand{\BRST}{{}_{_{\text{\tiny BRST}}}\,}

\newcommand{\DSK}{\mathcal{D}_{_\text{SK}}}
\newcommand{\DSKb}{\overline{\mathcal{D}}_{_\text{SK}}}

\newcommand{\psiW}{\psi_{_{\text{\tiny W}}}}
\newcommand{\psiC}{\psi_{_{\text{\tiny C}}}}

\newcommand{\Op}[1]{\mathbb{#1}}
\newcommand{\OpH}[1]{\widehat{\mathbb{#1}}}

\newcommand{\SKR}[1]{\mathbb{#1}_{\skR}}
\newcommand{\SKL}[1]{\mathbb{#1}_{\skL}}

\newcommand{\SKRel}[1]{\mathbb{#1}_{{dif}}}

\newcommand{\SKAdv}[1]{\mathbb{#1}_{adv}}
\newcommand{\SKRet}[1]{\mathbb{#1}_{ret}}

\newcommand{\SKG}[1]{\mathbb{#1}_{_G}}
\newcommand{\SKGb}[1]{\mathbb{#1}_{_{\overline{G}}}}

\newcommand{\FSgn}[1]{(-1)^{F_{\mathbb{#1}}}}

\newcommand{\gh}[1]{\text{gh}(#1)}

\newcommand{\gradcomm}[2]{ \brk{ #1, #2 }_{\scriptscriptstyle \pm} }
\newcommand{\comm}[2]{ \brk{ #1, #2 }}

\newcommand{\bcomm}[2]{ \prn{ #1, #2 }_\Kref}

\newcommand{\thb}{{\bar{\theta}} }
\newcommand{\thetab}{\bar{\theta} }
\newcommand{\SF}[1]{\mathring{#1}}
\newcommand{\Dsf}{\SF{\mathcal D}}
\newcommand{\Ibar}{\overline{\mathcal{I}}}

\newcommand{\IbarEG}[1]{\overline{\mathcal{I}}_{#1}^{\,\text{\tiny{E}}}}
\newcommand{\lieEG}[1]{{\cal L}_{#1}^{^\text{\tiny{E}}}}
\newcommand{\LamS}{\SF{\Lambda}}
\newcommand{\As}{\SF{\mathscr{A}}}
\newcommand{\Ascr}{\mathscr{A}}

\newcommand{\Fs}{\SF{\mathscr{F}}}

\newcommand{\Bs}{\SF{\mathscr{B}}}

\newcommand{\psib}{\overline{\psi}}

\newcommand{\phiT}{\phi_{_{\,\smallT}}}
\newcommand{\phibT}{\overline{\phi}_{_{\,\smallT}}}
\newcommand{\GT}{G_{_{\,\smallT}}}
\newcommand{\GbT}{\overline{G}_{_{\,\smallT}}}
\newcommand{\BT}{B_{_{\,\smallT}}}
\newcommand{\etaT}{\eta_{_{\,\smallT}}}
\newcommand{\etabT}{\overline{\eta}_{_{\,\smallT}}}

\newcommand{\etab}{\overline{\eta}}

\newcommand{\tx}{\tilde{X}}
\newcommand{\xpsi}{X_\psi}
\newcommand{\xpsib}{X_{\psib}}

\newcommand{\deltaB}{\delta_{_ {\bm\beta}}}

\newcommand{\delKMS}{{\Delta}_{_ {\bm\beta}} }

\newcommand{\Kref}{{\bm \beta}}

\newcommand{\Lagref}{\mathscr L}
\makeatletter
  \newcommand\Ttiny{\@setfontsize\Ttiny{1pt}{2}}
\makeatother

\newcommand{\Cref}{{\sf C}}

\newcommand{\form}[1]{\bm{#1}}

\newcommand{\ic}{\form{i}}

\newcommand{\lieD}{\pounds}

\newcommand{\Kbeta}{{\bm{\beta}}}
\newcommand{\LambdaB}{\Lambda_{\bm{\beta}}}

\newcommand{\Lag}{{\mathcal L}}

\newcommand{\skR}{\text{\tiny R}}
\newcommand{\skL}{\text{\tiny L}}

\newcommand{\smallT}{{\sf \!{\scriptscriptstyle{T}}}}

\newcommand{\UT}{U(1)_{\scriptstyle{\sf T}}}

\usepackage{tikz}
\usetikzlibrary{shapes.geometric, arrows}
\tikzstyle{thread} = [rectangle, minimum width=.3\textwidth, minimum height=.6cm, text centered, text width=.3\textwidth, draw=black, fill=blue!10]
\tikzstyle{details} = [rectangle, minimum width=.3\textwidth, minimum height=.6cm, text centered, text width=.3\textwidth, draw=black, fill=green!5]
\tikzstyle{arrowT} = [ultra thick,->,>=stealth']
\tikzstyle{arrow2} = [ultra thick,dashed,->,>=stealth']

\title{Schwinger-Keldysh formalism II: \\ Thermal equivariant cohomology}

\author[a]{Felix M. Haehl}
\author[b]{\!, R.\ Loganayagam}
\author[c]{\!, Mukund Rangamani}
\affiliation[\,a]{Department of Physics and Astronomy, University of British Columbia,\\
6224 Agricultural Road, Vancouver, B.C.\ V6T 1Z1, Canada.}
\affiliation[\,b]{International Centre for Theoretical Sciences (ICTS-TIFR), \\
Shivakote, Hesaraghatta Hobli, Bengaluru 560089, India.}
\affiliation[\,c]{
Center for Quantum Mathematics and Physics (QMAP)  \\
Department of Physics, University of California, Davis, CA 95616 USA.}
\emailAdd{f.m.haehl@gmail.com}
\emailAdd{nayagam@gmail.com}
\emailAdd{mukund@physics.ucdavis.edu}

\abstract{
Causally ordered correlation functions of local operators in near-thermal quantum systems computed using the Schwinger-Keldysh formalism obey a set of Ward identities.  These can be understood rather simply as the consequence of a topological (BRST) algebra, called the universal Schwinger-Keldysh superalgebra, as explained in our companion paper  \cite{Haehl:2016pec}.  In the present paper we provide a mathematical discussion of this topological algebra. In particular, we  argue that the structures can be understood in the language of extended equivariant cohomology. To keep the discussion self-contained, we provide a basic review of the algebraic construction of equivariant cohomology and explain how it can be understood in familiar terms as a superspace gauge algebra. We demonstrate how the Schwinger-Keldysh construction can be succinctly encoded in terms a thermal equivariant cohomology algebra which naturally acts on the operator (super)-algebra of the quantum system. 
The main rationale behind this exploration is to extract symmetry statements which are robust under renormalization group flow and can hence be used to understand low-energy effective field theory of near-thermal physics. To illustrate the general principles, we focus on Langevin dynamics of a Brownian particle, rephrasing some known results in terms of thermal equivariant cohomology. As described elsewhere, the  general framework enables construction of effective actions for dissipative hydrodynamics and could potentially illumine our understanding of  black holes.}
\begin{document}
\maketitle

%%%%%%%%%%%%%%%%%%%%%%%%%%%%%%%%%%%%%%%%%%%%%%%%%%%%%%%%

\newpage
%~~~~~~~~~~~~~~~~~~~~~~~~~~~~~~~~~~~~~~~~~~~~~~~
\part{Introduction and background}
\label{part:intro}
%~~~~~~~~~~~~~~~~~~~~~~~~~~~~~~~~~~~~~~~~~~~~~~
\hspace{1cm}

%~~~~~~~~~~~~~~~~~~~~~~~~~~~~~~~~~~~~~~~~~~~~~~~
\section{Introduction}
\label{sec:intro}
%~~~~~~~~~~~~~~~~~~~~~~~~~~~~~~~~~~~~~~~~~~~~~~~

The Schwinger-Keldysh formalism  \cite{Schwinger:1960qe,Keldysh:1964ud} was developed to enable one to  compute temporally ordered correlation functions in arbitrary states (either pure or mixed)  of a relativistic QFT. One envisages perturbing the state by time dependent classical sources, which by virtue of relativistic causality,   
 should only affect the correlations to the causal future of the disturbance. The key question is how does one ensure appropriate causal ordering in the path integral formalism? The essential idea in the Schwinger-Keldysh formalism is to work directly with density matrices, which evolve unitarily under the time-dependent Heisenberg evolution.  This is achieved by doubling the state space, viewing the density matrix as a pure state, and thence writing the path integral as a functional integral over this enlarged Hilbert space. 

In the doubled state space one can formulate a set of time-ordering prescriptions that lead to physical response and fluctuation functions which have been well studied in the literature. An excellent introduction to these techniques can be found for instance in the reviews \cite{Chou:1984es,Landsman:1986uw,Maciejko,Kamenev:2009jj}.  Examination of the time-ordered correlation functions reveals certain interesting identities, see e.g., \cite{Weldon:2005nr} for a concise summary. One learns that a class of correlation functions involving only the differences of operators acting independently on each component of the enlarged Hilbert space vanishes  independent of dynamics. Furthermore, such operators are prevented from being futuremost in the temporal ordering; one may say that the correlators satisfy a version of the largest time equation \cite{tHooft:1973pz}. These facts are well known to experts, but the novel aspect which was first previewed in \cite{Haehl:2015foa} and reviewed in our companion paper \cite{Haehl:2016pec} is that one can understand the results systematically in terms of a symmetry principle (see also \cite{Crossley:2015evo}).

The essential point is that the Schwinger-Keldysh construction embodies a field redefinition symmetry that can be recast as a redundancy on the enlarged configuration space. Phrased this way it is clear that these redundancies can be understood in terms of a set of topological BRST charges that act on the Schwinger-Keldysh functionals. The charges being topological, are nilpotent. The symmetry they generate ensures the constraints on the correlation functions described above. The main message of \cite{Haehl:2016pec} was to explain this structure and to demonstrate that one can equivalently recast the Schwinger-Keldysh functional integral into an elegant superspace formalism. To avoid causing confusion, let us remark at the outset that when we refer to superspace or supersymmetry we refer to structures governed by topological BRST symmetries; our supercharges are simply Grassmann-odd generators. In particular, they transform as scalars under the spacetime Lorentz group. Any Schwinger-Keldysh path integral ends up having a pair of BRST supercharges $\QSK$ and $\QSKb$ which suffice to enforce the aforementioned constraints on the physical correlation functions.

One can abstractly describe the Schwinger-Keldysh functional integral in terms of upgrading the operator algebra to an operator superalgebra. Each operator $\OpH{O}$ in the basic QFT gets embedded into a topological supermultiplet, which comprises of a quartet of fields. This structure is naturally realized in terms of a superspace generated by two Grassmann odd elements; we will use $\theta, \thb$ to denote the two supercoordinates.  

Observables in a generic density matrix are primarily constrained by these supercharges.  It is  however interesting to examine the class of Gibbsian density matrices, which describe the dynamics of a quantum system in a thermal or near-thermal state. Such density matrices (unnormalized) take the form $\rho = e^{-\beta (\Op{H} - \mu_I\, \Op{Q}^I)}$, where we have allowed for the possibility of turning on chemical potentials for conserved charges $\Op{Q}^I$. The special feature of these mixed states is that they are further constrained to satisfy the so-called KMS condition \cite{Kubo:1957mj,Martin:1959jp}. One can understand this condition from the viewpoint of writing an equilibrium thermal QFT as a statistical mechanical model with Euclidean time evolution. Periodicity of the Euclidean time circle demands a certain analytic property of thermal correlation functions in an imaginary time-strip of width $\beta$. This is the physical origin of the KMS condition, which has proven useful not only in understanding equilibrium thermal dynamics, but also plays an important role in algebraic formulations of QFT \cite{Haag:1967sg}.

The KMS condition turns out to provide an interesting upgrade of the Schwinger-Keldysh BRST symmetry.  Consider the difference between an  operator and its KMS conjugate, i.e., the operator obtained by Hamiltonian evolution by a thermal period around the Euclidean thermal circle.  One may view this operation in terms of a differential operator $\deltaB$ acting on the algebra of operators. This transformation is a thermal time translation; one is essentially evaluating departures from the KMS condition by this operation. In fact, it suffices to focus on the action of an infinitesimal generator of thermal translations which we denote as  $i\,\delKMS = 1 - \FSgn{} e^{-i\,\deltaB}$. The second term achieves thermal translations taking into account operator statistics.\footnote{ Thermal boundary conditions demand the bosons are periodic around the Euclidean thermal circle, while fermions are anti-periodic. $\FSgn{}$ is the ${\mathbb Z}_2$ valued fermion number operator.} In this paper, we will always work in the high temperature regime such that $\delKMS \approx \Kref\, \frac{d}{dt}$, which acts naturally as a derivation on the operator algebra.

Since the Schwinger-Keldysh construction endows the operator algebra with a grading structure, it follows that the KMS derivation should also be upgraded into a KMS super-operation. The Grassmann odd elements correspond to KMS supercharges, while the Grassmann even elements comprise of the KMS derivation $\delKMS$ which we denote as the operator $\Qbeta$ and another generator whose origins will become clearer in due course. All told, thermal density matrices end up having a total of six symmetry generating operations which we chose to denote as $\{\QSK,\QSKb, \QKMS,\QKMSb,\Qbeta,\Qzero\}$ in \cite{Haehl:2016pec}. The four ${\cal Q}$ operators are Grassmann-odd with the barred operators having opposite ghost number to the unbarred ones. We refer to this algebraic structure as the SK-KMS superalgebra (or SK-KMS algebra).

In the current paper we explore this algebra  and argue that it bears the hallmark of a well known algebraic structure, viz., the extended equivariant cohomology algebra. While the latter structure originally was discovered in the context of understanding cohomological properties of group actions on manifolds, it has played an important role in physical applications. As is well known starting from the seminal work of Witten \cite{Witten:1982im}, there is a natural link between cohomological constructions and topological supersymmetry. This has formed the backbone of investigations in the context of topologically twisted quantum field theories \cite{Witten:1988ze,Witten:1988xj}. These developments are nicely reviewed in \cite{Birmingham:1991ty}.

 Much of the early discussion focuses on the situation with one topological charge, as exemplified by the Donaldson-Witten theory. The case of interest to us is the situation with two topological charges which was first encountered in the context of topological twists of extended supersymmetric field theories like ${\cal N} =4$ SYM \cite{Vafa:1994tf}. The algebraic formalism we require was developed following this work in \cite{Dijkgraaf:1996tz}, whose terminology we adopt. More specifically, we will refer to theories with $k$ topological supercharges as $\mathcal{N}_\smallT =k$ topological field theories. We will argue that the Schwinger-Keldysh construction specifically requires $k=2$.\footnote{ As described in \cite{Haehl:2016pec} the higher $\mathcal{N}_\smallT =2n$ algebras with $n>1$ appear to be naturally realized by generalizations of Schwinger-Keldysh construction for computing out-of-time-order (OTO) correlation functions.} 

 In our discussion, the concept of equivariance will play a crucial role. Formally, this alludes to situations where the associated structures are unaffected by a group action. Of course, in physical terms this simply refers to the familiar notion of gauge invariance. The natural question we should ask ourselves is: if the SK-KMS algebra admits an interpretation in terms of an equivariant cohomological algebra, then what exactly is the symmetry group? 

 To address this  question we need to extend considerations beyond the equilibrium Gibbs density matrix. It will suffice to pay attention to the algebraic structures as we allow small departures from equilibrium. In fact,  all we need to do is to ask how could one implement KMS transformations if such deviations had non-trivial spatial and temporal profiles. The former case can be analyzed by considering equilibrium in non-trivial stationary curved space, with locally varying temperatures (and chemical potentials) following \cite{Banerjee:2012iz,Jensen:2012jh}.  We argue that one generically needs to allow for local (continuous) KMS transformations, which has the effect of gauging thermal translations. 

 This symmetry, which we call $\UT$ KMS symmetry, was first articulated in the context of analyzing hydrodynamic effective fields theories using standard phenomenological axioms in \cite{Haehl:2014zda,Haehl:2015pja}. An earlier indication of connection between thermal translations and gauge fields appeared in the discussion of Lorentz anomalies \cite{Jensen:2013rga}.  Our discussion provides a microscopic rationale for this structure.\footnote{ We have hitherto used the equivariant cohomological algebra based on this $\UT$ to construct effective actions for dissipative hydrodynamics in \cite{Haehl:2015uoc}. Independently, \cite{Crossley:2015evo} also discovered the supersymmetry structure to be useful in aiding the construction of hydrodynamic effective actions.}  As explained in \cite{Haehl:2016pec} the motivation behind our analysis of this algebraic structure of the Schwinger-Keldysh formalism was to exploit its robustness under the renormalization group to understand the origins of the low energy dynamics.

For the most part this paper focuses on setting the stage for describing the basic mathematical framework, amplifying on the prose of \cite{Haehl:2015foa} and filling in some of the essential details used in our construction of hydrodynamic effective actions \cite{Haehl:2015uoc}. To keep the discussion manageable we have chosen to eschew a more comprehensive analysis of hydrodynamics and other applications. Rather we describe in some detail the much simpler problem of a Brownian particle undergoing Langevin dynamics using the machinery of thermal equivariant cohomology. Our aim is to provide a simple example illustrating all the basic points in the construction without getting too tangled up in technicalities that are specific to applications. Further elaborations and applications of this formalism will appear in due course. Readers are also invited to consult \cite{Haehl:2016pec} in tandem for an explanation of the physical context and elaboration of the SK-KMS algebra.

\tikzset{
     block/.style={rectangle, draw, text width=40em,
                   text centered, rounded corners, minimum height=3em},
     arrow/.style={-{Stealth[]}}
     }
     
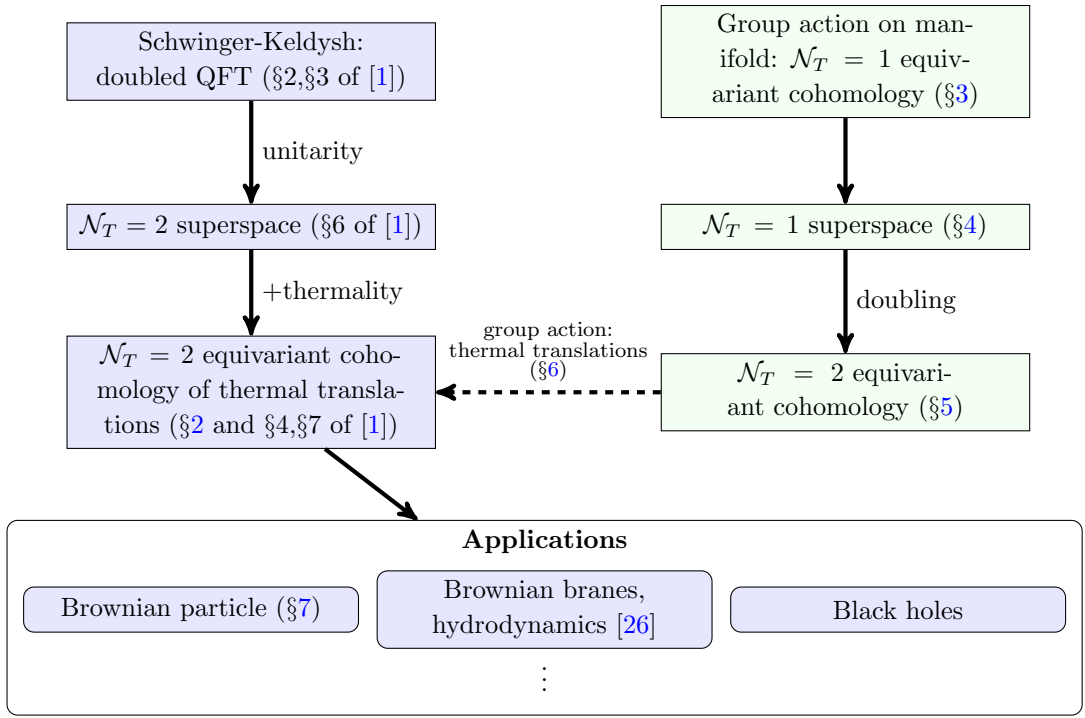
\begin{figure}
\centering
\small{
\begin{tikzpicture}[node distance=1.2cm]
\node (SK) [thread] {Schwinger-Keldysh:\\ doubled QFT (\S2,\S3 of \cite{Haehl:2016pec})};
\node (Nt2super) [thread, below of = SK, yshift = -1cm] {$\mathcal{N}_T=2$ superspace (\S6 of \cite{Haehl:2016pec})};
\node (NT2equiv) [thread, below of = Nt2super, yshift = -1cm] {$\mathcal{N}_T=2$ equivariant cohomology of thermal translations (\S\ref{sec:quadruplet} and \S4,\S7 of \cite{Haehl:2016pec})};
\node (appl) [block, below of = Nt2super, yshift = -4cm, xshift=3.9cm] {{\bf Applications}\\\vspace{.2cm}
\begin{tikzpicture}
\node (brown) [thread] {Brownian particle (\S\ref{sec:langevin})};
\node (hydro) [thread,right of = brown, xshift=3.5cm] {Brownian branes,\\hydrodynamics \cite{Haehl:2015uoc}};
\node (BH) [thread,right of = hydro, xshift=3.5cm] {Black holes};
\end{tikzpicture}
$\mathbf{\vdots}$\vspace{.2cm}
};

\node (NT1) [details, right of = SK, xshift = 6.7cm] {Group action on manifold: $\mathcal{N}_T=1$ equivariant cohomology (\S\ref{sec:equivariance})};
\node (NT1super) [details, below of = NT1, yshift = -1cm] {$\mathcal{N}_T=1$ superspace (\S\ref{sec:nt1super})};
\node (NT2math) [details, below of = NT1super, yshift = -1cm] {$\mathcal{N}_T=2$ equivariant cohomology (\S\ref{sec:extequivariance})};

\draw [arrowT] (SK) -- node [right] {unitarity} (Nt2super);
\draw [arrowT] (Nt2super) -- node [right] {+thermality} (NT2equiv);
\draw [arrowT] (NT1) -- (NT1super);
\draw [arrowT] (NT1super) -- node [right] {doubling} (NT2math);
\draw [arrow2] (NT2math) -- node [above] {{\large $\substack{\text{group action:}\\\text{thermal translations}\\\text{(\S\ref{sec:sknt2})}}$}} (NT2equiv);
\draw [arrowT] (NT2equiv) -- (appl);
\end{tikzpicture}
}
\caption{An illustration of the dependencies of various ideas presented here and in \cite{Haehl:2016pec}. The left hand side shows how basic ingredients of Schwinger-Keldysh formalism lead to an equivariant cohomology algebra; these are largely the content of \cite{Haehl:2016pec} and reviewed here only briefly. The present paper follows the logic on the right hand side. Part \ref{part:maths} reviews the mathematical structure of equivariant cohomology and derives from there the algebras uncovered in \cite{Haehl:2016pec} as a special case of the general formalism when the group acts by thermal translations. For a considerably more extensive list of applications, see \S11 of \cite{Haehl:2016pec}.}
\label{fig:flowchart}
\end{figure}

The outline of this review is as follows. {\bf Part \ref{part:intro}} is meant to orient the reader by providing a brief synopsis of our analysis of the SK-KMS algebra derived in \cite{Haehl:2016pec} in \S\ref{sec:quadruplet}. The diagram in Fig.~\ref{fig:flowchart} should help the reader putting this discussion into context. In {\bf Part \ref{part:maths}} we put aside the physics of SK theories and turn to a review of the equivariant cohomology in \S\ref{sec:equivariance}. This material is quite standard and is well explained in other sources, but we choose to present it for completeness in our terms. We emphasize in \S\ref{sec:nt1super} a superspace construction for the equivariant cohomology. This superspace structure allows one to immediately generalize the discussion to multiple cohomology generators; we focus on the case of two generators, and explain the structure of the extended $\mathcal{N}_\smallT =2$ algebra in \S\ref{sec:extequivariance}. This will finish our discussion of the general mathematical structures. 

In {\bf Part \ref{part:thermal}} we return to SK quantum field theories. The established mathematical framework allows us to give a more precise characterization of the SK-KMS algebraic structures, and we also explore basic properties of the thermal translational symmetry that is crucial in this scheme, see \S\ref{sec:sknt2}. At this point, we are ready for physical applications. We have described elsewhere \cite{Haehl:2015uoc} how these structures help in the construction of hydrodynamic effective field theories and could naturally segue into that discussion. However, in order to keep the material accessible for non-experts, we have chosen to illustrate the general construction using a very simple system of dissipative dynamics: the stochastic motion of a Brownian particle as described by Langevin dynamics. This is reviewed in some detail in \S\ref{sec:langevin}, both from the traditional viewpoint on the topological supersymmetry as presented in standard textbooks and in the language of the equivariant cohomology, amplifying our discussion in the Appendix A of \cite{Haehl:2015foa}. We conclude with some discussion of other applications and generalizations in \S\ref{sec:discussion}. A treatment of thermal equivariant cohomology as applied to other problems is deferred to a separate publication. The appendices contain some further elaboration of some of the mathematical results, in particular, the Mathai-Quillen and Kalkman isomorphisms between the various algebraic constructions.

%~~~~~~~~~~~~~~~~~~~~~~~~~~~~~~~~~~~~~~~~~~~~~~~
\section{BRST symmetries in the Schwinger-Keldysh formalism}
\label{sec:quadruplet}
%~~~~~~~~~~~~~~~~~~~~~~~~~~~~~~~~~~~~~~~~~~~~~~ 

In \S\ref{sec:intro} we have explained somewhat briefly that the Schwinger-Keldysh formalism is naturally viewed as an operator superalgebra.  The set of operators involved are naturally built out of the operator algebra by extending them into a cohomological supermultiplet. We will begin with a short summary of the results explained in  \cite{Haehl:2016pec}. Specifically we review the BRST symmetries inherent in the Schwinger-Keldysh formalism and elaborate on the cohomological superalgebra structure. The discussion will be brief, as it is mainly for orientation and setting up necessary notation; for details we refer the reader to \cite{Haehl:2016pec}. 

%~~~~~~~~~~~~~~~~~~~~~~~~~~~~~~~~~~~~~~~~~~~~~~~
\subsection{The SK-KMS algebra}
\label{sec:skkms}
%~~~~~~~~~~~~~~~~~~~~~~~~~~~~~~~~~~~~~~~~~~~~~~

The Schwinger-Keldysh formalism sets up the computation of time-ordered correlation functions by working with a doubled Hilbert space. Rather than viewing a density matrix as acting a physical Hilbert space ${\cal H}$, we double the latter into ${\cal H}_\skR \otimes {\cal H}_\skL^*$, which refer to the space of `kets 'and `bras' respectively. The labels right ($\text{R}$) and  left ($\text{L}$) are for reference and useful to denote the time-ordering prescription. Each element of the operator algebra $\OpH{O}$ naively gets duplicated into a pair,  $\OpH{O}\rightarrow \{\SKR{O},\SKL{O}\}$. 

The  primary object of interest is the generating functional for the time ordered correlators, which is a functional of two sets of sources. Given an initial density matrix $\rhoi$, one can write
\begin{equation}
\begin{split}
\mathscr{Z}_{SK}[{\cal J}_\skR,{\cal J}_\skL] &\equiv \Tr{\ U[{\cal J}_\skR]\ \rhoi\ (U[{\cal J}_\skL])^\dag\ } \,,\\
&= \int [{\cal D}\Phi] \, \exp\left(i\, S[\Phi_\skR] - i\,S[\Phi_\skL] + i\,\int \, {\cal J}_\skR \SKR{O} -{\cal J}_\skL \SKL{O} \right)
\end{split}
\label{eq:ZSKdef}
\end{equation}
$U[{\cal J}_{\skL/\skR}]$ is the unitary evolution operator with (a priori different) source deformations for the forward 
and backward evolution (for the kets and bras) respectively. In the second line we have given a schematic expression for the functional integral in terms of the fundamental fields $\Phi$, which have been doubled. The relative sign follows from the forward evolution of the right copy and the backwards evolution for the left fields.  For the rest of our discussion we will concentrate on Gibbsian density matrices, i.e.,
\begin{equation}
\rhoi = \rhoT = e^{-\beta (\Op{H} - \mu_I\, \Op{Q}^I)} \,.
\label{eq:rhoT}
\end{equation}	

A path integral that implements \eqref{eq:ZSKdef} has to satisfy some constraints arising from unitary evolution. For one, the alignment of right and left sources, collapses it onto the initial state: $\mathscr{Z}_{SK}[{\cal J}_\skR={\cal J}_\skL] = \Tr\rhoi$. This demands that a host of correlators of $\SKRel{O} = \SKR{O}-\SKL{O}$ vanishes. This is straightforward to infer from \eqref{eq:ZSKdef}, but its genericity suggests a deeper reason. We argued that a natural way to implement this in a manner that is dynamics agnostic, i.e., does not depend on choice of unitaries $U$, is to infer the existence of Grassmann-odd BRST supercharges $\{\QSK,\QSKb\}$. These follow from the fact that the double path integral comes with field redefinition symmetry, which can be used to construct these BRST charges. They carry non-vanishing conserved ghost number 
$\gh{\QSK} = +1$ and $\gh{\QSKb} =-1$.

One consequence of this is that we can give a covariant presentation of the Schwinger-Keldysh formalism, by making this BRST symmetry explicit. Rather than working with the doubled operator algebra, we upgrade to an operator superalgebra. This entails the introduction of BRST ghost and anti-ghost degrees of freedom associated with each operator $\OpH{O}$; we  effectively end up with a quadruplet of operators:
\begin{equation} 
\OpH{O} \quad \longrightarrow \quad \{\SKR{O},\SKL{O},\SKG{O},\SKGb{O}\} \,.
\label{eq:quadruple}
\end{equation}

Focusing in particular, on thermal correlation functions, we need to impose KMS condition which demands that correlation functions be analytic in the imaginary time strip of width $\beta = \frac{1}{T}$. Rather than view this as a statement about observables, we found it convenient to phrase it directly as an action on the operator algebra. We define a derivation action $\deltaB$ that relates an operator $\OpH{O}(t)$ to its KMS conjugate:
\begin{equation}
e^{-i\,\deltaB} \OpH{O}(t)  = \OpH{O}(t-i\,\beta) \equiv \rhoT ^{-1}\, \OpH{O}(t) \, \rhoT \,.
\end{equation}	
The advantage of introducing this derivation action $\deltaB$ is that we can simply imagine Lie transporting operators around the Euclidean thermal circle to obtain their thermal KMS conjugates.

The KMS condition demands that the operators $e^{-i\,\deltaB} \OpH{O}(t) $ and $\OpH{O}(t)$ be equivalent within correlation functions, modulo a statistics factor. Recalling that bosons are periodic and fermions anti-periodic under rotation by a period $\beta$, we can write:
\begin{equation}
\delKMS  \OpH{O} = 0 \,, \quad \text{where}\quad i \delKMS \equiv 1-\FSgn{} e^{-i\deltaB}  \,,
\label{eq:kmsDel}
\end{equation}	
where $(-1)^F$ denotes the fermion number operator. We will frequently refer to $\delKMS$ as a {\it thermal translation} operator; it measures deviation from the KMS condition. We also found it useful to define an operator $\Qbeta$ which acts through a commutator action on the operator algebra to implement $\delKMS$, viz., $\gradcomm{\Qbeta}{\OpH{O}} = \delKMS \OpH{O}$.\footnote{ Our conventions are described in \cite{Haehl:2016pec}. In particular, we will often make use of the graded commutator defined as 
$\gradcomm{A}{B} = AB - (-)^{AB}\, BA$, where $(-)^{AB}$ denotes the mutual Grassmann parity of the entries $A$ and $B$ respectively. }

The KMS condition leads to a set of Ward identities for thermal correlation functions, which can be understood in terms of a 
second pair of  BRST charges $\{\QKMS,\QKMSb\}$.  These are  nilpotent carrying ghost numbers $\gh{\QKMS} = +1$ and $\gh{\QKMSb} = -1$. They generate imaginary time thermal translations and can be thought of as the Grassmann-odd superpartners of $\Qbeta$. Rounding off the structure is a fourth Grassmann-even generator $\Qzero$. 

These four KMS charges are easily understood as a super-derivation extending the operator $\delKMS$ to be compatible with the SK BRST symmetry. Indeed, while the four Grassmann-odd generators $\{\QSK,\QSKb,\QKMS,\QKMSb\}$ follow from simple considerations of SK path integrals and the KMS condition, the remaining operators $\{\Qzero,\Qbeta\}$ simply ensure closure of the thus generated algebra. The action of these charges can be understood easily by basis rotating the quadruplet in \eqref{eq:quadruple} to define the retarded and advanced combinations
\begin{equation}\label{eq:RADef}
\begin{split}
\SKAdv{O} \equiv  \SKR{O}-\SKL{O}\ , \qquad 
\SKRet{O} &\equiv \frac{1}{1-\FSgn{O} e^{-i\deltaB} }  \prn{ \SKR{O}-\FSgn{O} e^{-i\deltaB} \SKL{O} } \,.
\end{split}
\end{equation}
In this basis, the action of the fermionic supercharges can be summarized schematically as
\begin{equation}
\begin{tikzcd}
&\SKRet{O} \arrow{ld}{\QSK} \arrow{rd}[below]{\QSKb}    &   \\
\SKG{O}\arrow{rd}{\QSKb} & & \SKGb{O} \arrow{ld}[above]{\!\!\!\!\!\!\!\!\!\!\!\!\!\!-\QSK}\\
&   \SKAdv{O} &
\end{tikzcd}
\qquad
\begin{tikzcd}
&\quad\SKAdv{O}\quad \arrow{ld}{\!\!\!\QKMS} \arrow{rd}[below]{\!\!\!\!\QKMSb}   &   \\
\delKMS \SKG{{O}}\arrow{rd}{\!\!\!\QKMSb \quad\;-\QKMS} & & -\delKMS\SKGb{{O}} \arrow{ld}\\
&  \delKMS\delKMS \SKRet{O} &
\end{tikzcd}
\label{eq:qskkmsaction}
\end{equation}
where arrows indicate action via graded commutator, e.g., $\gradcomm{\QSK}{\SKRet{O}} = \SKG{O}$ etc.. We refer the reader to \cite{Haehl:2016pec} for derivations and details. 

The six operators $\{\QSK,\QSKb,\QKMS,\QKMSb,\Qbeta,\Qzero\}$ satisfy the following algebra (which one can deduce from the above diagrams):
\begin{gather}
\QSK^2 =\QSKb^2 = \QKMS^2=\QKMSb^2 = 0\ , \nonumber\\
\gradcomm{\QSK}{\QKMS} =  \gradcomm{\QSKb}{\QKMSb} = \gradcomm{\QSKb}{\QSK} = \gradcomm{\QKMS}{\QKMSb} = 0\ ,\nonumber \\
\gradcomm{\QSK}{\QKMSb} =  \gradcomm{\QSKb}{\QKMS} = \Qbeta\,, \label{eq:kmsalg}\\
\gradcomm{\QKMS}{\Qzero} = \gradcomm{\QKMSb}{\Qzero} = 0 \,,\nonumber\\
\gradcomm{\QSK}{\Qzero} =  \QKMS\,, \qquad \gradcomm{\QSKb}{\Qzero} = -\QKMSb\,.\nonumber
\end{gather}
We refer to this as the \emph{SK-KMS superalgebra}. The goal of our present discussion is to obtain insight into this algebraic structure. We will see that the natural language for this exploration is terms of a graded algebra, with the grading being provided by the ghost number charge. This leads us then into the study of equivariant cohomological algebras which arise in this context, and extend the above structure mildly by making the KMS symmetries act locally. 

%~~~~~~~~~~~~~~~~~~~~~~~~~~~~~~~~~~~~~~~~~~~~~~~
\subsection{A superspace description}
\label{sec:sspace}
%~~~~~~~~~~~~~~~~~~~~~~~~~~~~~~~~~~~~~~~~~~~~~~

An extremely convenient way to view the KMS superalgebra \eqref{eq:kmsalg} is to express the operations directly in superspace. The superspace we need is a simple one with two Grassmann odd coordinates denoted $\theta,\thb$. We will take them to have equal and opposite ghost number, with $\gh{\theta} = +1$.\footnote{ We emphasize that the BRST symmetries are cohomological in nature. When we refer to superspace or supersymmetry we refer to such structures and not to standard supersymmetric field theories. Our supercoordinates therefore are Lorentz scalars and carry no Lorentz spin labels.} 

The quadruplet of operators $\{\SKRet{O}, \SKG{O}, \SKGb{O}, \SKAdv{O}\}$ associated with a single-copy operator $\OpH{O}$ are encapsulated into a single SK-superfield. Working in the adv-ret basis introduced above, we define the associated super-operator as:\footnote{ Following \cite{Haehl:2015uoc} we will use an accent ``$\; \SF{}$\;'' to denote superfields. We will elaborate on superspace conventions further in \S\ref{sec:nt1super} and \S\ref{sec:nt2}.}
\begin{equation}\label{eq:OpO}
\SF{\Op{O}} \equiv \SKRet{O} + \theta \, \SKGb{O} + \bar\theta \, \SKG{O} + \bar\theta\theta \, \SKAdv{O} \,.
\end{equation}

In the previous subsection we have reviewed the BRST charges $\{\QSK,\QSKb\}$ and a quadruplet of KMS translation generators $\{\QKMS,\QKMSb,\Qbeta,\Qzero\}$. All these act on the operator algebra via graded commutators.  We will now describe the identical structure in terms of superspace derivations, which act on components of covariant superfields. As will become clear later, superspace provides a very efficient tool for encoding the relevant algebraic structures. 
\begin{itemize}
\item In superspace, the  action of $\{\QSK,\QSKb\}$ is realized as derivations along the Grassmann-odd directions; these operators implement super-translations:
\begin{equation}
\QSK \;\longrightarrow\;  \partial_{\thb}\,, \qquad \QSKb \; \longrightarrow \;  \partial_\theta\,,
\end{equation}
which induce the action of $\{\QSK,\QSKb\}$ component-wise on superfields; for example, we have $\gradcomm{\QSK}{\SF{\mathbb{O}}} = \partial_\thb \SF{\mathbb{O}} = \SKGb{O} + \theta\, \SKAdv{O}$. We refer to \cite{Haehl:2016pec} for more details. 
\item Similarly, while we can think of $\{\QKMS,\QKMSb,\Qbeta,\Qzero\}$ as acting on superfields component-wise, it will be convenient to lift each of them to a super-operator.
To this end, we introduce the following quadruplet of thermal translation operators: 
\begin{equation}\label{eq:SKsuperops}
\begin{split}
   \IKMSzero &= \Qzero + \thb \, \QKMS - \theta \, \QKMSb + \thb \theta \, \Qbeta \,,\\
   \IKMS &= \QKMS + \theta \, \Qbeta \,,\\
   \IKMSb &= \QKMSb + \thb \, \Qbeta \,,\\
   \LKMS &=  \Qbeta \,.
\end{split}
\end{equation}
These are defined such that 
\begin{equation}
\IKMSzero \, \SF{\mathbb{O}} = \IKMS \, \SF{\mathbb{O}} = \IKMSb \, \SF{\mathbb{O}}= 0 \,.
\end{equation}
We will see later the rationale behind this choice in terms of constructing gauge invariant observables.
\item This quadruplet of thermal super-translations can be checked to  be related by the following diagram:
\begin{equation}
\begin{tikzcd}
&\IKMSzero \arrow{ld}{\partial_\thb} \arrow{rd}[below]{\partial_\theta}    &   \\
\IKMS\arrow{rd}{\partial_\theta} & &  -\IKMSb \arrow{ld}[above]{\!\!\!\!\!\!\!\!\!\!\!\!\!\!-\partial_\thb}\\
&   \LKMS &
\end{tikzcd}
\label{eq:QzeroDiag}
\end{equation}
Further, these superspace operators are clearly related to our previous parametrization as 
\begin{equation}\label{eq:Sops}
\begin{split}
\Qzero \simeq (\IKMSzero)| &\equiv {\cal I}^{\text{\tiny KMS}}_0 \,,\qquad
\QKMS \simeq (\IKMS)| \equiv {\cal I}^{\text{\tiny KMS}}  \,,\\
\Qbeta \simeq (\LKMS)| &\equiv {\cal L}^{\text{\tiny KMS}} \,,\qquad
\QKMSb \simeq (\IKMSb)| \equiv \overline{{\cal I}}^{\text{\tiny KMS}}  \,,
\end{split}
\end{equation}
where the symbol ``$\simeq$" means identification in superspace with the understanding that the action on superfields is component-wise and the vertical slash $|$ refers to projection onto $\theta =\thb =0$ subspace.
\end{itemize}

The analogue of the algebra \eqref{eq:kmsalg} can now be written in terms of superspace differentials as follows:
\begin{gather}
\partial_\thb^2 =\partial_\theta^2 = (\IKMS)^2=  (\IKMSb)^2= 0\ , \nonumber\\
\gradcomm{\QSK}{\IKMS} =  \gradcomm{\QSKb}{\IKMSb} = \gradcomm{\QSKb}{\QSK} = \gradcomm{\IKMS}{ \IKMSb}= 0\ ,\nonumber \\
\gradcomm{\QSK}{ \IKMSb} =  \gradcomm{\QSKb}{\IKMS} = \LKMS \,, \label{eq:kmsalg2}\\
\gradcomm{\LKMS}{ \IKMSzero} = \gradcomm{ \IKMSb}{ \IKMSzero} = 0 \,,\nonumber\\
\gradcomm{\QSK}{ \IKMSzero} = \IKMS\,, \qquad \gradcomm{\QSKb}{ \IKMSzero} =  -\IKMSb\,.\nonumber
\end{gather}
A more compact way of stating this algebra is simply by regarding the operators \eqref{eq:Sops} as superoperators which can be commuted in the obvious way, and on which $\{\QSK,\QSKb\}$ act as $\{\partial_\thb,\partial_\theta\}$. For instance, we have 
\begin{equation}\label{eq:Sops3}
 \gradcomm{\QSK}{{\bf I}} = \partial_\thb {\bf I} \,,\qquad \gradcomm{\QSKb}{{\bf I}} = \partial_\theta {\bf I} \,,\qquad \gradcomm{{\bf I}}{{\bf I}'} = 0 \
\end{equation}
for any superoperators ${\bf I},{\bf I}'\in \{\IKMSzero,\IKMS,\IKMSb,\LKMS\}$.  \\

While we motivated the above structures purely from an analysis of Schwinger-Keldysh path integrals in \cite{Haehl:2016pec}, we will now proceed to argue that these algebraic structures are encountered in the context of equivariant cohomology. More precisely, we will need to invoke the notion of an  extended equivariant cohomology algebra with two super-generators, often denoted as  $\mathcal{N}_\smallT =2$ following \cite{Dijkgraaf:1996tz}.

 As we shall see, the idea of equivariance is to look for structures that are compatible with a group action (such as a gauge symmetry). We review the salient features of such structures in Part \ref{part:maths} focusing on explaining the algebraic framework for obtaining cohomological data in more familiar gauge theoretic language. 

 Of course, in our above discussion, we never mentioned any group action, but simply found the structure of a super algebra by studying features of SK theories. If we wish to identify our KMS superalgebra with that of an extended equivariant cohomology, we not only need to give an identification of the defining operations and their relations, but also give a physical rationale for why these structures are appearing and what the associated gauge symmetry is. In \S\ref{sec:sknt2} we will provide this missing link and will explain that the KMS superalgebra can indeed be understood by thinking about a $\UT$ group action, which parametrizes thermal diffeomorphisms. The analog of an infinitesimal `gauge transformation' will turn out to be the thermal translation parameterized by $\delKMS$. For the present, the reader can keep in mind, the heuristic picture that we aim to upgrade the discrete KMS translations to  continuous local symmetries. This picture works well at high temperatures, and thus is relevant for the construction of low energy effective actions. 

At the end of the day, we will establish the notion of thermal equivariant cohomology, where the spacetime thermal translations are gauged.  Before we get there however, we will first provide in Part \ref{part:maths} the necessary mathematical tools, explaining both the algebraic and the superspace constructions of equivariant cohomology.

\newpage
%~~~~~~~~~~~~~~~~~~~~~~~~~~~~~~~~~~~~~~~~~~~~~~~
\part{Mathematical review: equivariant cohomology}
\label{part:maths}
%~~~~~~~~~~~~~~~~~~~~~~~~~~~~~~~~~~~~~~~~~~~~~~
\hspace{1cm}

%~~~~~~~~~~~~~~~~~~~~~~~~~~~~~~~~~~~~~~~~~~~~~~~
\section{An introduction to equivariant cohomology}
\label{sec:equivariance}
%~~~~~~~~~~~~~~~~~~~~~~~~~~~~~~~~~~~~~~~~~~~~~~

In \S\ref{sec:quadruplet} we followed the logic of the Schwinger-Keldysh construction to argue that for thermal density matrices we have a quartet of fermionic BRST supercharges $\{\QSK, \QSKb,\QKMS, \QKMSb\}$ which form an algebra \eqref{eq:kmsalg}. These operators are all nilpotent and involve an interesting commutation relation
\begin{align}
\gradcomm{\QSK}{\QKMSb} &=  \gradcomm{\QSKb}{\QKMS} = \Qbeta \,,
\label{eq:kmsnontr}
\end{align}
where $\Qbeta$ plays the role of a Hamiltonian which commutes with all other operators via $\gradcomm{\Qbeta}{\, \cdot \,} \equiv \delKMS$. 
It behooves us to understand better this algebraic structure, which is obviously reminiscent of supersymmetry. We will argue below that the algebra of these generators has been previously encountered in the topological field theory literature and goes by the name of the extended $\mathcal{N}_\smallT =2$ equivariant cohomology algebra.

 In the string theory literature  the algebra first arose in the context of topologically twisted ${\cal N}=4$ Super-Yang Mills theory \cite{Vafa:1994tf} and the name was coined subsequently by \cite{Dijkgraaf:1996tz} who gave a unified description of such topological symmetries.  Further useful analyses which we will exploit in the course of our discussion can be found in \cite{Blau:1996bx,Zucchini:1998rz}. However, to the best of our knowledge, first occurrence of these algebraic structures predates its appearance in the string theory literature; it seems to have been independently discovered in \cite{Gozzi:1989vv} where the authors were exploiting these structures to present a functional integral formulation of classical mechanics. The mathematical framework we need is embodied in the notion of equivariant cohomology which has played an important role in the construction of topological field theories, cf., \cite{Witten:1988ze}. We refer the  reader to  \cite{Birmingham:1991ty,Cordes:1994fc} for good introductory reviews aimed at physicists and \cite{Guillemin:2013aa} for a clear discussion of the associated mathematical framework.

Before delving into the details let us understand the physical picture. The basic idea is to examine gauge invariant structures in field theories where the action of the gauge group naturally leads to the mathematical concept of equivariance. The notion of equivariance refers to some group action which commutes across maps between topological spaces, so that one is interested in structures that are suitably immune to the transformation engendered by the group elements. In physical language we are after gauge invariant structures. 

Suppose we have the action of a gauge group on some field theory, and wish to focus on the space of gauge invariant states. The idea is to work off the gauge invariant subspace and demand that the configurations of interest satisfy the gauge constraints. The key is to implement this covariantly, which immediately requires us to introduce the BRST symmetry associated with the gauge group. 
Formally thinking of the configuration space as a manifold ${\cal M}$ with a group action $\group$, we wish to induce non-trivial cohomology on ${\cal M}/\group$. One can then view the directions normal to the gauge invariant space as gauge orbits and only consider structures that have neither support along these orbits, nor undergo any variation in these directions. It is then natural to induce the group action onto differential forms of ${\cal M}$ and define a class of gauge invariant forms (called basic forms). This construction can be carried out explicitly algebraically, and forms the essential basis of equivariant cohomology. 

The key point of the above discussion is that we need to identify an appropriate gauge symmetry that acts on our system. The BRST charges in the equivariant cohomology language can be viewed as a combination of ordinary cohomology operators (e.g., the exterior derivation of de Rham theory) suitably combined with the Faddeev-Popov BRST symmetry of the gauge sector. This construction was first appreciated in the early works of Weil  and Cartan, who formulated two local models of such construction, which carry their respective names. The relation between these two models of the cohomology was initially explained in a seminal work of Mathai and Quillen \cite{Mathai:1986tc}. Further details along with an interpretation in the language of topological supersymmetry was  explained by Kalkman \cite{Kalkman:1993zp}. We will review this basic construction first and then proceed to explain the extension we seek. The essential thesis of our argument will be that at the end of the day $\{\QKMS,\QKMSb\}$ should be viewed as the standard BRST charges of an abelian gauge symmetry, called $\UT$ KMS gauge symmetry. This was first introduced by us in \cite{Haehl:2014zda,Haehl:2015pja} and its role in constructing hydrodynamic effective actions was explained in terms of equivariance in \cite{Haehl:2015foa}. The following can be seen as an elaboration of our brief discussion in the latter paper. 

In the following we review the two local models of equivariant cohomology. First we discuss the Weil model which is simple to state algebraically, but more involved to work with if we wish to construct explicitly the space of gauge invariant operators. Then by a suitable change of basis (the so called Kalkman automorphism) we will pass to the Cartan model, which makes better contact with familiar constructions in gauge theories. 
Kalkman also introduced a third model, called the BRST model, which should be familiar to physicists from gauge theory literature. We won't dwell on this construction in the main text, but provide some basic elements of the discussion in Appendix \ref{sec:MQ}.

%~~~~~~~~~~~~~~~~~~~~~~~~~~~~~~~~~~~~~~~~~~~~~~
\subsection{The Weil model}
\label{sec:weiln1}
%~~~~~~~~~~~~~~~~~~~~~~~~~~~~~~~~~~~~~~~~~~~~~~

Consider a topological space $\mathcal{M}$ on which we have the action of a group $\group$. We will need also the associated algebra $\mathfrak{g}$ and its dual $\mathfrak{g}^*$.  The group action means that a point $p$ on the manifold gets shifted to a new point $g(p)$ under a group element $g \in \group$. Since all we do is permuting the points, we can equivalently view the transformation as a particular diffeomorphism.\footnote{ It will be useful for the sake of presentation to pretend that $\mathcal{M}$ is a manifold with coordinates $X^\mu$ to build some intuition, despite it being straightforward to work with abstract topological spaces. After all one of the circumstances of interest for us is to understand similar operations on operator algebras acting on a quantum Hilbert space.}

The infinitesimal group elements can be identified with elements of the Lie algebra $\mathfrak{g}$ by the usual expansion of the exponential map. Let  $t_i$ be the generators of the Lie algebra, satisfying commutation relations, 
\begin{equation}
[t_i, t_j] = f_{ij}^k\, t_k  \,, \qquad i,j = 1,2,\cdots, \text{dim}(\mathfrak{g})\,.
\label{eq:lieal}
\end{equation}	
We can equivalently think of the infinitesimal group action as being generated by vector fields 
$\xi_i^\mu \in {\cal T}{\cal M}$. These spacetime vectors satisfy an algebra of Lie derivations consistent with \eqref{eq:lieal}, viz.,\footnote{ It is straightforward to generalize this to the case where we have further internal structure such as a flavour symmetry acting on our fields, where the Lie algebra generators would also implement a flavour transformation. This would be relevant when we  consider the action of $\group$ on a bundle over ${\cal M}$. In fact, our choice of signs below differs from standard literature; it is engineered to allow interpreting $f_{ij}^k$ as the structure constants of the flavour symmetry when $\group$ acts as a global symmetry.\label{fn:signs}} 
\begin{equation}
\begin{split} 
\lieD_{\xi_i}\xi_j^\mu= - \lieD_{\xi_j}\xi_i^\mu &= -f^k_{ij} \xi_k^\mu \,.
% \lieD_{\xi_i} \Lambda_j - \lieD_{\xi_j} \Lambda_i - [\Lambda_i,\Lambda_j] &=-  f^k_{ij} \Lambda_k 
\end{split}
\end{equation}

We are interested in looking at  the cohomology of the orbifold $\mathcal{M}/\group$, which may in general be complicated by the fact that the group action may not be free and one has to deal with various singularities. The trick to address the general case algebraically is to introduce some redundancy to make the group action free and induce the cohomology from the larger space. To this end we need a topological space, which has trivial cohomology and is cognizant only of the group structure. 

The mathematical fact one employs is the existence of a contractible space $\text{E}_\group$ called the {\em universal $\group$ bundle} on which $\group$ acts freely and the quotient space $\text{B}_\group = \text{E}_\group/\group$ (called the {\em classifying space}) which is a manifold that has a universal $\group$-connection on it. The simplest example of this is to take $\group = {\mathbb Z}$ whence $\text{E}_\group = {\mathbb R}$ and $\text{B}_\group = {\mathbb R}/{\mathbb Z} = {\bf S}^1$. A more familiar physics example is to think of $\text{E}_\group$ as the space of all gauge connections, and $\text{B}_\group$ as the space of gauge orbits. 

In general, we can think of the connection as coming from a Kaluza-Klein style reduction over  $\text{E}_\group$. This universal connection can be  pulled back onto $\mathcal{M}$. The cohomology of the orbifold $\mathcal{M}/\group$ can then sensibly be defined as the cohomology of the manifold $\prn{ \text{E}_\group \times \mathcal{M}}/\group$, with the equivalence following from
contractibility of $\text{E}_\group$.\footnote{ Thinking of $\group$-action on $\mathcal{M}$ by thinking about $ \text{E}_\group\times \mathcal{M}$ is  called the Borel construction of the $\group$-action on $\mathcal{M}$.}

For clarity, we start with a discussion of the relevant structures on the universal 
$\group$-bundle $\text{E}_\group$ alone. We will subsequently return to the actual problem of interest and study algebraic structures on $\text{E}_\group\times {\cal M}$. The construction will be phrased in an algebraic fashion, but the similarities with construction of gauge theories are worth bearing in mind.

%~~~~~~~~~~~~~~~~~~~~~~~~~~~~~~~~~~~~~~~~~~~~~~~
\subsubsection{The Weil complex}
%~~~~~~~~~~~~~~~~~~~~~~~~~~~~~~~~~~~~~~~~~~~~~~

To define the space of forms on $\text{E}_\group$, we employ the so called Weil complex, which involves the tensor product of the exterior algebra of $\mathfrak{g}^*$ with the symmetric algebra of $\mathfrak{g}^*$. Recall that the exterior algebra is generated by the wedge product of forms, while the symmetric algebra is generated by symmetric polynomials with variates being the elements of the underlying vector space (here $\mathfrak{g}^*$). 

From a physical viewpoint (see \cite{Witten:1982im}) it is useful to think of the Weil complex as being generated by a set of Grassmann-odd ghosts $G^i$ and a Grassmann-even curvature 2-form called the ghost of ghost $\phi^i$. We will assign these objects definite ghost number and require that 
\begin{equation}
\gh{G^i} = 1 \,, \qquad \gh{\phi^i} = 2\,.
\label{eq:Gphignos}
\end{equation}	

 To define a cohomology we need a nilpotent operator $\QWEG$ acting on forms on $\text{E}_\group$ which we will take to be Grassmann-odd by virtue of its grading ($\gh{\QWEG} =1$). Since the action on $\text{E}_\group$ is supposed to mimic gauge transformations, we define $\QWEG$ action on $G^i, \phi^i$ to be of the familiar form for a gauge potential and field strength, viz., 
\begin{equation}\label{eq:WeilAlg}
\begin{split} 
\QWEG  G^i +\half  f^i_{jk} G^j G^k &=  \phi^i  \,, \\
\QWEG  \phi^i +  f^i_{jk} G^j \phi^k &=  0  \,. 
\end{split}
\end{equation}

The $\QWEG$-cohomology in $\text{E}_\group$ is by construction trivial. The only elements of the Weil complex  which are sensitive to the $\QWEG$ action are of the form ${\cal P}(\phi^i)\, \wedge_{k=1}^n\, G^i_k$, with ${\cal P}(\phi^i)$ being symmetric polynomials of the $\phi$s. This is of course consistent with the contractibility of the space. 

To define something interesting we need to isolate forms that are non-trivial in the quotient $\text{B}_\group= \text{E}_\group/\group$. To this end, let us parameterize the space of normals to $\text{B}_\group$, which is nothing but the space of fibres normal to the gauge orbits. Define then an operator that projects us along the $k^{\rm th}$ direction along the fibres. This can be achieved by contracting with the generator of the dual algebra. One can use this concept to introduce a set of Grassmann odd operators $\IbarEG{k}$ which satisfy
\begin{equation}\label{eq:WeilAlg2}
\begin{split} 
\delta^i_j +  \IbarEG{j} G^i  &=  0\ ,\qquad  
\IbarEG{j}   \phi^i  =  0   \,.
\end{split}
\end{equation}
The operator $\IbarEG{k}$ is simply an interior contraction in the $k^{\rm th}$ direction and we have aligned the basis of forms to it in \eqref{eq:WeilAlg2}. 

Readers familiar with these structures might find our choice of signs unconventional. The group action on the manifold results in a diffeomorphism by $-\xi_k^\mu$ which is the vector field corresponding  to the generator $t_k$. Usually the sign is absorbed into the definition of the interior contraction and one defines \eqref{eq:WeilAlg2} with an opposite sign. As explained in footnote \ref{fn:signs} we wish to adhere to conventions where the group action would correspond to an active flavour rotation (for global symmetries). Hence w.l.o.g. we have chosen a less familiar sign convention, which will keep manifesting itself in various formulae, cf., \eqref{eq:WeilAlg4} and \eqref{eq:LieSAlg}.

Once we have an analog of exterior derivation $\QWEG$ and an interior contraction $\IbarEG{k}$, we can immediately use them to construct a Lie derivative on $\text{E}_\group$, 
\begin{equation}
\lieEG{j} \equiv \gradcomm{\QWEG}{\IbarEG{j} } \,.
\label{eq:liederID}
\end{equation}	
Using the definitions we can check that
\begin{equation}\label{eq:WeilAlg3}
\begin{split} 
\lieEG{j} G^i = \gradcomm{\QWEG}{\IbarEG{j} } G^i  &=  f^i_{jk} G^k\ ,\qquad  
 \lieEG{j}  \phi^i =\gradcomm{\QWEG}{\IbarEG{j}  }   \phi^i  =   f^i_{jk} \phi^k   \,.
\end{split}
\end{equation}
Other useful relations are
\begin{equation}\label{eq:WeilAlg4}
\begin{split} 
\gradcomm{ \gradcomm{\QWEG}{\IbarEG{j} } }{ \IbarEG{k} } &=  -f^i_{jk} \IbarEG{i} \,,\\
\gradcomm{ \gradcomm{\QWEG}{\IbarEG{j} }}{ \gradcomm{\QWEG}{\IbarEG{k} }}
 &=  -f^i_{jk}   { \gradcomm{\QWEG}{\IbarEG{i} } } \,.\\
 \end{split}
\end{equation}

The equations \eqref{eq:WeilAlg}-\eqref{eq:WeilAlg4}  define the universal algebra on the Weil complex, involving the exterior derivative $\QWEG$, contraction $\IbarEG{i} $,  curvature  form $\phi^i$ and connection form  $G^i$ with gauge group $\group$.  We often like to pull-back nice form-like objects  (e.g., characteristic classes and their descendants such as Chern-Simons forms) from $\text{E}_\group$ which have useful geometric or cohomological 
interpretation. Such objects are usually made of exterior derivative, curvature form or connection form in $\text{E}_\group$, and after pullback  they can be written in terms of $\{\QWEG,\phi^i,G^i\}$: the mathematically precise form of this statement is called the Chern-Weil homomorphism.  

More pertinently for our discussion, $\{\QWEG, \IbarEG{i}, \lieEG{i}\}$ form a Lie superalgebra with the three operators having ghost numbers $\gh{\QWEG} = 1$, $\gh{\IbarEG{i}} =-1$ and $\gh{\lieEG{i}} =0$. Formally this superalgebra can be expressed through the following relations\footnote{ One can w.l.o.g. extend this algebra to include the ghost number operator $\text{G}_h$ as well for each of the other operators carry definite ghost charge.} which follow from  \eqref{eq:WeilAlg}-\eqref{eq:WeilAlg4}:
\begin{equation}
\begin{aligned}
\gradcomm{\IbarEG{i} }{\IbarEG{j} } &= 0 \,, &   \qquad
\gradcomm{\QWEG}{\IbarEG{j}} &= \lieEG{j}  \,,
 \\
\gradcomm{\lieEG{i} }{\IbarEG{j}} &=-f_{ij}^k\, \IbarEG{k} \,,  &  \qquad
\gradcomm{\QWEG}{\lieEG{j} } &= 0 \,,
 \\
\gradcomm{\lieEG{i} }{\lieEG{j} } &= -f_{ij}^k\,\lieEG{k}  \,,  &  \qquad
\gradcomm{\QWEG}{\QWEG} &= 0 \,.
\end{aligned}
\label{eq:LieSAlg}
\end{equation}
We will see in the sequel that the algebra of our SK and KMS supercharges is a suitable extension of such a structure. In Appendix \ref{app:gauge} we give a few details for extending these structures to involve explicit gauge parameters $\alpha = \alpha^i t_i$.

Once we have this algebraic structure, it is straightforward to write down the non-trivial elements generating the cohomology of $\text{B}_\group = \text{E}_\group/\group$. To this end, realize that we would like no support for objects on $\text{B}_\group$ nor have variations in the normal directions to $\text{B}_\group$. Since the normal directions are the gauge orbits we can simply use  $G^i$ to parameterize them. As a result we seek gauge invariant forms, which would be built purely out of invariant combinations of the field strengths $\phi^i$. A formal way of saying this is to define:
\begin{itemize}
\item {\em Horizontal forms:} belonging to the kernel of $\IbarEG{}$, viz., $\eta$ such that $\IbarEG{j} \eta =0$ for $j=1,2,\cdots , \dim(\mathfrak{g})$.
\item {\em Invariant forms:} belonging to the kernel of $\lieEG{}$, viz., $\eta$ such that $\lieEG{j}\eta =0$ for $j=1,2,\cdots , \dim(\mathfrak{g})$.
\item{\em Basic forms:} forms that are both horizontal and invariant.
\end{itemize} 
Non-trivial elements of the cohomology are defined by basic forms. Using \eqref{eq:WeilAlg2} one can convince oneself that they have to be built out of the field strengths $\phi^i$ alone, and furthermore the invariance condition can be solved simply by forming polynomial invariants which correspond to the group Casimirs. This structure is of course very familiar from the viewpoint of gauge theories.

%~~~~~~~~~~~~~~~~~~~~~~~~~~~~~~~~~~~~~~~~~~~~~~~
\subsubsection{Graded Weil algebra on $\text{E}_\group \times {\cal M}$}
%~~~~~~~~~~~~~~~~~~~~~~~~~~~~~~~~~~~~~~~~~~~~~~

Our interest was, of course, not in the properties of the classifying space $\text{B}_\group$, but rather in ${\cal M}/\group$. We therefore need to extend the discussion above to $\text{E}_\group \times {\cal M}$. Let us pick some coordinates $X^\mu$ on ${\cal M}$ and introduce a set of Grassmann-odd vectors $\psiW^\mu$. The latter are a convenient way to encode the space of one-forms on ${\cal M}$ as first understood in \cite{Witten:1982im} and serve to bring out the connection to the physical field theories we will wish to construct quite efficiently. Essentially the Grassmann nature ensures that $\psiW^\mu$ behave effectively as one-forms $d_{_{\cal M}}X^\mu$ as far as the cohomological properties are concerned. The subscript on the fields is to remind us that these are related to the Weil model. Later when we pass to the Cartan model we will change basis to a different set of fermions.

The question now is to extend the action of the operators  $\QWEG$ and $\IbarEG{i}$ to also capture  de Rham cohomology of $\mathcal{M}$. We will simply identify the Weil differential $\QW$ to act as 
\begin{equation}
\QW \equiv (\QW)_{_{\text{E}_\group \times \mathcal{M }}} \equiv \QWEG\otimes \mathbb{1} _{_{\mathcal{M}}}+ \mathbb{1}_{_{\text{E}_\group}} \otimes d_{_{\mathcal{M}}}\,,
\end{equation}	
where $d_{_{\mathcal{M}}}$ is the standard exterior derivative on $\mathcal{M }$. This allows us to pick contributions from the differential forms on the manifold of interest. Using the identification of one-forms and $\psiW^\mu$ we simply have 
\begin{equation}
\QW X^\mu = \psiW^\mu \,, \qquad \QW\psiW^\mu =0 \,.
\end{equation}	
We also view the Lie algebra valued vector fields $\xi^\mu_i$ as functions of $X^\mu$. The simplest way to think about these is to view $X^\mu$ as coordinates on a target space manifold in the language of sigma models. We will see explicitly in \S\ref{sec:langevin} that this perspective is extremely efficacious.

To finish the characterization of the algebra we need to define the action of the extended interior contraction 
\begin{equation}
\Ibar_k \equiv (\Ibar_k)_{_{\text{E}_\group \times {\cal M}}} \equiv \IbarEG{k} \otimes\mathbb{1} _{_{\mathcal{M}}} + \mathbb{1}_{_{\text{E}_\group}} \otimes (\Ibar_k)_{_{{\cal M}}} \,,
\end{equation}
where $(\Ibar_k)_{_{{\cal M}}}$ denotes interior contraction on forms on ${\cal M}$ along the vector field $\xi^\mu_k$. 
This may be done by realizing that we wish it to act on forms on the manifold ${\cal M}$ and pick out the gauge directions. This suggests that we should define the action to satisfy:
\begin{equation}
\Ibar_k X^\mu = 0 \,, \qquad \Ibar_k \psiW^\mu = \xi^\mu_k\,.
\end{equation}	
We can check that this gives the correct action for the Lie derivative ${\cal L}_k \equiv ({\cal L}_k)_{_{\text{E}_\group \times {\cal M}}} \equiv \gradcomm{\QW}{\Ibar_k}$:
\begin{equation}\label{eq:lieX}
\begin{split} 
\mathcal{L}_k X^\mu &= \xi_k^\mu\ ,\qquad  
\mathcal{L}_k  \psiW^\mu =   \QW \xi_k^\mu = \psiW^\sigma \,\partial_\sigma \xi_k^\mu \,,
\end{split}
\end{equation}
where we used the definition inherent in  \eqref{eq:LieSAlg}, which ensures that the superalgebra acts nicely on ${\cal M}$ as well. Indeed, using the isomorphism between the one-form basis $d_{_{\cal M}}X^\mu$ and $\psiW^\mu$  we identify $p$-forms via\footnote{ It is convenient think of $V_{\psiW}$ as an object pulled back onto  world-volume in the sigma model sense.}
\begin{align}
V_{\psiW}\equiv  \frac{1}{p!}\  V_{\mu_1 \mu_2\ldots \mu_p}(X)\ \psiW^{\mu_1} \psiW^{\mu_2} \ldots  \psiW^{\mu_p}\,.
\label{eq:Vpdef}
\end{align}
Given the action of the various generators defined above, we obtain by direct evaluation
\begin{equation}
 \QW V_{\psiW}  = (dV)_{\psiW} \,, \qquad \Ibar_k V_{\psiW}  =  (\ic_{\xi_k}V)_{\psiW} \,, \qquad 
 \mathcal{L}_k V_{\psiW} = (\lieD_{\xi_k}V)_{\psiW}\,,
\end{equation}	
where $\ic_\xi$ denotes the usual interior contraction with a vector field $\xi$. 

The operations $\{\QW,\Ibar_k,{\cal L}_k\}$ defined on forms on $\text{E}_\group \times {\cal M}$ comprise the so-called {\it Weil model} of equivariant cohomology. 
With these definitions in place, we once again restrict attention to the space of $\group$-invariant basic forms to obtain non-trivial elements of the cohomology on $\mathcal{M}/\group$. The complication we face is that the interior contraction operators $\Ibar_k$ act non-trivially on all the fields on ${\cal M}$. Simplifying this action, involves a change of basis, thus leading to the Cartan construction, which we now turn to. In the language of gauge theory, the idea of the Cartan construction is to focus on gauge covariant structures.

%~~~~~~~~~~~~~~~~~~~~~~~~~~~~~~~~~~~~~~~~~~~~~~
\subsection{The Cartan model}
\label{sec:cartann1}
%~~~~~~~~~~~~~~~~~~~~~~~~~~~~~~~~~~~~~~~~~~~~~~

Recall that we are interested in the space of basic forms which are both horizontal and invariant. While the degrees of freedom introduced to parameterize the gauge sector (the space $\text{E}_\group$) have been trivialized by construction, the same is not true for the physical degrees of freedom which are tensor valued fields in ${\cal M}$. To achieve this simplification we realize that we should modify the Grassmann-odd partners of $X^\mu$ so as to annihilate them by the interior contraction operators. Let us trade  $\psiW^\mu$ for  
a new field $\psiC^\mu$ via\footnote{ Kalkman \cite{Kalkman:1993zp} introduces a one-parameter family of fields $\psi^\mu_t$ defined to satisfy $\psi_t^\mu = \QW X^\mu + t \, G^k \, \xi_k^\mu $ which allows for other choices. We have taken the interpolating parameter to unity which corresponds at the end of the day to the Cartan model. The original Weil model is obtained by setting $t=0$.} 
\begin{equation}
\begin{split} 
\psiC^\mu &\equiv \psiW^\mu +  G^k \xi_k^\mu= (\QW + G^k \xi_k^\sigma \partial_\sigma) X^\mu \,.
% \chi &\equiv (\Q g)g^{-1}+t\ G^k \Lambda_k
\end{split}
\end{equation}
%
%Of course, with this defintion we still have $\QW^2 X^\mu =0$. 
The advantage of this shift is apparent when we examine the interior contraction, for now:
\begin{equation}
 \Ibar_k  \psiC^\mu =  0 \,.
\end{equation}
 We immediately see that in the new basis any combination of $\psiC^\mu$ is horizontal. With this redefinition, the only field not annihilated by $\Ibar_k$ would be the ghost $G^i$ and the horizontality is 
easily imposed by considering only combinations with no ghosts $G^i$. So any function of the coordinates $X^\mu$, their Grassmann partners $\psiC^\mu$, and the field strength $\phi^k$, can be taken in an ansatz for non-trivial elements of the cohomology. One can check further that the Lie derivatives act in this basis as 
\begin{equation}
 \mathcal{L}_k\psiC^\mu = \psiC^\sigma\,  \partial_\sigma \xi_k^\mu \,,
\end{equation}
which we now have to exploit to construct invariant forms.

Note that the action of the cohomology operator $\QW$ on $p$-forms is much more involved. One can evaluate the action on $V_{\psiC}$, defined in analogy with \eqref{eq:Vpdef}, to arrive at:
\begin{equation}\label{eq:dform}
\begin{split} 
 \QW V_{\psiC}  & =  (dV)_{\psiC}- G^k\, (\lieD_{\xi_k} V)_{\psiC} + \phi^k\  (\ic_{\xi_k} V)_{\psiC}\,,
 \end{split}
\end{equation}
which as before does satisfy $\mathcal{L}_k V_{\psiC} = (\lieD_{\xi_k} V)_{\psiC}$ as expected once we use 
\eqref{eq:WeilAlg4}.

As explained before, horizontality reduces to removing the ghost fields $G^i$, and it is hence convenient to go one step further and define a new cohomology generator, $\QC$, called the \emph{Cartan charge}, which amounts to effectively dropping all the ghost contributions from the Weil charge $\QW$. Its action on the fields that appear in our ansatz for basic forms, viz., $\{X^\mu, \psiC^\mu, \phi^k\}$ is given as:
\begin{equation}\label{eq:dCX}
\begin{split} 
 \QC X^\mu &\equiv \QW X^\mu + G^k \xi_k^\mu  = \psiC^\mu \,, \\
 \QC \psiC^\mu   &\equiv \QW \psiC^\mu +G^k (\partial_\nu \xi_k^\mu)  \psiC^\nu =
  \phi^k  \xi_k^\mu  \,,
  \\
\QC \phi^k  & \equiv \QW \phi^k + f^k_{ij} G^i \phi^j =  0\ .
\end{split}
\end{equation}
This defines the action of $\QC$ on the elements of the symmetric algebra. This precludes the action of $\QC$ on $G^k$ (in gauge theories this is the familiar statement that the covariant derivative $\QC$ does not act on gauge potentials). Eventually, however, we wish to employ a gauge where $G^k = 0$. In order for this to be consistent with the $\QC$ action, we wish to formally define $\QC$ on the full Weil complex and in particular demand $\QC G^k = 0$. We can summarize this condition consistently with \eqref{eq:dCX} as 
\begin{equation}
 \QC = \QW + G^i \, {\cal L}_i + \left( \frac{1}{2} \, f^k_{ij} G^i G^j + \phi^k \right) \, \Ibar_k \,,
 \label{eq:CarK}
 \end{equation}
 which makes explicit that $\QC = \QW$ on the subspace of basic forms. From this and \eqref{eq:dform} we conclude that the Cartan charge acts on $p$-forms in the Cartan basis as
\begin{equation}
 \QC V_{\psiC} = \left( \QW + G^i \, {\cal L}_i + \left(\frac{1}{2} \, f^k_{ij} G^i G^j  + \phi^k \right) \, \Ibar_k \right) V_{\psiC}
 = (dV)_{\psiC} + \phi^k\  (\ic_{\xi_k} V)_{\psiC} \,.
\end{equation}

There is a simple physical interpretation of the new cohomology generator $\QC$: it is a gauge covariant derivation.  If $\QW$ represents the ordinary exterior derivative, then $\QC$ can be thought of as a covariant exterior derivative with $G^k$ as the connection. Like any other covariant exterior derivative, $\QC$ is not nilpotent -- in fact, $\QC^2$ generates the group action along the `curvature' $\phi^k$:
\begin{equation}
 \QC^2=  \phi^k\, {\cal L}_k - \comm{G}{\phi}^k \, \Ibar_k  \,.
\end{equation}	
While we still have explicit appearances of $G^k$ in the above equations, we can now pass to the {\it Cartan model} of equivariant cohomology. This amounts to restricting to the symmetric algebra of $\mathfrak{g}^*$ (generated by $\phi^k$) instead of the whole Weil complex (generated by $\{G^k,\phi^k\}$). Effectively, this means restricting to the subspace generated by $\phi^k$ and setting $G^k = 0$. Setting $G^k =0$, the above equations yield the following Cartan model expressions:
\begin{equation}
 \QC = \QW + \phi^k \Ibar_k \qquad \Rightarrow \qquad \QC^2 = \phi^k \, {\cal L}_k \,.
 \end{equation}
This is the statement that the Cartan differential squares to a gauge transformation along $\phi^k$. more precisely, on restricting to this space $\QC$ action takes the form:
\begin{align}
\QC =\mathbb{1}_{_{\text{E}_\group}} \otimes d_{_{\cal M}}+ \phi^k \otimes (\Ibar_k)_{\cal M}\,,
\label{eq:CartanWeil1}
\end{align}
which algebraically encodes the fact that we have a gauge-covariant exterior derivation operator, since 
$\QC^2  = \phi^k \otimes ({\cal L}_k)_{_{\cal M}} \equiv \mathcal{L}_\phi$. The standard presentation is to follow this route and define the Cartan model only on the gauge invariant subspace. We find it useful however to follow the intuition of Kalkman's construction \cite{Kalkman:1993cm} and define the Cartan model on the tensor product of the  Weil algebra and exterior algebra on $\mathcal{M}$ as in \eqref{eq:CarK}.

Irrespective of the formalism, within a subset of field combinations which are invariant under $\group$-action, we have $\QC^2=0$. The cohomology on this invariant subset is the invariant cohomology of $\QC$ which in turn, gives the equivariant cohomology we are after. This formalism with the horizontality condition built in, is called the Cartan model for equivariant cohomology. By construction, it is equivalent to the original definition through Weil model and the equivalence can be explicitly shown by Kalkman automorphism mentioned earlier.\footnote{ One may argue that $\QC$ is the natural cohomological charge in topologically twisted gauge theories for the twisting is cognizant of the underlying BRST structure of the gauge theory. Indeed, in explicit constructions as in the Donaldson-Witten theory \cite{Witten:1988ze} one naturally ends up with the Cartan charge via twisting.} 

In physical applications, while dealing with sigma models, in addition to the fields above, we have a $\group$-connection on the worldvolume which we will denote by $\Ascr^i_a$.\footnote{ Our notation for sigma models is the following: we will continue to use lower-case Greek indices to correspond to the physical spacetime ${\cal M}$. We assume that there is some worldvolume with intrinsic coordinates $\sigma^a$ with lower-case Latin indices denoting the tangent space indices from the early part of the alphabet ($a,b,\cdots$ etc).  Hopefully our use of intermediate alphabet Latin alphabets $i,j,k, \cdots$ to index the generators of $\mathfrak{g}$ does not cause confusion. }  To complete the story we need to introduce its superpartner $\lambda^i_a$ such that 
\begin{equation}
\begin{split} 
 \QW \Ascr^i_a &=  \lambda^i_a + \partial_a G^i + f^i_{jk} \Ascr^j_a G^k\ , \\
 \QW \lambda^i_a  &= - f^i_{jk} G^j \lambda^k_a -\partial_a \phi^i - f^i_{jk} \Ascr^j_a \phi^k .
\end{split}
\label{eq:WAsf1}
\end{equation}
We will denote the corresponding field strength by $\mathscr{F}^i_{ab}  \equiv \partial_a  \Ascr^i_b-\partial_b  \Ascr^i_a + f^i_{jk}  \Ascr^j_a   \Ascr^k_b$. 
Note that  $\Ibar_k$ annihilates both the worldvolume gauge field and its superpartner, viz.,
\begin{equation}
 \Ibar_k \Ascr^i_a =\Ibar_k \lambda^i_a =0\,.
\label{eq:IAaact}
\end{equation}	
 This follows from the absence of any ghost number $-1$ field that these objects can transform into.
The Cartan model analogue of the above equations is 
\begin{equation}
\begin{split} 
 \QC \Ascr^i_a &\equiv \QW \Ascr^i_a -  \partial_a G^i - f^i_{jk} G^j \Ascr^k_a = \lambda^i_a\ , \\
 \QC \lambda^i_a &\equiv  \QW \lambda^i_a +  f^i_{jk} G^j \lambda^k_a =  -\partial_a \phi^i - f^i_{jk} \Ascr^j_a \phi^k .
\end{split}
\label{eq:CAsf1}
\end{equation}
Note that this shows that $\lambda^i_a$ is like a field strength for $\Ascr^i_a$. This analogy can be made precise in a superspace description of equivariant cohomology which we now turn to.

For completeness, let us also note that there is a third model of equivariant cohomology called the BRST model. 
The idea introduced by Kalkman \cite{Kalkman:1993cm} was to parallel the discussion of BRST symmetries in gauge theories, generalizing the algebra isomorphism of Mathai-Quillen. We will not need this construction in our discussion, but for completeness discuss the construction in Appendix \ref{sec:MQ}.

%~~~~~~~~~~~~~~~~~~~~~~~~~~~~~~~~~~~~~~~~~~~~~~
\section{Equivariant cohomology in superspace}
\label{sec:nt1super}
%~~~~~~~~~~~~~~~~~~~~~~~~~~~~~~~~~~~~~~~~~~~~~~

It is useful to rewrite the formalism discussed in the preceding subsections directly in superspace, where all the transformations have a natural interpretation. We will follow the general framework introduced originally in \cite{Horne:1988yn}.\footnote{ Various authors have described elements of the superspace formalism e.g., \cite{Dijkgraaf:1996tz,Blau:1996bx}, but there does not appear to be a single comprehensive discussion readily available in the literature which covers all that we want. So the discussion here contains some new elements interspersed with the review of known material.}  The idea is to combine the various fields into supermultiplets whose components transform into each other under the action of the cohomology charge.
To this end, we introduce a Grassmann odd coordinate $\thb$ such that $\QW$ acts as the nilpotent supertranslation operator  $\partial_{\thb}$. 

As we are eventually going to be interested in sigma models, let us also introduce worldvolume coordinates $\sigma^a$ which parameterize the Grassmann even section of the superspace. The Grassmann odd section is parameterized by $\thb$.\footnote{ We choose to denote the supercoordinate by $\thb$ to notationally convey that its ghost number is negative $\gh{\thb} = -1$. This is in keeping with the identification with the $\QW$ action.} We will promote all the fields to superfields, i.e., given a field $\Phi(\sigma^a)$ we will denote the corresponding superfield as $\SF{\Phi}(\sigma, \thb)$. We also denote projection using the notation $|$, which projects the superfield to its bottom component by setting $\thb = 0 $; to wit,
\begin{equation}
\SF{\Phi}| \equiv \SF{\Phi}(\sigma,\thb ) |_{\thb=0} = \Phi(\sigma)\,.
\end{equation}	
It is also convenient in writing various formulae to use upper-case Latin indices (from the middle of the alphabet) to denote the superspace directions -- we thus let $z^I = \{\sigma^a, \thb\}$, and let the index 
$I \in \{a,\thb\}$ for simplicity. To keep the discussion simple, we will work with flat worldvolume superspace and will hence not worry about supercurvatures. It is possible to extend our analysis to curved superspace following \cite{DeWitt:1992cy}; we defer this to a future discussion.

Let us take stock of the fields we have in the equivariant cohomology construction. The matter fields parameterizing the target space assemble into a matter superfield $\SF{X}^\mu(\sigma, \thb)$. The ghost fields $\{G^i,\phi^i\}$ form a second multiplet, but it is more efficacious to combine them with the gauge multiplet $\{\Ascr^i_a, \lambda^i_a\}$ to construct a gauge superfield 1-form $\As$:
\begin{equation}
\As \equiv \As_I\, dz^I =  \As_a \, d\sigma^a + \As_{\thb}\, d\thb\,.
\end{equation}	
These superfields admit a Taylor expansion which is adapted to the definition of the Weil charge $\QW \equiv \partial_{\thb}(\ldots)|$ as follows:
\begin{equation}\label{eq:TaylorN1}
\begin{split} 
\SF{X}^\mu &\equiv X^\mu + \thb (\psiC^\mu-G^k \xi_k^\mu) \ ,\\
\As_a^i &\equiv \Ascr_a^i +  \thb \prn{\lambda_a^i+  \partial_a G^i + f^i_{jk} \Ascr^j_a G^k} 
\,,\\ 
\As^i_{\thb} & \equiv G^i + \thb \prn{\phi^i-\half f^i_{jk} G^j G^k} \,.
\end{split}
\end{equation}
The target space coordinates $\SF{X}^\mu$ are viewed as fields which map points and indices on the worldvolume superspace to ordinary target space. We effectively work with an RNS like formalism where the worldvolume supersymmetry is manifest, but the target space supersymmetry is hidden. 

We now should understand the action of various operators, such as the Weil and Cartan derivatives and the interior contraction operation on the fields in superspace. Let us first discuss this for covariant adjoint-valued fields and then understand the action on the other superfields in \eqref{eq:TaylorN1} and objects derived therefrom.

%~~~~~~~~~~~~~~~~~~~~~~~~~~~~~~~~~~~~~~~~~~~~~~~
\subsection{Adjoint superfields} 
\label{sec:adjsf}
%~~~~~~~~~~~~~~~~~~~~~~~~~~~~~~~~~~~~~~~~~~~~~~

Consider first a superfield $\SF{\mathfrak{F}}$ that is adjoint-valued under $\mathfrak{g}$. We can write this in terms of its components along the lines of \eqref{eq:TaylorN1} as 
\begin{equation}
\SF{\mathfrak{F}} =\mathfrak{F} + \thb  \Big(\mathfrak{F}_\psi  -\comm{G}{\mathfrak{F}} \Big) \;\; \Longrightarrow \;\;
\SF{\mathfrak{F}}^i =\mathfrak{F}^i + \thb  \Big(\mathfrak{F}^i_\psi  - f^i_{jk}  \, G^j \, \mathfrak{F}^k \Big) \,.
\label{eq:adjsf}
\end{equation}	
In what follows we will often suppress the Lie algebra index and resort to the usual commutator notation for terms involving the structure constants, e.g., $[\mathfrak{H}, \,\mathfrak{F}]^i = f^i_{jk} \; \mathfrak{H}^j\;  \mathfrak{F}^k$ etc., as indicated.
We first need to understand how the various operators $\QW$, $\Ibar$, etc., act on these adjoint superfields. Let us do this in a somewhat elaborate manner, and then try to distill the essentials into a simple computational algorithm. 

The Weil derivative $\QW$ as presaged should act on superspace as $\partial_{\thb}$. However, this simple rule of thumb makes sense only for the bottom-component of the superfield. To ascertain the action on the top component (the coefficient of $\thb$), we need to construct a new superfield. Let us see this explicitly; first we compute
\begin{align}
\QW \mathfrak{F} =   {\mathfrak F}_\psi - \comm{G}{\mathfrak{F}} \,.
\label{eq:QWF1}
\end{align}
Then requiring that $\QW^2=0$, we learn that 
\begin{equation}
\QW\mathfrak{F}_\psi  = \QW\comm{G}{\mathfrak{F}} = \comm{\QW G}{\mathfrak{F} } - \comm{G}{\QW \mathfrak{F}} = 
\comm{\phi}{\mathfrak{F}} - \comm{G}{\mathfrak{F}_\psi}
\label{eq:QWF2}
\end{equation}	
where we used \eqref{eq:WeilAlg} to evaluate some of the terms. 

To represent \eqref{eq:QWF2} as the action of $\partial_{\thb}$, we need to first construct a superfield $\SF{\mathfrak{F}}_\psi$ whose bottom component is $\mathfrak{F}_\psi$. We cannot do this by ordinary partial differentiation along the supercoordinate $\thb$, since that does not lead to a covariant superfield. We need to construct a derivative that maps covariant objects to covariant ones. To do so, we define the gauge covariant derivative:
\begin{equation}
\Dsf_I = \partial_I + \comm{\SF{\Ascr}_I}{ \,\cdot\;}\, ,
\label{eq:gaugecov}
\end{equation}	
which supplements the action of the usual partial derivatives in superspace with the commutator action of the gauge potential. 

Now consider the superfield $\Dsf_{\thb}$ acting on $\SF{\mathfrak{F}}$, which evaluates to
\begin{equation}
\begin{split}
\Dsf_{\thb} \SF{\mathfrak{F}} 
&= 
	\partial_{\thb}  \left( \mathfrak{F} + \thb  \Big(\mathfrak{F}_\psi  -\comm{G}{\mathfrak{F}} \Big)\right)
	 + \comm{G + \thb \Big(
	 \phi- \half\,\comm{G}{G} \Big)}{\mathfrak{F} + \thb  \Big(\mathfrak{F}_\psi  -\comm{G}{\mathfrak{F}}\Big)} \\
&= 
	\mathfrak{F}_\psi  -\comm{G}{\mathfrak{F}} + \comm{G}{\mathfrak{F}} 
	+ \thb\; \Big(
	\comm{\phi}{\mathfrak{F}} - \half\comm{\comm{G}{G}}{\mathfrak{F}} -\comm{G}{\mathfrak{F}_\psi} 
	+ \comm{G}{\comm{G}{\mathfrak{F}}}
	 \Big)\\
&=
	\mathfrak{F}_\psi  + \thb\; \Big(
	\comm{\phi}{\mathfrak{F}} -\comm{G}{\mathfrak{F}_\psi} \Big) .
\end{split}
\end{equation}	
We then see that we can write:
\begin{equation}
\QW \SF{\mathfrak{F}}|  = \partial_{\thb} \SF{\mathfrak{F}} \,, \qquad 
\QW (\Dsf_{\thb}\SF{\mathfrak{F}}|) =  \partial_{\thb} (\Dsf_{\thb} \SF{\mathfrak{F}}) .
\end{equation}	
This should hopefully make clear how the Weil differential acts on the superfields that transform covariantly as adjoints under $\mathfrak{g}$.

The interior contraction operation $\Ibar$ has a very simple action in that it annihilates the components of the superfield $\mathfrak{F}$, viz., 
\begin{equation}
\Ibar_\alpha \mathfrak{F}  =\Ibar_\alpha \mathfrak{F}_\psi = 0\,,
\end{equation}	
where we recall that $\Ibar_\alpha = \alpha^k \, \Ibar_k$  as defined in \eqref{eq:Ibaral}.

%~~~~~~~~~~~~~~~~~~~~~~~~~~~~~~~~~~~~~~~~~~~~~~~
\subsection{Gauge and matter multiplets} 
\label{sec:gmul}
%~~~~~~~~~~~~~~~~~~~~~~~~~~~~~~~~~~~~~~~~~~~~~~

Having understood the adjoint superfields, we can now turn to the gauge and matter multiplets. Their superfield expansions have already been described in \eqref{eq:TaylorN1}. It is useful to get some intuition for the gauge multiplet which has a four distinct fields. Clearly, $\Ascr_a$ and $G$ are not gauge invariant, as they transform as a gauge potential and gauge transformation parameter respectively, with appropriate inhomogeneous pieces. Their superpartners are however gauge covariant. To see this, let us first define using the gauge covariant derivative  the field strengths. In superspace it is simplest to talk about the super-field strength which can be defined as
\begin{equation}
\Fs_{IJ}  \equiv(1-\frac{1}{2}\, \delta_{IJ}) \left(\partial_I\, \As_J - (-)^{IJ} \,\partial_J\, \As_I  + [\As_I,\As_J]\right)\,.
\label{eq:fdef0}
\end{equation}	
The factor of half is included to keep normalizations simple in future computations.

We can now identify various fields using the projection to the $\thb = 0 $ slice of superspace. To wit, 
\begin{equation}
\begin{aligned} 
\SF{X}^\mu | &\equiv X^\mu\ , &\; 
 \Dsf_{\thb} \SF{X}^\mu | &\equiv \left(\partial_{\thb} \SF{X}^\mu +\As^i_{\thb} \;\xi_i^\mu(\SF{X}) \right)|
 	\equiv \psiC^\mu\ , \\
\As_a |&\equiv  \Ascr_a \ , &  \;
\Fs_{\thb a}|  &\equiv \left( \partial_{\thb} \As_a-\partial_a \As_{\thb}
+ \comm{\As_{\thb}}{ \As_a } \right)| \equiv \lambda_a\ , \\
\As_{\thb}|&\equiv G \ , & \; 
\Fs_{\thb\thb} |  &\equiv \left( \partial_{\thb} \As_{\thb}
+\half \,\comm{ \As_{\thb} }{ \As_{\thb}} \right)| \equiv \phi\ .
\end{aligned}
\end{equation}
As advertised earlier, this presentation of our basic fields makes it manifest that all the fields barring $G$  and $A_a$ are covariantly defined. In particular, the Grassmann partner of the gauge field $\lambda_a$ and the ghost for ghost $\phi$ are components of the super-covariant field strength. The other super-field strength component which is often useful is 
 \begin{equation}
\begin{split} 
\Fs_{ab} |&\equiv  \left(\partial_a  \As_{b}-\partial_b  \As_{a} + \comm{\As_{a}} {  \As_{b} }\right)|
= \mathscr{F}_{ab} \,.
\end{split}
\end{equation}

Using these definitions, the action of $\QW$ can be worked out via $\QW=\partial_{\thb} (\ldots)|$. As before to perform this computation, we have to construct the desired superfield before taking the derivative with respect to $\thb$.

The expressions for $\QW X^\mu$,  $\QW  \Ascr_a$ and $\QW G$ can then be read of from the superspace Taylor expansion which reproduces  the definitions from before (by construction). For the remainder, we have to work out the action of two super-covariant derivatives on superfields, since various components of the transformations as we have seen above involve gauge covariant expressions.

Let us begin by defining
\begin{equation}
\begin{split} 
\Dsf_{\thb} (\Dsf_{\thb} \SF{X}^\mu)  &\equiv  \partial_{\thb} (\Dsf_{\thb} \SF{X}^\mu) + 
 \As^k_{\thb}\ (\Dsf_{\thb} \SF{X}^\nu) \partial_\nu \xi_k^\mu(\SF{X})\ \\
&=  \partial_{\thb} (\Dsf_{\thb} \SF{X}^\mu) +
  \Gamma^\mu{}_{\nu\lambda}(\SF{X})\  (\partial_{\thb} \SF{X}^\lambda) (\Dsf_{\thb} \SF{X}^\nu)  + 
 \As^k_{\thb}\  \Dsf_{\thb} \SF{X}^\nu\ \nabla_\nu \xi_k^\mu(\SF{X}) \ .
\end{split}
\end{equation}
where in the second line, we have covariantized the expression using a Christoffel connection on $\mathcal{M}$. The equality follows from fact that $\Gamma^\mu{}_{\nu\lambda}(\SF{X})\  \Dsf_{\thb} \SF{X}^\lambda\;  \Dsf_{\thb} \SF{X}^\nu =0 $.

A short computation then gives $\Dsf_{\thb}^2 \SF{X}^\mu = \Fs^k_{\thb\thb} \xi_k^\mu(\SF{X})$. Upon projection down onto the $ \thb =0$ component, we get
 \begin{equation}
\begin{split} 
  \partial_{\thb} \Dsf_{\thb} \SF{X}^\mu + 
 \As^k_{\thb}\  \partial_\nu \xi_k^\mu(\SF{X})\  \Dsf_{\thb} \SF{X}^\nu | &= \mathscr{F}^k_{\thb\thb} \ \xi_k^\mu(\SF{X})|  \\
\qquad \Longrightarrow \qquad  \QW \psiC^\mu + G^k \psi^\nu  \partial_\nu \xi_k^\mu &=  \phi^k  \xi_k^\mu\,,
\end{split}
\end{equation}
reproducing the expression in our previous subsections.

The transformation of $\lambda_a^i$ and $\phi^i$ can similarly be deduced from the super-covariant derivatives acting on the super-field strengths. Based on \eqref{eq:gaugecov} we define:
 \begin{equation}
\begin{split} 
\Dsf_c  \Fs_{ab} &\equiv  \partial_c  \Fs_{ab} + \comm{ \As_c}{ \Fs_{ab} }\ ,\qquad
\Dsf_{\thb}  \Fs_{ab} \equiv  \partial_{\thb}  \Fs_{ab} + \comm{ \As_{\thb} }{ \Fs_{ab}}\,, \\
\Dsf_c \Fs_{\thb a} &\equiv  \partial_c \Fs_{\thb a} + \comm{ \As_c} {\Fs_{\thb a} }\ ,\qquad
\Dsf_{\thb} \Fs_{\thb a} \equiv  \partial_{\thb} \Fs_{\thb a} + \comm{ \As_{\thb}}{ \Fs_{\thb a}} \,,\\
\Dsf_c \Fs_{\thb\thb} &\equiv  \partial_c \Fs_{\thb\thb} + \comm{ \As_c} {\Fs_{\thb\thb}} \ ,\qquad
\Dsf_{\thb} \Fs_{\thb\thb} \equiv  \partial_{\thb} \Fs_{\thb a} + \comm{ \As_{\thb} }{\Fs_{\thb\thb}} \ .
\end{split}
\end{equation}
Not all of these field strengths are independent. They satisfy super-Bianchi identities of the form
\begin{equation}
\begin{split} 
\Dsf_c  \Fs_{ab} + \Dsf_a  \Fs_{bc}+ \Dsf_b  \Fs_{ca} &= 0 \,,\qquad
\Dsf_{\thb}  \Fs_{ab} - \prn{\Dsf_a \Fs_{\thb b}-\Dsf_b \Fs_{\thb a}}  =0\,,\\
\Dsf_{\thb} \Fs_{\thb a} +  \Dsf_a \Fs_{\thb\thb}  &= 0\,,\qquad 
\Dsf_{\thb} \Fs_{\thb\thb} =0\ .
\end{split}
\end{equation}
Using these, $\Dsf_{\thb}$ acting on any  super-field strength can be traded for $\Dsf_a$ acting on  super-field strengths. Under the  restriction $|$ , $\Dsf_a$ then reduces to ordinary covariant derivative. The restriction of the last three  identities then yields
\begin{equation}
\begin{split} 
\QW \mathscr{F}_{ab} +\comm{ G }{\mathscr{F}_{ab}}- \prn{ \partial_a \lambda_b +\comm{ \Ascr_a}{ \lambda_b} - (a\leftrightarrow b)} & =0\,,\\
\QW \lambda_a +\comm{ G^j}{ \lambda^k_a}+ \partial_a \phi +\comm{\Ascr_a}{ \phi} &= 0\,,\\
\QW  \phi + \comm{ G}{ \phi} &=  0 \ .
\end{split}
\end{equation}
This completes the superspace derivation of expressions we had before. 

\paragraph{Cartan model in superspace:}
While the equations above were derived for the Weil model, shifting to Cartan model of equivariant cohomology is extremely straightforward. One can check from the above expressions that the construction would be equivalent to shifting to a description only in terms of super-covariant derivative  $\Dsf_{\thb}$ eschewing the use of $\partial_{\thb}$ everywhere. This is one of the many advantages of the superspace formulation, which makes things a lot more simple than working with the component fields where the gauge symmetry is not clearly manifest.

We define $\QC=\Dsf_{\thb}(\ldots)|$. The non-nilpotence of $\QC$ then is simply the  familiar non-nilpotence of covariant exterior derivatives which square to curvatures instead. Going through the exercise we obtain the expressions quoted earlier, viz.,
\begin{equation}
\begin{split} 
\QC X^\mu &= \Dsf_{\thb} \SF{X}^\mu | = \psiC^\mu\,, \\  
 \QC \psiC^\mu &= \Dsf^2_{\thb} \SF{X}^\mu | = \Fs^k_{\thb\thb} \, \xi_k^\mu(\SF{X})|=  \phi^k \xi_k^\mu \\
 \QC \Ascr_a &=  \Fs_{\thb a}| =  \lambda_a  \,, \\
 \QC \lambda_a &=  \Dsf_{\thb} \Fs_{\thb a} |  = -  \Dsf_a \Fs_{\thb\thb} |
 =  -\partial_a \phi - \comm{\Ascr_a}{ \phi}\,.
\end{split}
\end{equation}
It is then clear from $\Dsf_{\thb}^2 \sim \SF{\mathscr{F}}_{\thb\thb}$ that $\QC^2$ evaluates to a $\group$-action along $\phi$ as expected for the Cartan supercharge, cf., \eqref{eq:CartanWeil1}.

%~~~~~~~~~~~~~~~~~~~~~~~~~~~~~~~~~~~~~~~~~~~~~~~
\subsection{Graded operations in superspace}
\label{sec:superLie}
%~~~~~~~~~~~~~~~~~~~~~~~~~~~~~~~~~~~~~~~~~~~~~~

We have thus far understood the superspace counterparts of the various statements reviewed in \S\ref{sec:weiln1}. In particular, we found that on covariant superfields, the Weil and Cartan charges act as 
\begin{equation}
\QW = \partial_\thb \,, \qquad \QC = \SF{\mathcal{D}}_\thb \,.
\end{equation}
It is useful to have an independent discussion of the superspace analog of Lie derivation and interior contraction, which can be combined in a single Lie super-operator.  
Consider the following super-interior contraction and super-Lie derivative:
\begin{equation}\label{eq:superLie}
\begin{split}
  \SF{\Ibar}_k & \equiv \Ibar_k + \thb {\cal L}_k  \,, \qquad
  \SF{\cal L}_k \equiv {\cal L}_k \,,
\end{split}
\end{equation}
where the operators on the right hand side are understood to act component-wise on superfields. This super-operator satisfies
\begin{equation}
\begin{split}
   \SF{\Ibar}_k \SF{\mathfrak{F}}^i = \SF{\Ibar}_k \SF{X}^\mu= \SF{\Ibar}_k \As_a^i=0 \,, \qquad
   \SF{\Ibar}_k \As_\thb^i = -  \delta_k^i \,.
\end{split}
\end{equation}
and hence generalizes $\Ibar_k$ in a sensible way to superspace. 
Furthermore, it is immediate to see the following commutation relations:
\begin{equation}
\begin{split}
&\QW^2 = \partial_\thb^2= 0\,,\qquad  \gradcomm{\SF{\Ibar}_i}{\SF{\Ibar}_j} = 0 \,, \qquad\qquad\quad\,  \gradcomm{\QW}{\SF{\Ibar}_k} = \partial_\thb \SF{\Ibar}_k = \SF{\cal L}_k \,,\\
 &\;\;\gradcomm{\gradcomm{\QW}{\SF{\Ibar}_i}}{\SF{\Ibar}_j} = - f^k_{ij} \, \SF{\Ibar}_{k} \,, \qquad\;\;
  \gradcomm{\QW}{\gradcomm{\QW}{\SF{\Ibar}_k}} = \partial_\thb^2 \SF{\Ibar}_k = 0 \,,
 \end{split}
 \end{equation}
which we recognize as the superspace version of the Weil algebra \eqref{eq:LieSAlgParam}.

Further, we can study the algebra involving super-covariant derivatives. It is an easy exercise to check the following on covariant superfields:
\begin{equation}
\gradcomm{\Dsf_{\thb}}{\SF{\mathcal{I}}_k} = 0 \,.
\label{eq:sCa}
\end{equation}	
%

%~~~~~~~~~~~~~~~~~~~~~~~~~~~~~~~~~~~~~~~~~~~~~~~
\subsection{Super-gauge transformations and Wess-Zumino gauge}
\label{sec:n1wz}
%~~~~~~~~~~~~~~~~~~~~~~~~~~~~~~~~~~~~~~~~~~~~~~

The passage from the Weil to the Cartan algebra involves eschewing working with gauge potentials and instead using gauge covariant data which naturally incorporate the space of basic forms. In terms of the action of various operators the main non-trivial element in the Weil algebra is the presence of the ghost $G$, which picks out a direction of gauge transformation (recall that it has a non-trivial interior contraction). It is efficacious to pass onto  a gauge which fixes this Grassmann odd field to zero, and then work with gauge covariant objects. We will refer to this gauge fixing choice as  the 
\emph{Wess-Zumino (WZ) gauge}.

Before explaining the WZ gauge, let us briefly introduce the concept of a super-gauge transformation. By this we mean the most general gauge transformation allowed by the superspace structure, i.e., one where the gauge parameter itself is an adjoint superfield. To this end, let us lift an arbitrary gauge parameter, which has so far played a passive role, to an adjoint superfield
\begin{equation}
\SF{\Lambda} = \Lambda + \thb \, (\QW \Lambda) \equiv \Lambda + \thb \, \left( \Lambda_\psi - \comm{G}{\Lambda} \right) \,.
\end{equation}
An infinitesimal gauge transformation along $\SF{\Lambda}$ can more formally be defined as simply a Lie derivative of the form
\begin{equation}
\begin{split}
 \SF{{\cal L}}_{\SF{\Lambda}} \SF{\mathfrak{F}} &\equiv \gradcomm{\QW}{\SF{\Lambda}^k\,\SF{\Ibar}_{k}} \SF{\mathfrak{F}}  = \comm{\SF{\Lambda}}{\SF{\mathfrak{F}}} \,,\\
 \SF{{\cal L}}_{\SF{\Lambda}} \As_\thb &\equiv \gradcomm{\QW}{\SF{\Lambda}^k\,\SF{\Ibar}_{k}} \As_\thb = \comm{\SF{\Lambda}}{\As_\thb} - (-)^\Lambda \, \left(\Lambda_\psi-\comm{G}{\Lambda}\right) = - (-)^\Lambda \, \SF{\cal D}_\thb \SF{\Lambda}\,,\\
 \SF{{\cal L}}_{\SF{\Lambda}} \As_a &\equiv \gradcomm{\QW}{\SF{\Lambda}^k\,\SF{\Ibar}_{k}} \As_a = \comm{\SF{\Lambda}}{\As_a} - (-)^\Lambda \, \partial_a \SF{\Lambda} = -(-)^\Lambda\, \SF{\cal D}_a \SF{\Lambda}\,,
 \end{split}
\end{equation}
where we allowed for the possibility of $\Lambda$ being Grassmann-odd. 

It will sometimes be useful to have a version of the commutator relations for Lie derivatives along adjoint superfields as defined above. Clearly, because various operators now act on the gauge parameter itself, the algebra changes. We give some of the covariant identities here for reference:
\begin{equation}
\label{eq:CommsCartan}
\Dsf_{\thb}^2 = \SF{\cal L}_{\SF{\mathscr{F}}_{\thb\thb}} \,,\qquad
\gradcomm{\Dsf_{\thb}}{\SF{\mathcal{L}}_{\SF{\Lambda}}} = \SF{\mathcal{L}}_{\Dsf_\thb \SF{\Lambda}}  \,, 
\qquad
\gradcomm{ \SF{\cal L}_{\SF{\Lambda}_1}}{\SF{\cal L}_{\SF{\Lambda}_2}} = \SF{\mathcal{L}}_{ \comm{ \SF{\Lambda}_1 }{ \SF{\Lambda}_2}  } \,.
\end{equation}	

\paragraph{Transformation of adjoint superfields:}
Let us consider such gauge transformations in more detail for the case of an adjoint superfield $\SF{\mathfrak{F}}$. Under a super-gauge transformation as above, it transforms as 
\begin{equation}
\begin{split}
\SF{\mathfrak{F}} &\mapsto  \SF{\mathfrak{F}} + \comm{\SF{\Lambda}}{\SF{\mathfrak{F}} }  \\
&=  \SF{\mathfrak{F}} +
\comm{\Lambda + \thb\, \left(\Lambda_\psi - \comm{G}{\Lambda}\right)}{\mathfrak{F} + \thb\, \left(\mathfrak{F}_\psi - \comm{G}{\mathfrak{F}}\right)} \\
& = \mathfrak{F} + \comm{\Lambda}{\mathfrak{F}} + \thb \left\{ \mathfrak{F}_\psi - \comm{G-\Lambda_\psi}{\mathfrak{F}} - \comm{G}{\comm{\Lambda}{\mathfrak{F}}} + (-)^\Lambda \comm{\Lambda}{\mathfrak{F}_\psi} \right\} \,.
\end{split}
\label{eq:NT1gt}
\end{equation}	
Clearly, if we pick a super-gauge parameter with vanishing bottom component ($\Lambda=0$), then  the transformation can be expressed as 
\begin{equation}
\SF{\mathfrak{F} }\mapsto \SF{\mathfrak{F}} \Big\{ G \mapsto G - \Lambda_\psi\Big\}  \,, \qquad 
 \prn{\begin{array}{c} \mathfrak{F} \\ \mathfrak{F}_{\psi}  \end{array}}  \; \; \text{fixed}
\label{eq:FFP}
\end{equation}	
i.e., we simply replace $G$ by $G-\Lambda_\psi$ in the superspace expansion $\SF{\mathfrak{F}} = \mathfrak{F} + \thb (\mathfrak{F}_\psi - \comm{G}{\mathfrak{F}})$.  Thus the super-gauge transformed superfield can be obtained by absorbing the gauge transformation into the ghost field $G$  keeping the fundamental components fixed. 

We find it intuitive to adopt standard language and think of $G$ as a Faddeev-Popov (FP) gauge parameter field.  The subset of gauge transformations which shifts this field and leaves the fundamental components intact can then we viewed as \emph{FP boosts}. These are generated by $\SF{\Lambda} = \thb\, \Lambda_\psi$. 
On the other hand, the class of non-trivial gauge transformations that leave $G$ invariant, but change the building blocks of $\SF{\mathfrak{F}} $ can then be called as \emph{FP rotations}. Geometrically this is analogous to the usual decomposition of the Poincar\'e group on massive spin fields as comprising of a little group of inertial frame rotations combined with translations and boosts that help choose the rest frame.  More precisely, we can think of the transformations of superfields using the method of induced representations whereby FP rotations that leave $G$ invariant can be used to induce the whole group $\group$ of gauge transformations.

 While the above describes infinitesimal gauge transformations, we will shortly need finite gauge transformations, as well. These are obtained by exponentiation: 
\begin{equation}
 \SF{\mathfrak{F}} \, \mapsto \, 
 \exp\left({\cal L}_{\SF{\Lambda}}\right) \SF{\mathfrak{F}} = \SF{\mathfrak{F}}+\comm{\SF{\Lambda}}{\SF{\mathfrak{F}}} + \frac{1}{2!}\comm{\SF{\Lambda}}{\comm{\SF{\Lambda}}{\SF{\mathfrak{F}}}} + \frac{1}{3!} \comm{\SF{\Lambda}}{\comm{\SF{\Lambda}}{\comm{\SF{\Lambda}}{\SF{\mathfrak{F}}}}} + \ldots  \,,
\end{equation}
which is the usual adjoint action. For a FP boost with vanishing bottom component $\Lambda=0$, this truncates at first order and yields 
\begin{equation}
 \SF{\mathfrak{F}} \, \mapsto \,  \SF{\mathfrak{F}}+\comm{\SF{\Lambda}}{\SF{\mathfrak{F}}}  \,.
\end{equation}
Hence, for adjoint superfields, infinitesimal FP boosts are the same as finite boosts. This will not be quite true for other cases, as we will see below. 

\paragraph{Transformation of the gauge field:}
Similar discussions apply to the gauge and matter superfields, with the inhomogeneous transformation being important in the case of the gauge potentials. A finite transformation of the gauge field now reads
\begin{equation}
 \As_J \, \mapsto \, 
 \exp\left({\cal L}_{\SF{\Lambda}}\right) \As_J = \As_J - (-)^\Lambda \, \SF{\cal D}_\thb \SF{\Lambda} - \frac{1}{2!} \, \comm{\SF{\Lambda}}{(-)^\Lambda \, \SF{\cal D}_\thb \SF{\Lambda}} + \frac{1}{3!} \, \comm{\SF{\Lambda}}{\comm{\SF{\Lambda}}{(-)^\Lambda \, \SF{\cal D}_\thb \SF{\Lambda}}} + \ldots \,,
\end{equation}
In particular, for the FP boost generated by $\SF{\Lambda} = \thb\, \Lambda_\psi$ this truncates at second order. The FP boost again induces a simple transformation, which only shifts the ghosts, but leaves fixed the covariant components of the gauge superfield one-form:
\begin{equation}
\As \mapsto \As\Big\{ G \mapsto G - \Lambda_\psi\Big\}  \,, \qquad 
 \prn{\begin{array}{c} \Ascr_a \\ \lambda_a  \end{array}}  \ \& \ \phi
 \; \; \text{fixed}
\label{eq:AsFP}
\end{equation}	

\paragraph{Wess-Zumino gauge:}
We can now use these insights to fix a gauge by performing a boost transformation along the ghost $G$, i.e., we pick a gauge parameter $\SF{\Lambda}_{_{FP}} = \thb\, G$. According to Eqs.\ \eqref{eq:FFP} and \eqref{eq:AsFP} this has the effect of setting the ghost to zero.
The superfield expansions for  various fields in the WZ gauge are simple:
\begin{equation}
\begin{split}
(\SF{\mathfrak{F}})_{_{WZ}} &= \mathfrak{F} + \thb \, \mathfrak{F}_\psi \,,\\
(\As)_{_{WZ}} &= (\As_a)_{_{WZ}} \, d\sigma^a + (\As_\thb)_{_{WZ}} \, d\thb = \left( \Ascr_a + \thb \, \lambda_a \right) \, d\sigma^a + \left( \thb \, \phi \right) \, d\thb \,, \\
(\SF{X}^\mu)_{_{WZ}} &= X^\mu + \thb\, \psiC^\mu\,.
\end{split}
\end{equation} 
The passage to the WZ gauge (analogous to the inertial frame choice) makes it clear that it is sufficient to restrict attention to covariant building blocks  $\mathfrak{F}, \mathfrak{F}_\psi$ of  $ \SF{\mathfrak{F}}$ or the corresponding elements in the other fields. 
Once we know how they transform we can always obtain the contribution to the $\thb$ term arising from ghosts as above. We can use this to introduce some convenient notation, where we give the action of various operations on superfields as maps between covariant building blocks. 

\paragraph{Component maps:}
To give a simple representation of various operations as maps between covariant components of superfields, we first assume WZ gauge as explained above. To distinguish these maps from the operators introduced so far, we use a slightly different notation. Schematically, we replace the maps $\{\QW,\QC,\Ibar_k,{\cal L}_k\}$ defined hitherto by a set of equivalent operations as follows:
\begin{equation}
 \QW \longrightarrow \dSK \,,\qquad \QC \longrightarrow \DSK \,,\qquad \Ibar_k \longrightarrow \ibmap_k \,,\qquad {\cal L}_k \longrightarrow \lmap_k \,.
\end{equation}
The label `${\rm SK}$' does not have a particular meaning at this point. The logic behind this name will become clear later. 
To define the action of these operators, we use a convenient notation where we abbreviate any superfield by a component vector of its covariant components. For instance, the superfield $\SF{\mathfrak{F}}$ would just be abbreviated as a column vector $\prn{\begin{array}{c} \mathfrak{F} \\ \mathfrak{F}_{\psi} \end{array}}$. The Weil/Cartan derivatives, interior contraction, and the Lie derivation can now be represented as maps from covariant superfields to covariant superfields, which act on the building blocks of as follows:
\begin{subequations}
\begin{align}
\dSK : \prn{\begin{array}{c} \mathfrak{F}^i \\ \mathfrak{F}_{\psi}^i  \end{array}} &\mapsto
\prn{\begin{array}{c}  \mathfrak{F}_\psi^i   \\ 0 \end{array}}\ , 
\qquad\quad
\DSK:
	\prn{\begin{array}{c} \mathfrak{F}^i \\ \mathfrak{F}_{\psi}^i   \end{array}}
	\mapsto
	\prn{\begin{array}{c}  \mathfrak{F}_\psi^i \\\relax f^i_{jk}\phi^j\mathfrak{F}^k
 \end{array}} 
 \label{eq:N1QWadj} \\ 
 \ibmap_k:
	\prn{\begin{array}{c}  \mathfrak{F}^i\\\relax \mathfrak{F}_{\psi}^i  \end{array}} 
	&\mapsto
	\prn{\begin{array}{c}  0 \\\relax - f^i_{kj} \mathfrak{F}^j \end{array}} , 
\qquad
\lmap_k :\ 
	\prn{\begin{array}{c}  \mathfrak{F}^i\\\relax \mathfrak{F}_{\psi}^i  \end{array}} 
	\mapsto
	\prn{\begin{array}{c} f^i_{kj} \mathfrak{F}^j \\\relax f^i_{kj} \mathfrak{F}_{\psi}^j 
	\end{array}}\,,
\label{eq:N1ILadj} 
\end{align}
\end{subequations}
 In Wess-Zumino gauge, these maps simply implement the action of $\{\QW,\QC,\Ibar_k,{\cal L}_k\}$ component-wise.
 We can similarly write down expressions for the rest of the fields.\footnote{ The difference between $\{\QW,\QC,\Ibar_k,{\cal L}_k\}$ and $\{\dSK,\DSK,\ibmap_k,\lmap_k\}$ becomes only clear if one computes commutation relations: since the maps in \eqref{eq:N1QWadj}, \eqref{eq:N1ILadj} implement a `left' action, their commutators will differ by an overall sign from those of $\{\QW,\QC,\Ibar_k,{\cal L}_k\}$, which act from the `right'. For example, $\lmap_k = - \gradcomm{\dSK}{\ibmap_k}$.} 
We note the following useful relations:
\begin{equation}\label{eq:DSKexpl}
\DSK = \dSK - \phi^k\, \ibmap_k \,, \qquad \DSK^2 = \phi^k \, \lmap_k\,.
\end{equation}

For the standard $\mathcal{N}_T=1$ equivariant cohomology that we have been discussing, we don't seem to gain much with this presentation. The pay-offs are greater in the case of extended equivariant cohomology when the WZ gauge allows much simpler treatment of the algebraic construction. We turn to this subject now.

%~~~~~~~~~~~~~~~~~~~~~~~~~~~~~~~~~~~~~~~~~~~~~~~
\section{Extended equivariant cohomology}
\label{sec:extequivariance}
%~~~~~~~~~~~~~~~~~~~~~~~~~~~~~~~~~~~~~~~~~~~~~~

We have thus far seen the construction of the equivariant cohomology algebra with a single supercharge $\QW$ or equivalently $\QC$. While this suffices for some physical problems, we have seen that the Schwinger-Keldysh construction equips us with a bigrading, since for every supercharge like $\QSK$ we have a conjugate operation $\QSKb$ with opposite ghost number. The natural setting for such a graded algebra is the so-called {\it extended equivariant cohomology algebra}, which has more than one generator.  We will now review aspects of this story for a pair of generators which corresponds to the case of interest.

%~~~~~~~~~~~~~~~~~~~~~~~~~~~~~~~~~~~~~~~~~~~~~~~
\subsection{Basic field content}
\label{sec:nt2}
%~~~~~~~~~~~~~~~~~~~~~~~~~~~~~~~~~~~~~~~~~~~~~~

Let us start by attempting to double the number of BRST generators in equivariant cohomology construction of \S\ref{sec:equivariance}. We would like a pair of Weil supercharges $\QW$ and $\QWb$ which are individually nilpotent and anti-commuting with each other.  To get a sense of the structure and the algebra of the generators we will work with the gauge sector first, i.e., construct appropriate representatives for the fields we wish to use in the universal $\group$-bundle $\text{E}_\group$ and subsequently add in the various matter fields. The main difference from the earlier presentation is that we are going to directly pass onto superspace where the action of the charges is much simpler to explain. The  algebraic construction underlying our description is explained quite clearly in the original discussion of \cite{Dijkgraaf:1996tz}. The superspace discussion we employ below for motivating this construction was described  in \cite{Blau:1996bx} building on their earlier work. We  also refer the reader \cite{Zucchini:1998rz} for a useful perspective on the algebraic construction.

Since we have a natural interpretation of the action of $\QW$ in superspace, as the derivative along the super-coordinate $\thb$, it follows that inclusion of a second supercharge necessitates enlarging the superspace to include another Grassmann-odd direction. We will coordinatize the second direction by $\theta$ so that
\begin{equation}
\QW = \partial_{\thb}(\ldots)| \,, \qquad \QWb = \partial_\theta(\ldots)|\,.
\end{equation}	
Thus our superspace has two Grassmann-odd directions and so we upgrade all superfields to be functions of the Grassmann-even coordinates $\sigma^a$ as well as $\theta, \thb$. We then require by obvious extension of the notation introduced in \S\ref{sec:nt1super} that our superfields satisfy:
\begin{equation}
\SF{\mathfrak{F}}(z) = \SF{\mathfrak{F}} | +\theta\, (\QWb \SF{\mathfrak{F}})| + \thb \, (\QW \SF{\mathfrak{F}})| + \thb \theta \,
(\QWb \QW \SF{\mathfrak{F}})| 
\label{eq:nt2sf}
\end{equation}	
where we have upgraded $z = \{\sigma^a, \theta, \thb\}$ and will use upper-case mid-alphabet Latin indices $I,J,\cdots$ to take values in $\{a,b,\cdots \} \cup \{\theta,\thb\}$.

Given this superspace we can now proceed to analyzing the gauge sector of the theory, i.e., find a useful parameterization of the universal $\group$-bundle $\text{E}_\group$. We will proceed algebraically and construct local representatives for the Weil and Cartan models.  A necessary consequence of the doubling of the supercharges is that we have many more component fields. In particular, if we consider the  super-gauge field which is the primary character in the construction of 
equivariant cohomology, we have a dodecuplet of fields, since the full gauge superfield one-form  can be written as\footnote{ To keep notational clutter to a minimum we are suppressing the Lie algebra index in the following.}
\begin{equation}
 \As = \As_I \, dz^I = \As_a \, d\sigma^a + \As_\theta \, d\theta + \As_{\thb} \, d\thb\,.
\label{eq:A1fsf}
\end{equation}	
Each of the $\As_I$ admits a superfield expansion as in \eqref{eq:nt2sf} with four components, which altogether gives us the desired dodecuplet.

To understand the structure it is useful to once again introduce the gauge covariant derivative which allows the definition of the field strength. Firstly, as earlier, we pick the gauge covariant derivative to act as 
\begin{equation}
 \Dsf_I = \partial_I + [\As_I,\ \cdot\;] \,,
\label{eq:covDI}
\end{equation}	
which implies that the field strength is given by
\begin{equation}
\Fs_{IJ}  \equiv(1-\frac{1}{2}\, \delta_{IJ}) \left(\partial_I\, \As_J - (-)^{IJ} \,\partial_J\, \As_I  + [\As_I,\As_J]  
\right) .
\label{eq:fdef}
\end{equation}	
In addition it is also convenient to define the following non-covariant object:
\begin{align}
\Bs _{\theta\thb} \equiv \partial_\theta\As_{\thb} + \frac{1}{2}\, [\As_\theta,\As_{\thb}] \,.
\end{align}
This will be useful to pick out a particular non-gauge covariant field. Our definition explicitly breaks the symmetry between $\theta$ and $\thb$, for under an exchange of these super-coordinates we have 
$\Bs_{\theta\thb}  \mapsto \Fs_{\theta\thb} - \Bs_{\theta\thb}$.

As in our discussion of $\mathcal{N}_T = 1$ superalgebra, the  covariant derivatives along the super-coordinates $\{\theta,\thb\}$ can be treated as the Cartan charges: $\QC = \Dsf_{\thb}(\ldots)|\,, \QCb = \Dsf_\theta(\ldots)|$. This  enables us to use the  field strengths to define the covariant fields that we will employ to parameterize the extended cohomological structure. Paralleling that analysis the bottom component of the superfields $\As_\theta$ and $\As_{\thb}$ will be taken to be ghost fields (with opposite ghost numbers). The other components of the gauge multiplets can be filled out in terms of the field strengths.  

We find it useful to parameterize the super-components (the ghosts and  ghosts for ghosts) in the following fashion. First we pick out the gauge non-invariant combinations and use them to define the various ghost fields (i.e., ghost valued connection forms):
\begin{align}
\As_{\thb}|  \equiv  G \,, \qquad \As_\theta| \equiv \bar{G} \,, \qquad \Bs_{\theta\thb}| \equiv B \,.
\label{eq:triplet}
\end{align}
We will refer to these fields as the {\em Faddeev-Popov ghost triplet}. These fields will be gauge parameters, generalizing the single Faddeev-Popov ghost field of the $\mathcal{N}_T =1$ superalgebra.

The remaining five fields which make up the superfields $\As_\theta$ and $\As_{\thb}$ are captured and denoted as follows:
\begin{align}
& \Fs_{\thb\thb}| \equiv \phi\,,\qquad \Fs_{\theta\theta}|\equiv \overline{\phi} \,,\qquad
  \Fs_{\theta\thb}| \equiv \phi^0 \,,
\nonumber \\
&\qquad\Dsf_{\thb} \Fs_{\theta\thb}| \equiv \eta \,, \qquad
\Dsf_{\theta} \Fs_{\theta\thb}| \equiv \overline{\eta} \,.
\label{eq:quintet}
\end{align}
We will henceforth refer to them as the {\em Vafa-Witten ghost of ghost quintet} following \cite{Vafa:1994tf} where this structure was first described. Finally, we have four more (vector) fields in the gauge potential $\SF{\mathscr{A}}_a$, making up the {\em vector quartet}
 which we parameterize in a covariant fashion as 
\begin{align}
\As_a| \equiv \Ascr_a \,, \qquad \Fs_{\thb a}| \equiv \lambda_a \,, \qquad \Fs_{\theta a}| \equiv \overline{\lambda}_a \,, \qquad
\Dsf_\theta \Fs_{\thb a}| \equiv \mathcal{F}_a \,.
\label{eq:quartet}
\end{align}

The ghost charge assignments for these fields can be worked out once we pick a convention that assigns
\begin{align}
\gh{\theta} = +1 \,, \qquad \gh{\thb } = -1 \,,
\label{eq:ghththb}
\end{align}
which leads to the following ghost numbers as indicated in Table \ref{tab:multiplets}.
\begin{table}[h!]
\centering
\begin{tabular}{|| c || c | c | c ||c|| }
\hline\hline
  {\shadeR{ghost}} & {\shadeB{Faddeev-Popov}}& {\shadeB{Vafa-Witten ghost}} & {\shadeB{Vector}} \\
  {\shadeR{charge}} & {\shadeB{ghost triplet}}  & {\shadeB{of ghost quintet}} & {\shadeB{quartet}} \\
  \hline
  {\shadeR{2}} &          & $\qquad\phi\qquad\qquad$ & \\
  {\shadeR{1}} & $G\qquad$                          & $\qquad\eta$      & $\lambda_a$ \\
  {\shadeR{0}} & $\qquad B$ & $\phi^0\qquad$       & $\quad\mathscr{A}_a \qquad \mathcal{F}_a\quad$\\
  {\shadeR{-1}}& $\bar G\qquad$               & $\qquad\bar\eta$& $\bar{\lambda}_a$       \\
  {\shadeR{-2}}&      & $\bar\phi\qquad$ & \\
\hline\hline
\end{tabular}
\caption{Dodecuplet of basic fields in the $\mathcal{N}_T = 2$ superalgebra and their respective ghost number assignments.}
\label{tab:multiplets}
\end{table}

This collection of fields parameterizes for us the extended gauge multiplet \eqref{eq:A1fsf} which embodies the action of the two supercharges, and can be thought of as the algebraic parameterization of the classifying space subject to such a bi-grading.  We will work out the complete parameterization for the superfields once we understand the superalgebra structure.

% We leave it as an easy exercise to take the above covariant parameterization of the twelve component fields and work out from there the superspace Taylor expansion of $\SF{\mathscr{A}}_I$ (c.f., the $\mathcal{N}_T=1$ version \eqref{eq:TaylorN1}). 

%~~~~~~~~~~~~~~~~~~~~~~~~~~~~~~~~~~~~~~~~~~~~~~~
\subsection{The $\mathcal{N}_\smallT =2 $ superalgebra}
\label{sec:nt2alg}
%~~~~~~~~~~~~~~~~~~~~~~~~~~~~~~~~~~~~~~~~~~~~~~

We are now in a position to describe the so called $\mathcal{N}_\smallT = 2$ superalgebra which generalizes  \eqref{eq:LieSAlg}.
Firstly, realize that we would a-priori expect to have two interior contraction operators $\mathcal{I}_k$ and 
$\Ibar_k$ which would be required to satisfy (reinstating the Lie algebra index):
\begin{align}
\Ibar_k G^j = - \delta^j_k  \,, \qquad {\mathcal{I}}_k \bar{G}^j = - \delta^j_k \,,
\end{align}
generalizing \eqref{eq:WeilAlg2} in an obvious manner. These two interior contraction operators are as before Grassmann odd and we choose to give them ghost charge $\gh{\Ibar}  =-1$ and $\gh{{\mathcal{I}}} = +1$ respectively.

However, among the dodecuplet of fields introduced in the gauge superfield $\As_I$ we also encounter a third gauge parameter, which we have captured in the gauge non-covariant field $B$. We therefore have a third interior contraction operator which we will denote as $\mathcal{I}^0$ which we choose to act non-trivially on the $B$ fields, viz., 
\begin{align}
\mathcal{I}^0_k B^j  = -\delta^j_k \,,\qquad \mathcal{I}^0_k G^j = 0 = \mathcal{I}^0_k \bar{G}^j \,.
\end{align}
We also find that the other two interior contractions act non-trivially on the field $B$:
\begin{align}
\Ibar_k B^j = -\frac{1}{2}\, f^j_{k\ell} \,\bar{G}^\ell  \,, \qquad  {\mathcal{I}}_k B^j = \frac{1}{2} \, f^j_{k\ell}\, G^\ell \,.
\end{align}
One can check that these definitions lead to the expected identity
\begin{equation}
{\cal L}_i B^k \equiv \gradcomm{\QW}{\Ibar_i} B^k \equiv \gradcomm{\QWb}{{\cal I}_i} B^k = f^k_{ij} B^j \,.
\end{equation}

The triplet of operators $\{ \Ibar, \mathcal{I}^0, {\mathcal{I}}\} $ together generate an $\mathfrak{sl}(2)$ algebra. Our parameterization of the fields is not quite $\mathfrak{sl}(2)$ covariant, but this can easily be achieved by a simple change of basis.\footnote{ The description of the extended equivariant cohomology algebra in an $\mathfrak{sl}(2)$ covariant fashion can be found in \cite{Dijkgraaf:1996tz,Blau:1996bx,Zucchini:1998rz}. Our choices are dictated by simplifying some of the analysis in the context of physical applications. For example, in \S\ref{sec:langevin} we will argue that dissipative effects arise when the ghost number zero field in the Vafa-Witten quintet picks up a vacuum expectation value which is easier to implement in the $\mathfrak{sl}(2)$ non-covariant presentation; cf., footnote \ref{fn:cptphi0}.
 \label{fn:sl2}}

\paragraph{Weil model of $\mathcal{N}_T=2$ algebra:}
With the interior contraction operators at hand we can generate the Lie derivation $\mathcal{L}_k$ as in \eqref{eq:liederID} by commuting with the Weil charges (which as always act as exterior derivatives). This then generates the $\mathcal{N}_\smallT =2 $ extended equivariant cohomology algebra which we can abstractly write as   
\begin{equation}
\begin{split}
\QW^2 = \QWb^2  &=\gradcomm{\QW}{\QWb} = 0 \\ 
\gradcomm{\QW}{\Ibar_j} = \gradcomm{\,\QWb}{{\mathcal{I}}_j} 
= \mathcal{L}_j  \,, & \qquad 
\gradcomm{\QW}{{\mathcal{I}}_j} = \gradcomm{\,\QWb}{\Ibar_j} = 0\\
\gradcomm{\QW}{\mathcal{I}^0_j} = {\mathcal{I}}_j\,, & \qquad 
  \gradcomm{\,\QWb}{\mathcal{I}^0_j} =- {\Ibar}_j \\
 \gradcomm{\QW}{\mathcal{L}_j} &= \gradcomm{\,\QWb}{\mathcal{L}_j} = 0 \\
 \gradcomm{\Ibar_i}{{\mathcal{I}}_j} &= f_{ij}^k\, \mathcal{I}^0_k \\
 \gradcomm{\mathcal{L}_i}{\Ibar_j} = -f_{ij}^k\,\Ibar_{k}\,, \qquad 
   \gradcomm{\mathcal{L}_i}{{\mathcal{I}}_j} &= -f_{ij}^k\,{\mathcal{I}}_{k}\,, \qquad
   \gradcomm{\mathcal{L}_i}{\mathcal{I}^0_j} = - f_{ij}^k\,\mathcal{I}^0_{k} \,.
\end{split}
\label{eq:LieSAlg2}
\end{equation}	
While the last two lines simply describe the $\mathfrak{sl}(2)$ structure of interior contractions and the obvious action of Lie derivatives respectively, we find it again quite useful to illustrate the first four lines (i.e., the cohomology of Weyl charges) in a diagrammatic form:
\begin{equation}
\begin{tikzcd}
& \mathcal{L}_j    &   \\
{\mathcal{I}}_j\arrow{ru}[below]{\quad\;\;\QWb} & & \Ibar_j \arrow{lu}[below]{\!\!\!\!\!\QW}\\
&  \mathcal{I}^0_j \arrow{lu}[above]{\quad\;\;\QW} \arrow{ru}{\quad\;\;\;\;-\QWb}&
\end{tikzcd}
\label{eq:LieSAlgDiag}
\end{equation}
where the action of operators on arrows is understood to be via graded commutators.
The reader can ascertain that the structure of \eqref{eq:LieSAlg2} closely resembles that of the original construction in \eqref{eq:LieSAlg} with the increased supercharges and interior contractions. Generalizations to higher number of topological supercharges is straightforward and are discussed in \cite{Dijkgraaf:1996tz}.

\paragraph{Extended equivariant algebra in superspace:}
Let us now encode all of the above relation in a compact superspace notation. We start by defining the super-contraction operators (in analogy with \eqref{eq:superLie}):
\begin{equation}\label{eq:NT2superops}
\begin{split}
  \SF{{\cal I}}^0_k & \equiv {\cal I}^0_k  + \thb \, {\cal I}_k- \theta\, \Ibar_k + \thb \theta \, {\cal L}_k \,,\\
  \SF{\cal I}_k &\equiv {\cal I}_k + \theta \, {\cal L}_k \,,\\
  \SF{\Ibar}_k &\equiv \Ibar_k + \thb \, {\cal L}_k \,,\\
  \SF{{\cal L}}_k &\equiv {\cal L}_k \,.
\end{split}
\end{equation}
We have the following identities:
\begin{equation}
\begin{split}
 \gradcomm{\QWb}{\SF{{\cal I}}^0_k} &= \partial_\theta \SF{{\cal I}}^0_k = - \SF{\Ibar}_k \,,\qquad
 \gradcomm{\QW}{\SF{{\cal I}}^0_k} = \partial_\thb \SF{{\cal I}}^0_k = \SF{{\cal I}}_k \,, \quad\;\;\\
\gradcomm{\QW}{\SF{\Ibar}_k} &= \partial_\thb \SF{\Ibar}_k = \gradcomm{\QWb}{\SF{{\cal I}}_k} = \partial_\theta \SF{{\cal I}}_k = \SF{\cal L}_k \,, \\
  \gradcomm{\QWb}{\SF{\Ibar}_k} &= \gradcomm{\QW}{\SF{{\cal I}}_k} = \gradcomm{\SF{{\cal I}}^0_j}{\SF{{\cal I}}^0_k} = 0 \,,
\end{split}
\end{equation}
which could also be organized in a diagram similar to \eqref{eq:LieSAlgDiag}. 
By expanding out these relations component-wise, we can easily verify the algebra \eqref{eq:LieSAlg2}. 
 
\paragraph{Cartan model of $\mathcal{N}_T=2$ algebra:}
With these definitions of the $\mathcal{N}_\smallT=2$ algebra we can pass onto the Cartan construction, which eschews the gauge connections, parameterized here by the Faddeev-Popov triplet $\{\bar G, B , G\}$ in favour of the physical fields in the vector quartet and ghost of ghost quintet respectively. In analogy with Eq.\ \eqref{eq:CartanWeil1}, we start by defining Cartan differentials on the full Weil complex:\footnote{ These expressions can most easily be derived by demanding the consistency condition $\QC G^k = \QC \overline{G}^k = \QC B^k = \QCb G^k = \QCb \overline{G}^k = \QCb B^k = 0$.}
\begin{equation}
\begin{split}
\QC &= \QW + \left(\phi^k+\frac{1}{2}\comm{G}{G}^k\right) \, \Ibar_k + \left((\phi^0)^k-B^k + \frac{1}{2} \comm{\overline{G}}{G} \right) \, {\cal I}_k \\
&\qquad\;\; +  \left(\eta^k-\comm{\phi+\frac{1}{2}\comm{G}{G}}{\overline{G}}^k + \comm{G}{B-\phi^0}^k\right)\, \mathcal{I}^0_k + G^k \, {\cal L}_k \,,\\
\QCb &= \QWb +  \left(\overline{\phi}^k + \frac{1}{2} \comm{\overline{G}}{\overline{G}} \right)  \, {\mathcal{I}}_k  + \left( B^k + \frac{1}{2} \comm{\overline{G}}{G}^k \right)\Ibar_k \\
&\qquad\;\; + \left(\comm{\overline{\phi} + \frac{1}{2} \comm{\overline{G}}{\overline{G}} }{G}^k+ \comm{\overline{G}}{B}^k \right) {\cal I}^0_k + \overline{G}^k \, {\cal L}_k\,.
\end{split}
\label{eq:QCQCbfull}
\end{equation}
such that $\QC = \Dsf_\thb(\ldots)|$ and $\QCb = \Dsf_\theta (\ldots)|$. 
We can now again pass to the Cartan model by restricting the full Weil complex to the symmetric algebra of $\mathfrak{g}^*$. This amounts to setting $G^k=B^k=\overline{G}^k=0$ and the Cartan differentials then simply read
\begin{equation}
\begin{split}
\QC &= \QW + \phi^k \, \Ibar_k + (\phi^0)^k \, {\cal I}_k +  \eta^k\, \mathcal{I}^0_k \,,\\
\QCb &= \QWb +  \overline{\phi}^k \, {\mathcal{I}}_k \,.
\end{split}
\label{eq:QCQCb}
\end{equation}
Given the algebra \eqref{eq:LieSAlg2} we can immediately check that $\QC$ and $\QCb$ are no longer nilpotent, but rather generate gauge transformations as before along $\phi$ and $\overline{\phi}$ respectively:
\begin{equation}
\QC^2 = \phi^k \, {\cal L}_k \,,\qquad
\QCb^2 = \overline{\phi}^k \, {\cal L}_k \,.
\end{equation}
 This can be stated most compactly by passing to superspace, where we have the following relations:
\begin{equation}
\begin{split}
&\Dsf_\thb^2 = \SF{\mathcal{L}}_{\SF{\mathscr{F}}_{\thb\thb}} \,, \qquad \Dsf_\theta^2 = \SF{\mathcal{L}}_{\SF{\mathscr{F}}_{\theta\theta}}  \,, 
\qquad 
\gradcomm{\Dsf_\thb}{\Dsf_\theta} = \SF{\mathcal{L}}_{\SF{\mathscr{F}}_{\theta\thb}} \,.
\end{split}
\label{eq:sCa2}
\end{equation}	
We can extend these relations to involve the action of super-gauge transformations along arbitrary superfields in the spirit of \eqref{eq:CommsCartan}:
\begin{equation}
\gradcomm{\Dsf_\theta}{\SF{\mathcal{L}}_{\SF{\Lambda}}} = \SF{\mathcal{L}}_{\Dsf_\theta \SF{\Lambda}} \,, \qquad \gradcomm{\Dsf_\thb}{\SF{\mathcal{L}}_{\SF{\Lambda}}} = \SF{\mathcal{L}}_{\Dsf_\thb \SF{\Lambda}} \,, \qquad \gradcomm{ \SF{\cal L}_{\SF{\Lambda}_1}}{\SF{\cal L}_{\SF{\Lambda}_2}} = \SF{\mathcal{L}}_{ \comm{ \SF{\Lambda}_1 }{ \SF{\Lambda}_2 } } \,.
\end{equation}

\paragraph{Generalities:} 
A few words about this algebraic structure are in order. We have indicated that in general we can associate with any Lie algebra $\mathfrak{g}$ a Lie superalgebra $\hat{\mathfrak{g}}$ which is simply a graded structure with elements having grading $\{0,\pm1\}$, via the identification
\begin{equation}
\hat{\mathfrak{g}} = \mathfrak{g}_{-1}^{\scriptscriptstyle\Ibar} \oplus 
\mathfrak{g}_0^{\scriptscriptstyle\mathcal{L}} \oplus 
\mathfrak{g}_1^{\scriptscriptstyle{\QW}}\,,
\end{equation}	
 where  the generators of $\mathfrak{g}_{-1}$ are the interior contraction operations, the Lie derivations generate $ \mathfrak{g}_0$, and $\mathfrak{g}_1$ is thence generated by some nilpotent exterior derivation. This is the structure of the $\mathcal{N}_T=1$ algebra \eqref{eq:LieSAlg}. 

 This algebraic structure is being refined in the $\mathcal{N}_\smallT =2$ case  with a further sub-grading of the Grassmann-odd elements. To wit, one may formally think of the $\mathcal{N}_\smallT = 2$ algebra defined in 
 \eqref{eq:LieSAlg2} as an iterated superspace construction (cf., \cite{Dijkgraaf:1996tz})
\begin{equation}
\hat{\hat{\mathfrak{g}}} = \mathfrak{g}_{(-1,0,1)}^{\scriptscriptstyle{\Ibar} , \mathcal{I}^0, {\mathcal{I}} } \oplus 
\mathfrak{g}_0^{\scriptscriptstyle\mathcal{L}} \oplus \mathfrak{g}_{(-1,1)}^{\scriptscriptstyle{\QWb,\QW}} \,,
\end{equation}	
where we have indicated in order the gradings of the interior contractions, the Lie derivation, and finally the cohomological charges.

From this formal viewpoint any algebraic structure to which we can associate such a grading of generators allows an interpretation in terms of the extended equivariant cohomology algebra. In \S\ref{sec:sknt2} we proceed to argue that this is indeed the case for the Schwinger-Keldysh BRST algebra based on the field redefinition redundancies and the KMS conditions, as reviewed earlier in \S\ref{sec:quadruplet}.

%~~~~~~~~~~~~~~~~~~~~~~~~~~~~~~~~~~~~~~~~~~~~~~~
\subsection{$\mathcal{N}_T =2$ superfields}
\label{sec:nt2sf}
%~~~~~~~~~~~~~~~~~~~~~~~~~~~~~~~~~~~~~~~~~~~~~~

We have already indicated that we can express the fields of the $\mathcal{N}_T=2$ superalgebra in superspace parameterized by $\theta$ and $\thb$. What remains for us to describe is the explicit component forms of these fields. We will first do this explicitly without any gauge fixing and then indicate the simplifications engendered by passing onto the WZ gauge. The discussion for the most part will parallel our story in \S\ref{sec:nt1super}; the main novelty is the possibility of moving into two distinct (anti-commuting) super-directions.

Let us start with a gauge covariant superfield, whose general form is given in \eqref{eq:nt2sf}. We would like to identify the  fundamental building blocks of this superfield and isolate the contributions which arise from the Faddeev-Popov ghost triplet. For the $\theta$ and $\thb$ components, one can pretty much guess at the behaviour based on two independent Grassmann directions. The new ingredient is to ascertain the contribution to the $\thb \theta$ term at the top level. 
We can a-priori decide to designate the labels $\{ \mathfrak{F}_\psi, \mathfrak{F}_{\psib}, \tilde{\mathfrak{F}} \} $ for the covariant components carrying ghost charge $\{1,-1,0\}$ respectively.
We can write:
\begin{equation}
\mathfrak{F}_\psi  = \Dsf_{\thb} \SF{\mathfrak{F}}| \,, \qquad \mathfrak{F}_{\psib}  = \Dsf_{\theta} \SF{\mathfrak{F}}|
\,, \qquad \tilde{\mathfrak{F}} = \Dsf_{\thb} \Dsf_{\theta} \SF{\mathfrak{F}}|\,.
\end{equation}	
By examining our definitions of the FP triplet fields and a bit of algebra one can verify that the superfield expansion of such a covariant adjoint superfield takes the form:
\begin{equation}
\begin{split}
\SF{\mathfrak{F}} &= 
	\mathfrak{F}  + \thb \Bigl\{\mathfrak{F}_{\psi} - \comm{G}{\mathfrak{F}} \Bigr\}  
	+\theta \Bigl\{\mathfrak{F}_{\psib} - \comm{{\bar G}}{\mathfrak{F}} \Bigr\}  \\
& \qquad	+\; \thb \theta \Bigl\{\tilde{\mathfrak{F}} -\comm{B}{\mathfrak{F}} + \comm{G}{\mathfrak{F}_{\psib}} - \comm{{\bar G}}{\mathfrak{F}_\psi} + \frac{1}{2} \comm{{\bar G}}{\comm{G}{\mathfrak{F}}} 
	- \frac{1}{2} \comm{G}{\comm{{\bar G}}{\mathfrak{F}}} \Bigr\}\,.
\end{split}
\label{eq:F2SF}
\end{equation}	
Basically, all  terms with the correct ghost number involving the FP triplet appear in the above expression. One however needs to go through the algebra to ascertain the  coefficients. This exercise can be automated using symbolic packages; we have found it useful to carry out the exercise in Mathematica (using Matthew Headrick's Grassmann package).

One can similarly work out the expressions for the gauge potentials. This requires computing not just the curvatures but derivatives of the curvature since the fields that appear at the top level involve two covariant derivatives on the gauge field. A longer computational exercise results in (nb: we define the gauge covariant derivative $D_a \equiv \partial_a + \comm{\Ascr_a}{\cdot}$ )
\begin{subequations}
\begin{align} 
\As_a \equiv&\ 
	\Ascr_a + \thb  \bigbr{ \lambda_a + D_a G  } + \theta \bigbr{  \bar{\lambda}_a + D_a \bar{G}  } 
\nonumber \\	
& 
\qquad	+\;\thb   \theta \;\Bigl\{ \mathcal{F}_a +D_a B + \comm{G}{\bar{\lambda}_a}
	- \comm{\bar{G}}{\lambda_a}  
	+\half \comm{G}{D_a \bar{G} }-  \half \comm{\bar{G} }{D_a G} \Bigr\} ,
\label{eq:AaExp2} \\
\As_\thb \equiv&\ 
	 G + \thb  \bigbr{ \phi-  \half \comm{G}{G}  } + \theta \bigbr{ B-  \half \comm{\bar{G} }{G} } 
\nonumber \\
&
\qquad	 -\; \thb  \theta \bigbr{ \eta +  \comm{\bar{G} }{\phi} - \comm{G}{B }+ \half \comm{G}{\comm{\bar{G}}{G}}   } ,
\label{eq:AthbExp2}  \\
\As_\theta \equiv&\ 
	\bar{G}  + \theta  \bigbr{ \bar{\phi}-  \half \comm{\bar{G} }{\bar{G} }  } 
	+ \thb  \bigbr{\phi^0- B-  \half \comm{\bar{G} }{G} } 
\nonumber \\
& 
\qquad	+\;\thb   \theta \bigbr{\etab +  \comm{G}{\bar{\phi}} -  \comm{\bar{G} }{\phi^0- B}
	+ \half  \comm{\bar{G} }{ \comm{\bar{G} }{G}}  } .
\label{eq:AthExp2}  
\end{align}
\end{subequations}
We won't write out the position multiplet for the present; the reader can find explicit expressions later in \S\ref{sec:langevin} for a one-dimensional target space.

%~~~~~~~~~~~~~~~~~~~~~~~~~~~~~~~~~~~~~~~~~~~~~~~
\subsection{Wess-Zumino gauge and component maps}
\label{sec:WZ2}
%~~~~~~~~~~~~~~~~~~~~~~~~~~~~~~~~~~~~~~~~~~~~~~

Now that we have the superfields we can work out the gauge transformations. We will be brief since everything works in complete analogy to the discussion in \S\ref{sec:n1wz}. On an adjoint superfield we find the analog of \eqref{eq:NT1gt}, which now has a more elaborate expansion in components.  Despite their intricacy it nevertheless follows that a general gauge transformation can be decomposed as before into Faddeev-Popov boosts and rotations. The gauge transformation by superfield $\SF{\Lambda}$ (generically expanded as in \eqref{eq:F2SF}) with vanishing bottom component would act on covariant objects as
\begin{equation}
\SF{\mathfrak{F} }\mapsto \SF{\mathfrak{F}} 
\Bigg\{ 
\begin{array}{c}
G \mapsto G - \Lambda_\psi  \\
\bar{G} \mapsto \bar{G} - \Lambda_{\psib}  \\
B \mapsto B - \tilde{\Lambda} + \half\,\comm{\Lambda_{\psib}}{G} - \half\,\comm{\Lambda_{\psi}}{\bar{G}}
\end{array}
\Bigg\} 
 \,, \qquad 
 \prn{\begin{array}{c} 
 \mathfrak{F} \\ \mathfrak{F}_{\psi}  \\  \mathfrak{F}_{\psib} \\ \tilde{\mathfrak{F}} 
 \end{array}}  \; \; \text{fixed.}
\label{eq:FFP2}
\end{equation}	
Similarly we can check that 
\begin{equation}
\begin{split} 
\As_J &\mapsto \As_J
\bigbr{\begin{array}{c} 
G \mapsto G- \Lambda_\psi, \\ 
\bar{G} \mapsto \bar{G} - \Lambda_{\psib},\\ 
B \mapsto B- \tilde{\Lambda}
 + \half \comm{\Lambda_{\psib}}{G}- \half \comm{\Lambda_{\psi}}{\bar{G} }
 \end{array} } 
 \quad \text{with} \quad 
 \prn{\begin{array}{c} 
 \Ascr_a  \\ \lambda_a \\ \bar{\lambda}_a \\ \mathcal{F}_a 
 \end{array}} 
\& 
\prn{\begin{array}{c} 
\phi \\ \eta \\ \phi^0 \\ \etab \\ \bar{\phi} 
\end{array}}
\quad \text{fixed.}
\end{split}
\end{equation}
Therefore by choosing a gauge parameter 
\begin{equation}
\LamS_{_{FP}} = \thb \, G + \theta \, \bar{G} + \thb \theta\, B
\label{eq:fplam}
\end{equation}	
we can eliminate the contribution of the FP triplet $\{G,\bar{G} ,B\}$ appearing in the superfield expansions.

Thus as indicated we can absorb the gauge transformation by superfields with no bottom component into shifts of the FP triplet. Once again we may pass onto the WZ gauge by picking an appropriate gauge parameter to annihilate the FP triplet, based on the transformations  above. As one can readily appreciate modulo extra fields, the picture here is completely analogous to the discussion at the end of \S\ref{sec:superLie}.  In the WZ gauge, a generic covariant superfield and the connection 1-form take the simple forms
\begin{equation}
\begin{split}
(\SF{\mathfrak{F}})_{_{WZ}} &= \mathfrak{F} + \theta \, \mathfrak{F}_{\psib} + \thb \, \mathfrak{F}_\psi + \thb\theta \,  \tilde{\mathfrak{F}} \,, \\
(\As)_{_{WZ}} & \equiv (\As_a)_{_{WZ}} \, d\sigma^a + (\As_\theta)_{_{WZ}} \, d\theta + (\As_\thb)_{_{WZ}} \, d\thb \\
 &= \left( \Ascr_a + \theta \, \overline{\lambda}_a + \thb \, \lambda_a + \thb\theta \, \mathcal{F}_a \right) d\sigma^a
     + \left( \theta \, \overline\phi + \thb \, \phi^0 + \thb\theta\, \overline{\eta} \right) d\theta
     + \left( \thb \, \phi - \thb \theta \, \eta \right) d\thb \,.
\end{split}
\end{equation}
It is helpful write these out in component notation as a column vector as depicted in \eqref{eq:FFP2}. 

\paragraph{The field strength superfields:} For instance the building blocks of superfields which encode the vector quartet can be captured in the connection and field strength superfields as
\begin{subequations}
\begin{align}
&  
\As_a \ : \ 
\prn{\begin{array}{c}
\Ascr_a \\ \lambda_a \\ \bar{\lambda}_a \\ \mathcal{F}_a 
\end{array}} \qquad \qquad \qquad
 \Dsf_\theta\Fs_{\thetab a} \ :\ 
\prn{\begin{array}{c}
\mathcal{F}_a \\ \comm{\phi^0}{\lambda_a}-  D_a \eta -\comm{\phi}{\bar{\lambda}_a}\\  \comm{\bar{\phi}}{\lambda_a} \\ \comm{\phi^0}{\mathcal{F}_a } + \comm{\bar{\phi}}{D_a\phi} +\comm{\etab}{\lambda_a}  
\end{array}} 
\\
& 
\Fs_{\thetab a}\ :\ 
\prn{\begin{array}{c}  
\lambda_a \\ - D_a\phi  \\  \mathcal{F}_a \\ D_a \eta + \comm{\phi}{\bar{\lambda}_a}
\end{array}} \qquad \qquad
\Fs_{\theta a}\ :\ 
\prn{\begin{array}{c}  
\bar{\lambda}_a  \\ - D_a\phi^0- \mathcal{F}_a  \\  -D_a\bar{\phi} \\ \comm{\phi^0}{\bar{\lambda}_a} -  D_a \etab -
\comm{\bar{\phi}}{\lambda_a} 
\end{array}} 
 \end{align}
\end{subequations}
while the Vafa-Witten quintet fits into the super-components of the field strengths and derivatives thereof:
\begin{subequations}
\begin{align}
&  
\Fs_{\thetab \thetab }\ :\ 
\prn{\begin{array}{c}  
\phi \\ 0 \\  -\eta \\ 0 
\end{array}} \qquad \qquad \quad
\Fs_{\theta \thetab }\ :\ 
\prn{\begin{array}{c}  
\phi^0 \\ \eta \\  \etab \\ \comm{\phi}{\bar{\phi}}  
\end{array}} \qquad \qquad \quad
\Fs_{\theta \theta }\ :\ 
\prn{\begin{array}{c}  
\bar{\phi} \\   -\etab \\ 0 \\ \comm{\phi^0}{\bar{\phi}}
\end{array}}
\\
& 
\Dsf_{\thetab } \Fs_{\theta\thetab }= -\Dsf_{\theta } \mathscr{F}_{\thetab\thetab } \ :\ 
\prn{\begin{array}{c}  
\eta  \\ \comm{\phi}{\phi^0} \\  \comm{\phi}{\bar{\phi}} \\ \comm{\phi}{\etab}- \comm{\eta}{\phi^0} 
\end{array}} \qquad\qquad
\Dsf_{\theta} \Fs_{\theta\thetab }=-\Dsf_{\thetab } \Fs_{\theta\theta } :\ 
\prn{\begin{array}{c}  
\etab \\- \comm{\phi}{\bar{\phi}} \\  \comm{\bar{\phi}}{\phi^0} \\ -\comm{\bar{\phi}}{\eta}\ 
\end{array}}
 \end{align}
 \label{eq:Fvectors}
\end{subequations}
Finally the field strength itself has the component decomposition 
\begin{equation}
\Fs_{ab}\ :\ 
\prn{\begin{array}{c} 
\mathscr{F}_{ab} \\ 2 \,D_{[a}\lambda_{b]}  \\ 2\, D_{[a}\bar{\lambda}_{b]}  \\ 2 \,D_{[a} \mathcal{F}_{b]} +
\comm{\lambda_a}{\bar{\lambda}_b} -\comm{\bar{\lambda}_a}{\lambda}  
\end{array}} 
\\
\end{equation}	
These expressions should suffice for any further computation, since we can again encode the action of the derivations as maps on the component fields directly. 

\paragraph{Component maps for graded operators:}
We give here the maps that allow the evaluation of the Weil, super-Lie and Cartan operators on covariant superfields. As in \S\ref{sec:n1wz}, we wish to carefully distinguish the operator algebra of $\{\QW,\QWb,\QC,\QCb,\Ibar_k,{\cal I}_k,{\cal I}^0_k,{\cal L}_k\}$ from their representation as maps between covariant components of superfields. To this end we introduce equivalent avatars of these operators, which act more conveniently on superfields via
\begin{equation}
\begin{split}
 \QW &\longrightarrow \dSK \,,\qquad 
\QWb \longrightarrow \dSKb \,,\qquad 
 \QC \longrightarrow \DSK \,,\qquad 
\QCb \longrightarrow \DSKb \,,\\
{\cal I}^0_k &\longrightarrow \izmap_k \,,\qquad\;\;\;\,
{\cal I}_k \longrightarrow \imap_k \,,\qquad\;\;
\Ibar_k \longrightarrow \ibmap_k \,,\qquad\quad
{\cal L}_k \longrightarrow \lmap_k \,.
\end{split}
\end{equation}
The action on component fields of all the right hand side maps is just the one induced by the left hand side operations. For instance:
\begin{equation}
\begin{split}
\dSK\ :\ \prn{\begin{array}{c} \mathfrak{F} \\ \mathfrak{F}_{\psi} \\ \mathfrak{F}_{\psib} \\  \tilde{\mathfrak{F}} \end{array}} & \mapsto
\prn{\begin{array}{c}  \mathfrak{F}_\psi \\ 0\\  \tilde{\mathfrak{F}}\\   0
 \end{array}} \,,\qquad\,
 \DSK:\ \prn{\begin{array}{c} \mathfrak{F} \\ \mathfrak{F}_{\psi} \\ \mathfrak{F}_{\psib} \\  \tilde{\mathfrak{F}} \end{array}} \mapsto
\prn{\begin{array}{c}  \mathfrak{F}_\psi \\ \comm{\phi}{\mathfrak{F}} \\  \tilde{\mathfrak{F}}\\   \comm{\phi}{\mathfrak{F}_{\psib}} - \comm{\eta}{\mathfrak{F}}
 \end{array}}
 \,,
\\
\dSKb\ :\ \prn{\begin{array}{c} \mathfrak{F} \\ \mathfrak{F}_{\psi} \\ \mathfrak{F}_{\psib} \\  \tilde{\mathfrak{F}} \end{array}} &\mapsto
 \prn{\begin{array}{c}  \mathfrak{F}_{\psib} \\ - \tilde{\mathfrak{F}}\\ 0\\     0   \end{array}} \,,
 \qquad
  \DSKb:\ \prn{\begin{array}{c} \mathfrak{F} \\ \mathfrak{F}_{\psi} \\ \mathfrak{F}_{\psib} \\  \tilde{\mathfrak{F}} \end{array}} \mapsto
   \prn{\begin{array}{c}  \mathfrak{F}_{\psib} \\ \comm{\phi^0}{\mathfrak{F}}- \tilde{\mathfrak{F}}\\ \comm{\bar\phi}{\mathfrak{F}} \\    
\comm{\phi^0}{\mathfrak{F}_{\psib}}-\comm{\bar\phi}{\mathfrak{F}_{\psi}}  + \comm{\etab}{\mathfrak{F}}
  \end{array}} \,.
\end{split}
\end{equation}
Note that the superfield maps $\{\DSK,\DSKb\}$ are simply representations of the covariant derivatives $\{\Dsf_\thb,\Dsf_\theta\}$. In a similar spirit, we can write down the component map representation of $\Dsf_a$, which we denote as follows:
\begin{equation}\label{eq:DamapAdj}
\begin{split}
{\cal D}_a\ :\ \prn{\begin{array}{c} \mathfrak{F} \\ \mathfrak{F}_{\psi} \\ \mathfrak{F}_{\psib} \\  \tilde{\mathfrak{F}} \end{array}} &\mapsto
\prn{\begin{array}{c} D_a\mathfrak{F} \\ D_a\mathfrak{F}_{\psi}+\comm{\lambda_a}{\mathfrak{F}} \\ 
D_a \mathfrak{F}_{\psib}+\comm{\bar{\lambda}_a}{\mathfrak{F}}  \\ 
D_a  \tilde{\mathfrak{F}} + \comm{\mathcal{F}_a}{\mathfrak{F}} +\comm{\bar{\lambda}_a}{\mathfrak{F}_\psi}-\comm{\lambda_a}{\mathfrak{F}_{\psib}}
\end{array}} \ ,
\end{split}
\end{equation}
Similarly, the interior contraction and Lie derivation are captured by the maps:
\begin{equation}
\begin{split}
 \ibmap_k : 
	\prn{\begin{array}{c} \mathfrak{F}^i \\ \mathfrak{F}_{\psi}^i \\ \mathfrak{F}_{\psib}^i \\  \tilde{\mathfrak{F}}^i \end{array}}
	&\mapsto
	\prn{\begin{array}{c}  0 \\\relax -f^i_{kj}\mathfrak{F}^j \\ 0 \\ -f^i_{kj}\mathfrak{F}^j_{\psib} \end{array}}\ , 
\; \quad\;\;\,
 \izmap_k: 
	\prn{\begin{array}{c} \mathfrak{F}^i \\ \mathfrak{F}_{\psi}^i \\ \mathfrak{F}_{\psib}^i \\  \tilde{\mathfrak{F}}^i \end{array}}
	\mapsto
	\prn{\begin{array}{c}  0 \\\relax 0 \\ 0  \\ f^i_{kj}\mathfrak{F}^j  \end{array}} 
 \\
  \imap_k : 
	\prn{\begin{array}{c} \mathfrak{F}^i \\ \mathfrak{F}_{\psi}^i \\ \mathfrak{F}_{\psib}^i \\  \tilde{\mathfrak{F}}^i \end{array}}
	&\mapsto
	\prn{\begin{array}{c}  0 \\\relax 0 \\-f^i_{kj}\mathfrak{F}^j \\ f^i_{kj}\mathfrak{F}^j_\psi \end{array}}\ , 
 \qquad 
 \lmap_k: 
	\prn{\begin{array}{c} \mathfrak{F}^i \\ \mathfrak{F}_{\psi}^i \\ \mathfrak{F}_{\psib}^i \\  \tilde{\mathfrak{F}}^i \end{array}}
	\mapsto
	\prn{\begin{array}{c} f^i_{kj}\mathfrak{F}^j \\\relax f^i_{kj}\mathfrak{F}^j_\psi \\ f^i_{kj}\mathfrak{F}^j_{\psib} \\ f^i_{kj}\tilde{\mathfrak{F}}^j  \end{array}}
\end{split}
 \label{eq:N1LadjWZ}
\end{equation}	
These relations encode all the transformations of covariantly defined objects (including the Vafa-Witten quintet and the vector quartet via the definitions \eqref{eq:quintet} and \eqref{eq:quartet}) and hence complete our discussion of the $\mathcal{N}_T=2$ algebra. To summarize, the algebra \eqref{eq:LieSAlg2} now reads\footnote{ Note again opposite signs in commutators due to right vs.\ left action.}
\begin{equation}
\begin{split}
\dSK^2 = \dSKb^2  &=\gradcomm{\dSK}{\dSKb} = 0 \\ 
\gradcomm{\dSK}{\ibmap_j} = \gradcomm{\dSKb}{\imap_j} 
= - \lmap_j  \,, & \qquad 
\gradcomm{\dSK}{\imap_j} = \gradcomm{\dSKb}{\ibmap_j} = 0\\
\gradcomm{\dSK}{\izmap_j} = - \imap_j\,, & \qquad 
  \gradcomm{\dSKb}{\izmap_j} =  \ibmap_j \\
 \gradcomm{\dSK}{\lmap_j} &= \gradcomm{\dSKb}{\lmap_j} = 0 \\
 \gradcomm{\ibmap_i}{\imap_j} &= -f_{ij}^k\, \izmap_k \\
 \gradcomm{\lmap_i}{\ibmap_j} = f_{ij}^k\,\ibmap_{k}\,, \qquad 
   \gradcomm{\lmap_i}{\imap_j} &= f_{ij}^k\,\imap_{k}\,, \qquad
   \gradcomm{\lmap_i}{\izmap_j} = f_{ij}^k\,\izmap_{k}
\end{split}
\label{eq:LieSAlg3}
\end{equation}	

The algebra itself can be succinctly summarized by writing the gauge covariant derivations as combinations of ordinary (Weil) derivations and super-Lie derivations along the connection superfield in the WZ gauge. The analog of \eqref{eq:DSKexpl}, which defines the component map representation of the covariant (Cartan) differentials, now takes the form
\begin{equation}\label{eq:DSKexpl2}
\begin{split}
\DSK &= \dSK - \phi^k \,\ibmap_k  - \eta^k \,\izmap_k \,,\\
\DSKb &= \dSKb - (\phi^0)^k \,\ibmap_k- \overline{\phi}^k\, \imap_k + \etab^k \,\izmap_k \,.
\end{split}
\end{equation}
 These relations encode the same information as the Cartan differentials in \eqref{eq:QCQCb}.\footnote{ Note, however, that \eqref{eq:DSKexpl2} appears to differ from \eqref{eq:QCQCb} at a superficial level. The difference is, of course, again due to \eqref{eq:QCQCb} giving a relation on the operator algebra (i.e., operators with a `right' action), while \eqref{eq:DSKexpl2} relates component maps on superfields (i.e., operators with a `left' action).}

However, composing various maps to compute covariant derivatives is now much easier: we simply take the maps as defined above and compose them in the obvious way. As a simple example, we immediately find the following relations for the covariant differentials:\footnote{  In writing these expressions we have to pass the Grassmann-odd fields $\eta$ and $\etab$ outside the superfield they act on. This results in a change of sign, so we can mnemonically think of 
\begin{equation}
\eta^k \,\imap_k : \prn{\begin{array}{c} \mathfrak{F}^i \\ \mathfrak{F}_{\psi}^i \\ \mathfrak{F}_{\psib}^i \\  \tilde{\mathfrak{F}}^i \end{array}}
	\mapsto
	\prn{\begin{array}{c}  0 \\\relax 0 \\  \comm{\eta}{\mathfrak{F}}  \\ \comm{\eta}{\mathfrak{F}_\psi } \end{array}} 
\end{equation}	
}
\begin{equation}
\begin{split}
& \DSK^2 = \phi^k \, \lmap_k - \eta^k \, \imap_k \,,  \qquad 
\DSKb^2 = \overline{\phi}^k \, \lmap_k - \etab^k \, \ibmap_k + \comm{\phi^0}{\overline{\phi}}^k \, \izmap_k \,,\\
&\quad\gradcomm{\DSK}{\DSKb} = (\phi^0)^k \, \lmap_k + \eta^k \,\ibmap_k + \etab^k \, \imap_k + \comm{\phi}{\overline{\phi}}^k \, \izmap_k \,.
\end{split}
\end{equation}
This is equivalent to (but much easier to verify than) the operator algebra relations \eqref{eq:sCa2}. Once more, we recognize this as the statement that covariant derivatives square to gauge transformations along $\{\phi,\overline{\phi},\phi^0\}$ on the subspace of horizontal forms.

\newpage
%~~~~~~~~~~~~~~~~~~~~~~~~~~~~~~~~~~~~~~~~~~~~~~~
\part{Thermal equivariant cohomology}
\label{part:thermal}
%~~~~~~~~~~~~~~~~~~~~~~~~~~~~~~~~~~~~~~~~~~~~~~
\hspace{1cm}

%~~~~~~~~~~~~~~~~~~~~~~~~~~~~~~~~~~~~~~~~~~~~~~~
\section{SK-KMS thermal equivariant cohomology algebra}
\label{sec:sknt2}
%~~~~~~~~~~~~~~~~~~~~~~~~~~~~~~~~~~~~~~~~~~~~~~

Now that we have extensively reviewed the well-known framework of equivariant cohomology, the time has come to demonstrate our basic claim: {\it The SK-KMS superalgebra \eqref{eq:kmsalg} or equivalently \eqref{eq:kmsalg2} encoding symmetries of thermal Schwinger-Keldysh theories can be understood as a special instance of an extended equivariant cohomology algebra.} 

Rather crucially, the equivariant gauge symmetry, which acts on the algebra of quantum operators and their low energy realizations, acts differently from the action on spacetime coordinates $X^\mu$ as introduced in \eqref{eq:lieX}.  The distinction lies in the fact that the usual discussion gauges the group action in  target space, whereas we will end up gauging a worldvolume action in thermal equivariant cohomology.

%~~~~~~~~~~~~~~~~~~~~~~~~~~~~~~~~~~~~~~~~~~~~~~~
\subsection{Identifying SK-KMS symmetries as $\mathcal{N}_\smallT =2$ algebra }
%~~~~~~~~~~~~~~~~~~~~~~~~~~~~~~~~~~~~~~~~~~~~~~

The operators
involved are the BRST charges arising from the SK field redefinitions and KMS conditions,  $\{\QSK, \QSKb,\QKMS,\QKMSb\}$  which are all Grassmann-odd, along with the Grassmann-even generators of thermal translations $\{\Qbeta,\Qzero\}$. While the SK supercharges naturally behave as super-derivations, we have also explained how the KMS operations can be combined to act as superspace operators \eqref{eq:SKsuperops}, which express the action of $\{\QKMS,\QKMSb,\Qbeta,\Qzero\}$ conveniently. We  picked a rather suggestive notation for these  which we reproduce here for convenience: 
\begin{equation}\label{eq:SKKMSsuper0}
\begin{split}
   \IKMSzero &= \Qzero + \thb \, \QKMS - \theta \, \QKMSb + \thb \theta \, \Qbeta \,,\\
   \IKMS &= \QKMS + \theta \, \Qbeta \,,\\
   \IKMSb &= \QKMSb + \thb \, \Qbeta \,,\\
   \LKMS &=  \Qbeta \,.
\end{split}
\end{equation}
The SK-KMS superalgebra we have seen has the following defining graded commutation relations:
\begin{equation}
\begin{split}
\QSK^2 = \QSKb^2  &=\gradcomm{\QSK}{\QSKb} = 0 \\ 
\gradcomm{\QSK}{\IKMSb} = \gradcomm{\,\QSKb}{\IKMS}
= \LKMS  \,, & \qquad 
\gradcomm{\QSK}{\IKMS} = \gradcomm{\,\QSKb}{\IKMSb} = 0\\
\gradcomm{\QSK}{\IKMSzero} = \IKMS\,, & \qquad 
  \gradcomm{\,\QSKb}{\IKMSzero} =- \IKMSb\\
 \gradcomm{\QSK}{\LKMS} &= \gradcomm{\,\QSKb}{\LKMS} = 0 \\
 \gradcomm{\IKMSb}{\IKMS} =
 \gradcomm{\LKMS}{\IKMSb} &= 
   \gradcomm{\LKMS}{\IKMS} =
   \gradcomm{\LKMS}{\IKMSzero} = 0 \,.
\end{split}
\label{eq:SKKMSfinal}
\end{equation}	
where $\{\QSK,\QSKb\}$ are understood to act component-wise on the 
superoperators (which is equivalent to their acting as $\{\partial_\thb,\partial_\theta\}$). 

Inspection of the SK-KMS superalgebra \eqref{eq:SKKMSfinal} suggests an analogy with the general equivariant cohomology algebra \eqref{eq:LieSAlg2}. More precisely consider the superspace version  involving the operators \eqref{eq:NT2superops}, which reads
\begin{equation}
\begin{split}
\QW^2 = \QWb^2  &=\gradcomm{\QW}{\QWb} = 0 \\ 
\gradcomm{\QW}{\SF{\Ibar}_j} = \gradcomm{\,\QWb}{\SF{\mathcal{I}}_j}
= \SF{\mathcal{L}}_j  \,, & \qquad 
\gradcomm{\QW}{\SF{\mathcal{I}}_j} = \gradcomm{\,\QWb}{\SF{\Ibar}_j} = 0\\
\gradcomm{\QW}{\SF{\mathcal{I}}^0_j} = \SF{\mathcal{I}}_j\,, & \qquad 
  \gradcomm{\,\QWb}{\SF{\mathcal{I}}^0_j} =- {\SF{\Ibar}}_j \\
 \gradcomm{\QW}{\SF{\mathcal{L}}_j} &= \gradcomm{\,\QWb}{\SF{\mathcal{L}}_j} = 0 \\
 \gradcomm{\SF{\Ibar}_i}{\SF{\mathcal{I}}_j}= f_{ij}^k\, \SF{\mathcal{I}}^0_k \,,\quad
 \gradcomm{\SF{\mathcal{L}}_i}{\SF{\Ibar}_j} = -f_{ij}^k\,\SF{\Ibar}_{k}\,, &\quad 
   \gradcomm{\SF{\mathcal{L}}_i}{\SF{\mathcal{I}}_j} = -f_{ij}^k\,{\SF{\mathcal{I}}}_{k}\,, \quad
   \gradcomm{\SF{\mathcal{L}}_i}{\SF{\mathcal{I}}^0_j} = - f_{ij}^k\,\SF{\mathcal{I}}^0_{k} \,.
\end{split}
\label{eq:NT2final}
\end{equation}	
The similarity between \eqref{eq:SKKMSfinal} and \eqref{eq:NT2final} is quite obvious now. Let us record some salient features:
\begin{itemize}
\item Clearly we should identify the SK BRST charges with the Weil differentials.
\item The KMS charges conspire to become the super-interior contractions and super-Lie derivative operation.
\item The first four lines of the algebras obviously agree, while the last lines seems to constrain the structure constants $f_{ij}^k$.
\end{itemize}
Thus we schematically we have the obvious identifications:
\begin{equation}\label{eq:match1}
\begin{split}
\underline{{\cal N}_\smallT = 2 \text{ algebra}} \quad &\;\;\, | \quad\;\; \underline{\text{SK-KMS symmetries}}\\
 \{\QW\simeq\partial_\thb,\QWb\simeq\partial_\theta\} \quad &\leftrightarrow \quad \{\QSK\simeq\partial_\thb,\QSKb\simeq \partial_\theta\} \,,\\
 \{\SF{\mathcal{I}}_k,\,\SF{\overline{\mathcal{I}}}_k\} \quad &\leftrightarrow \quad \{  \IKMS, \IKMSb\} \,,\\
 \{\SF{\mathcal{L}}_k,\,\SF{\mathcal{I}}^0_k\} \quad &\leftrightarrow \quad \{\LKMS,\,\IKMSzero\} \,,
\end{split}
\end{equation}
By restriction to ordinary space, it is clear that there is a canonical map between the two sets of operations, which respects the bigrading structure:
\begin{equation}\label{eq:match2}
\begin{split}
\;\;\;
\underline{{\cal N}_\smallT = 2 \text{ algebra}} \quad &\;\;\, | \quad\;\; \underline{\text{SK-KMS symmetries}}\\
 \{\QW, \QWb\} \quad &\leftrightarrow \quad \{\QSK,\QSKb\} \,,\\
 \{\mathcal{I}_k,\,\overline{\mathcal{I}}_k\} \quad &\leftrightarrow \quad \{\QKMS,\QKMSb\} \,,\\
 \{\mathcal{L}_k,\,\mathcal{I}^0_k\} \quad &\leftrightarrow \quad \{\Qbeta,\Qzero\} \,.
\end{split}
\end{equation}
As we can readily see the KMS supercharges $\{\QKMS,\QKMSb\}$ along with the bosonic generator $\Qzero$ make up the interior contraction operations, while the SK supercharges $\{\QSK,\QSKb\}$ are the standard exterior derivations (the Weil charges). Finally, the Lie derivation can be naturally identified with $\Qbeta$, consistent with our observation that the operator provides a means of ascertaining the deviation from the exact KMS condition. 

Having established this algebra isomorphism, let us address the next question: what is the SK-KMS algebra equivariant with respect to? We should identify the gauge algebra to complete the specification of the symmetry of the thermal QFTs. Naively, based on the vanishing of the structure constants one might conclude that the symmetry algebra is Abelian. However, before we rush to this conclusion we should be aware that we obtained the SK-KMS algebra as acting on the operator superalgebra of our quantum system, which is analogous to $\mathcal{M}$ in our discussion of equivariant cohomology. We know from our review in Part \ref{part:maths} that the interior contraction operations do not act non-trivially on the topological space $\mathcal{M}$. So it is not entirely clear at this stage whether we should declare the algebra to be Abelian. More precisely, the superoperators \eqref{eq:SKKMSsuper0} annihilate the usual covariant superfields. Effectively, we appear to have managed to construct the overall algebraic operations without any information about the underlying gauge structure. Nevertheless, we can recover the requisite information if we examine the role played by the Lie derivation carefully.

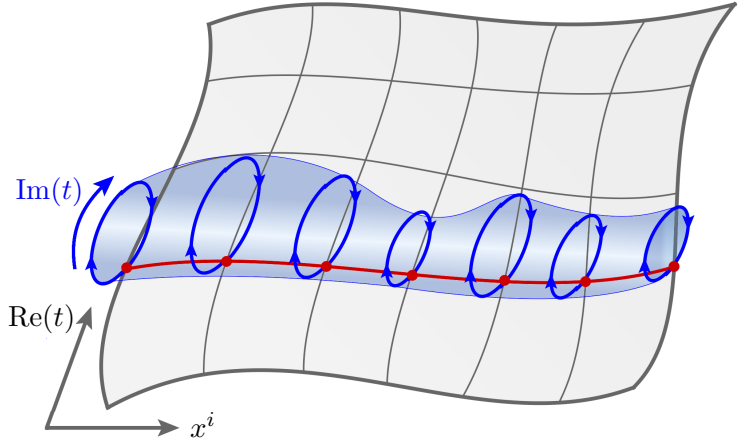
\begin{figure}[t!]
\begin{center}
\includegraphics[width=.6\textwidth]{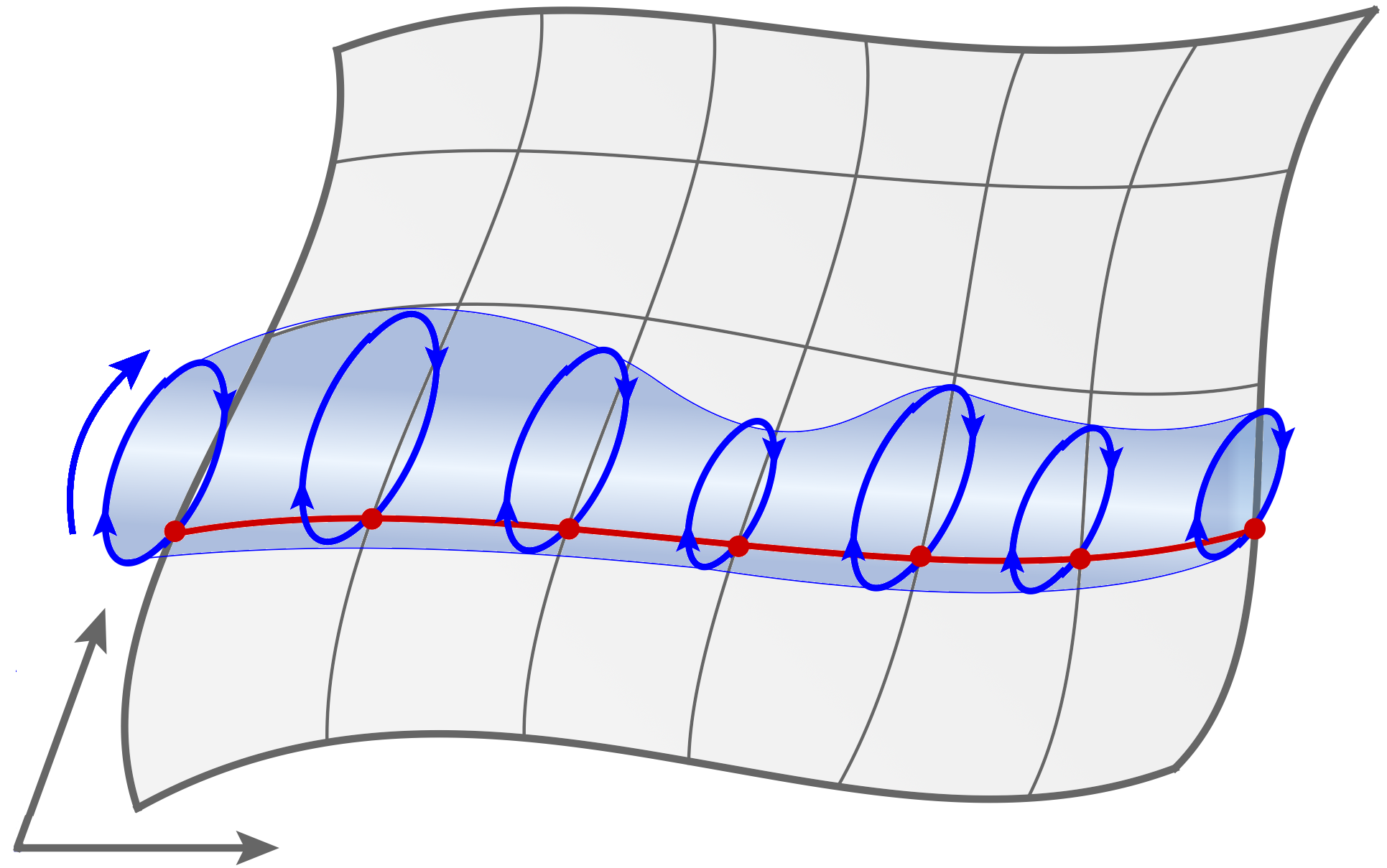}
\begin{picture}(0.3,0.4)(0,0)
\put(-265,93){\makebox(0,0){{\color{blue}Im$(t)$}}}
\put(-268,45){\makebox(0,0){{\color{black}Re$(t)$}}}
\put(-207,5){\makebox(0,0){{\color{black}$x^i$}}}
\end{picture}
\caption{Illustration of the spacetime picture as it emerges from the proposed KMS gauge invariance.
We upgrade the spacetime manifold on which our quantum system resides to a thermal fibre bundle.  
The grey manifold represents a Lorentzian spacetime with a typical Cauchy slice indicated in red. We assume  local thermal equilibrium (as in, e.g., hydrodynamics) at each spacetime point which guarantees a thermal vector $\Kref^\mu$. Geometrically we encode this vector field as a circle fibration with a thermal circle whose size is set by the local temperature. The KMS transformations we seek implement 
equivariance with respect to thermal translations along this local imaginary time circle. Restricting to a gauge slice corresponds picking a Lorentzian section of this fibration. Note that the size of the thermal circle is exaggerated; our arguments are clean in the high temperature limit where the size of the thermal circle is much smaller than the fluctuation scale.}
\label{fig:SKspacetime}
\end{center}
\end{figure}

The operator $\Qbeta$ generates infinitesimal gauge transformations. It acts on Hilbert space operators $\Op{O}$ and takes them around the thermal circle; its action is to compute the deviation between an operator and its KMS conjugate, for
\begin{equation}\label{eq:lieF}
\LKMS \SF{\Op{O}} = \delKMS \SF{\Op{O}} \,.
\end{equation}
As described so far we are considering a single traversal of the thermal circle. While we have also pretended for the most part to be in global thermal equilibrium, as argued in \cite{Haehl:2016pec} we will actually be interested in local thermal equilibrium where the temperature changes from spacetime-point to spacetime-point. In general we can view a non-equilibrium setting with a local temperature profile, as a thermal circle bundle over the spacetime manifold, cf., Fig.~\ref{fig:SKspacetime} for an illustration.

The picture may be understood as follows. At each spacetime point we have a local thermal vector $\Kref^\mu$ which picks out a local inertial frame and whose magnitude gives us the local temperature (see \cite{Haehl:2014zda,Haehl:2015pja} for details). If the variations are purely spatial, then we really have a direct product of Lorentzian time with a Euclidean geometry where the thermal circle is fibred over the spatial sections, as is appropriate for global thermal equilibrium. More generally, we simply have a non-trivial thermal fibration over the spacetime manifold where the QFT resides.\footnote{ We make no attempt here to define a pseudo-Riemannian fibre bundle, except to note that we want the sections to admit Lorentz signature metrics for physical applications.} We tend not to view finite temperature non-equilibrium QFT geometrically as described above, except in equilibrium, but we will argue that this perspective offers fresh and useful insight.

Given this set-up, the KMS Lie derivation $\Qbeta$ takes the operators around the fibres of the thermal fibration.  A-priori we only have discrete transformations along this circle. One can physically view $\delKMS \Op{O}$ as comparing $\Op{O}(t)$ against $\Op{O}(t-i\,\beta)$, with the latter as the consequence of the group action, see \eqref{eq:kmsDel}.\footnote{ In defining $\delKMS$ we have taken care to incorporate the  quantum numbers of   $\Op{O}$, e.g., spin, flavour etc. by associating with $\delta_\Kref$ a Lie derivative along a time-like vector field $\Kref^\mu$; see \cite{Haehl:2016pec} for further details. In the  Euclidean formulation of the theory thermal boundary conditions on operators involve twists along the thermal circle depending on the chemical potential for the flavour charge etc.. The action of $\delKMS$ is cognizant of these facts and we can think of it as a generalized Lie derivation along the Euclidean thermal circle.} While we initially infer the KMS operations by action along a single thermal period, we can upgrade the traversal of operators around this circle to involve a local winding number, i.e., consider $\OpH{O}(t,x) \to \OpH{O}(t- i\,m(t,x) \,\beta, x)$ for 
integer valued $m(t,x)$. The gauge group we seek for the microscopic theory is the one implementing these discrete thermal translations. Since there is a simple operation of  discrete winding it is not surprising that the gauge algebra is Abelian. 

%~~~~~~~~~~~~~~~~~~~~~~~~~~~~~~~~~~~~~~~~~~~~~~~
\subsection{Low energy implementation of SK-KMS superalgebra}
\label{sec:}
%~~~~~~~~~~~~~~~~~~~~~~~~~~~~~~~~~~~~~~~~~~~~~~

While it would be instructive to explore this discrete KMS structure in greater detail, we found it simpler to contemplate the limit where the local temperatures are  high compared to other physical scales of interest. This is the case for instance, if we are interested in the low energy dynamics of quantum systems at frequencies and momenta low compared to the thermal scale $\omega \,T \,, k \, T \ll 1$.
In such a limit, the local thermal circle of size $\beta$ is small since  we equivalently  can 
imagine being at $\beta \ll 1$, ensuring that there is a local realization of the SK-KMS superalgebra generators. In particular, the originally non-local operator $\delKMS$ becomes local in this limit, i.e., $\delKMS \approx\delta_\Kref$ (on bosonic fields). 

We physically imagine that the high-temperature limit corresponds to allowing the discrete KMS transformations to become continuous. In other words we propose to gauge thermal translations; identifying configurations on different Lorentzian sections of the local thermal fibration. The gauge symmetry we are after is then just the group of thermal translations along the Euclidean thermal circle. Said differently, we seek to implement the notion of thermal equivariance and assert that the low energy dynamics is governed by invariance under an emergent \emph{KMS gauge symmetry}. The associated gauge group we shall call $\UT$, to both capture the origins from a circle action and the thermal nature. This structure appears to play very well in the context of hydrodynamic effective field theories as we explain at the end. 

\paragraph{$\UT$ gauge transformations:}
At this point, it is useful to take this physical picture of thermal translations as a gauge symmetry and work out the consequences. Firstly, we will shortly see that it is very useful here to consider explicit gauge parameters. Note that for a typical Lie derivative in generic equivariant cohomology, the presence of a gauge parameter yields an anti-symmetric Lie derivation: 
\begin{equation}
{\cal L}_{\SF{\Lambda}} \, \SF{\mathfrak{F}}^k \equiv \SF{\Lambda}^i\,{\cal L}_i\, \SF{\mathfrak{F}}^k = f^k_{ij} \SF{\Lambda}^i \SF{\mathfrak{F}}^j \equiv \comm{\SF{\Lambda}}{\SF{\mathfrak{F}}}^k\,.
\end{equation}
We will view the action of the KMS Lie derivative \eqref{eq:lieF} as the fundamental action of the $\UT$ gauge transformation, which we can upgrade to involve an infinitesimal gauge parameter  and write 
\begin{equation}
{\cal L}_{\SF{\Lambda}} \SF{\Op{O}} =\SF{\Lambda}\, \delKMS \SF{\Op{O}} \approx
\SF{\Lambda}\, \deltaB \SF{\Op{O}} \,.
\end{equation}	
We have also accounted for the high temperature limit, and will now think of $\Op{O}$ as quantum Hilbert space operators in the low energy theory. However, we will continue to write $\delKMS$ as the derivation operation below, leaving implicit the approximation $\delKMS \approx \deltaB$.

Given this action, it is then easy to work out the adjoint action of $\UT$ which encodes the gauge algebra. On adjoint valued covariant superfields $\SF{\mathfrak{F}} $ the $\UT$ action can then be inferred to be
\begin{equation}\label{eq:lieF2}
 \LKMS_{\SF{\Lambda}} \SF{\mathfrak{F}} = \SF{\Lambda} \, \delKMS \SF{\mathfrak{F}} - \SF{\mathfrak{F}} \, \delKMS \SF{\Lambda} \equiv (\SF{\Lambda},\SF{\mathfrak{F}})_\Kref \,,
\end{equation}
where we defined a thermal commutator $(\,\cdot \,, \, \cdot\,)_\Kref $, which is anti-symmetric if evaluated on adjoint superfields, and will play the role of the Lie commutator. One can easily check that this anti-symmetrization is very sensible as it will, for example, lead to the familiar statement that the commutator of successive gauge transformations yields a gauge transformation along the (thermal) commutator:\footnote{ In fact we could have derived \eqref{eq:lieF2} by commuting two gauge transformations and reading off the commutator action of one adjoint gauge parameter acting on the other one.} 
\begin{equation}
 \gradcomm{\LKMS_{\SF{\Lambda}_1}}{\LKMS_{\SF{\Lambda}_2}} \SF{\mathfrak{F}} = \LKMS_{(\SF{\Lambda}_1,\SF{\Lambda}_2)_\Kref} \, \SF{\mathfrak{F}} \,.
\end{equation}
We summarize these findings by making the following schematic identification:
\begin{equation}
\begin{split}
\underline{{\cal N}_\smallT = 2 \text{ algebra}} \quad &\;\;\, | \quad\;\; \underline{\text{SK-KMS symmetries}}\\
 \comm{\SF{{\mathfrak{F}}}_1}{\SF{{\mathfrak{F}}}_2}^k = f^k_{ij}\,\SF{{\mathfrak{F}}}_1 \SF{{\mathfrak{F}}}_2 \quad &\leftrightarrow \quad (\SF{{\mathfrak{F}}}_1,\SF{{\mathfrak{F}}}_2)_\Kref = \SF{{\mathfrak{F}}}_1 \, \delKMS \, \SF{{\mathfrak{F}}}_2 - \SF{{\mathfrak{F}}}_2\, \delKMS \, \SF{{\mathfrak{F}}}_1
 \end{split}
\end{equation}
for adjoint superfields $\SF{\mathfrak{F}}_{1,2}$. 

Thus, while the microscopic discrete KMS symmetry ends up Abelian, the continuum version displays non-Abelian characteristics. This can be intuited from the fact that the underlying set of transformations are diffeomorphisms along the thermal circle. At some heuristic level we should view $\UT $ as a deformation of $\text{diff}({\bf S}^1)$. Relations of the KMS conditions to deformation quantization have indeed been explored in the past \cite{Basart:1984aa,Basart:1985aa,Bordemann:1998aa,Bordemann:1999aa} without incorporation of the extended symmetry structure we are exploring here. Note also that as remarked earlier,  the above discussion hides the fact that in general $\delKMS$ compares operators which are non-locally separated in Euclidean time (see the definition \eqref{eq:kmsDel}). Per se, we therefore expect that the $\mathcal{N}_\smallT =2 $ algebra is realized with a non-local action on the physical Schwinger-Keldysh theory. These ideas may help us unveil the discrete structure valid outside the low energy limit.

Returning to our main line of development, we can now take the existence of $\UT$ in the low energy limit to its logical conclusion. The simplest way to proceed is to introduce a gauge field for the $\UT$ symmetry of gauged thermal translations. The construction of the thermal gauge superfield one-form proceeds exactly as in the general discussion of equivariant cohomology. That is, we introduce a dodecuplet of fields and arrange them into gauge field components as in \eqref{eq:AaExp2}-\eqref{eq:AthExp2}. In the following we will denote the `thermal $\UT$ dodecuplet' by the same letters as in Table \ref{tab:multiplets}, but give them a subscript ``{\sf T}" in order to distinguish them from the general discussion of Part \ref{part:maths}.\footnote{ It is amusing to note that the authors of \cite{Dijkgraaf:1996tz} presciently labeled the number of topological symmetries with a subscript $T$ which does allow a dual interpretation as `topological' or `thermal' depending on one's inclination. }

As an illustration, we can, for example, define now covariant Cartan differentials of the thermal SK-KMS equivariant theory:
\begin{equation}
\begin{split}
 \Q &\equiv \QSK + \phiT \, \QKMSb + \phiT^0 \, \QKMSb + \etaT \, \Qzero \,,\\
 \Qb &\equiv \QSKb + \phibT\, \QKMS\,.
\end{split}
\label{eq:SKcartanDef}
\end{equation}
This follows from translating the definition of Cartan differentials as in \eqref{eq:QCQCb} to SK-KMS context using the identifications in \eqref{eq:match2}. From their naturalness in the theory of equivariant cohomology, we would expect these linear combinations to play a role in the physics of SK theories. As we will see later, this is indeed the case.\footnote{ See also Eq.\ (A.15) of \cite{Haehl:2015foa}, where the operators $\{\Q,\Qb\}$ have appeared before in this context. For detailed comparison, note the following difference in convention between here and reference \cite{Haehl:2015foa}:
\begin{equation}
 [\QKMS]_{_\text{here}} = i[\QKMS]_\text{\cite{Haehl:2015foa}} \,,\qquad
  [\QKMSb]_{_\text{here}} = -i[\QKMSb]_\text{\cite{Haehl:2015foa}} \,,\qquad
   [\Qzero]_{_\text{here}} = i[{\cal Q}_0]_\text{\cite{Haehl:2015foa}}\,.
 \end{equation}
}

\paragraph{$\UT$ interior contractions:}
To see the full equivariant structure emerge, we will now change perspective: instead of thinking about the SK superfield formalism as being one where all gauge field dependence has been dropped, we will think of the latter actually just being hidden. The identification of $\{\QKMS,\QKMSb,\Qzero\}$ with interior contractions in \eqref{eq:match2} forces us to demand that the covariant components of a SK superfield be annihilated by these operators. To implement this, let us re-write the basic SK superfields \eqref{eq:OpO}  of the low energy Hilbert space operators as follows:
\begin{equation}\label{eq:OpOnew}
\begin{split}
\SF{\Op{O}} &= \SKRet{O} + \theta \, \SKGb{O} + \bar\theta \, \SKG{O} + \bar\theta\theta \, \SKAdv{O} \\
&\equiv \SKRet{O} + \theta \, \left\{ \SKGb{O}^\cov - (\GbT,\SKRet{O})_\Kref \right\} + \thb \, \left\{ \SKG{O}^\cov - (\GT,\SKRet{O})_\Kref \right\} \\
&\qquad\quad +\; \thb \theta \Big\{ \SKAdv{O}^\cov-\bcomm{\BT}{\SKRet{O}} + \bcomm{\GT}{\SKGb{O}^\cov} - \bcomm{{\GbT}}{\SKG{O}^\cov} \\
&\qquad\qquad\qquad + \frac{1}{2} \bcomm{{\GbT}}{\bcomm{\GT}{\SKRet{O}}} 
	- \frac{1}{2} \bcomm{\GT}{\bcomm{{\GbT}}{\SKRet{O}}} \Big\}
\end{split}
\end{equation}
where the covariant components (with superscript `cov.') are annihilated by the interior contractions:
\begin{equation}
 \gradcomm{\QKMS}{\mathbf{O}^\cov} =  \gradcomm{\QKMSb}{\mathbf{O}^\cov} =  \gradcomm{\Qzero}{\mathbf{O}^\cov} = 0
\end{equation}
for all $\mathbf{O}^\cov \in \{ \SKRet{O}^\cov \equiv \SKRet{O},\SKG{O}^\cov,\SKGb{O}^\cov,\SKAdv{O}^\cov \}$. We can therefore now interpret the operators $\{\QKMS,\QKMSb,\Qzero\}$ as genuine interior contractions, which annihilate covariant superfield components and only act non-trivially on the ghost triplet:
\begin{equation}\label{eq:KMSG}
\begin{split} 
\gradcomm{\QKMS}{\GbT} &\equiv {\cal I}^{\text{\tiny KMS}} \, \GbT = -1 \,,\qquad\;\;\;
\gradcomm{\QKMSb}{\GT} \equiv \overline{\cal I}^{\text{\tiny KMS}} \, \GT = -1 \\
\gradcomm{\QKMS}{\BT} &\equiv {\cal I}^{\text{\tiny KMS}} \, \BT = \frac{1}{2} \,,\qquad\quad\;\;
\gradcomm{\QKMSb}{\BT} \equiv \overline{\cal I}^{\text{\tiny KMS}} \, \BT = -\frac{1}{2} \,,\\
&\qquad\gradcomm{\Qzero}{\BT} \equiv {\cal I}^{\text{\tiny KMS}}_0 \, \BT = -1 \,.
\end{split}
\end{equation}
We are thus  suggesting that the FP triplet of $\UT$ ghosts is already hidden in the superfield structure. We merely have to disentangle covariant and non-covariant pieces, such as $\SKG{O} \equiv \SKG{O}^\cov - \bcomm{\GT}{\SKRet{O}}$ etc.. After this shift in our point of view, the KMS symmetry generators have become genuine interior contractions of an equivariant cohomology. 
Note the following intriguing feature of this point of view: the non-locality of the $\delKMS$ action  (away from the high temperature limit) is now hidden within the SK superfield expansion itself. The interior contractions $\{\QKMS,\QKMSb,\Qzero\}$ have become simple algebraic operators which act locally on a (slightly non-local) field.  

For completeness we note that the computationally most efficient representation of the various operators in extended equivariant cohomology are the superfield maps described in \S\ref{sec:WZ2}. They all have a natural SK-KMS realization. This gives a third (and, again, equivalent) version of the correspondences \eqref{eq:match1}, \eqref{eq:match2}:
\begin{equation}\label{eq:match3}
\begin{split}
\underline{{\cal N}_\smallT = 2 \text{ algebra}} \quad &\;\;\, | \quad\;\; \underline{\text{SK-KMS symmetries}}\\
 \{\dSK, \dSKb\} \quad &\leftrightarrow \quad \{\dSK,\dSKb\} \,,\\
 \{\imap_k,\ibmap_k\} \quad &\leftrightarrow \quad \{\iKMS,\iKMSb\} \,,\\
 \{\lmap_k,\izmap_k\} \quad &\leftrightarrow \quad \{\lKMS,\iKMSzero\} \,.
\end{split}
\end{equation}
Given our discussions above, we can view this last diagram as the {\it definition} of the SK-KMS component map operators on the right hand side. The way they act on covariant superfields should now be clear and can be read off from the general discussion in \S\ref{sec:WZ2}.

\paragraph{Hydrodynamic effective field theories:} As argued above the structure of thermal equivariant cohomology appears to be simplest in the high energy limit, whence we obtain the $\UT$ gauge symmetry. We can think of this as being emergent in the low energy limit.  As initially proposed in \cite{Haehl:2014zda,Haehl:2015pja} this thermal gauge symmetry ensures that the theory respects the microscopic unitarity constraints. We explain these concepts in greater detail in \cite{Haehl:2016pec}, but it is useful to record here a few salient facts in the context of hydrodynamics. 

The dynamical infra-red variables of hydrodynamic effective field theories are Nambu-Goldstone bosons corresponding to spontaneously broken global diffeomorphism and flavour symmetries of the doubled Schwinger-Keldysh theory to their diagonal (average). Such a breaking pattern leads to a set of vector Goldstone modes for diffeomorphisms and scalar Goldstones for the flavour symmetry. These naturally admit an interpretation in the hydrodynamic theory as the velocity rescaled by temperature, $\Kbeta^\mu  = u^\mu/T$, called thermal vector in \cite{Haehl:2015pja} and thermal twists $\LambdaB = \mu/T - \Kbeta^\mu A_\mu$ which encode the chemical potentials for flavour. In this sense the low energy dynamics naturally comes equipped with the data that allows us to use the intuition suggested in Fig.~\ref{fig:SKspacetime}.

The hydrodynamic effective field theory is then simply a Landau-Ginzburg sigma model for the these light modes.  The construction of the theory proceeds naturally in terms of a worldvolume theory  (with coordinates $\sigma^a$) analogous to the Nambu-Goto or the RNS model for the string.   On this worldvolume we have a reference thermal vector $\Kref^a$ which keeps track of the local thermal vector, and the physical fields are maps $X^\mu(\sigma)$ from the worldvolume to the target space (the physical thermal vector is given as the push-forward of the worldvolume data). On the worldvolume we have a realization of the $\mathcal{N}_\smallT =2$ algebra, which means that in addition to the dodecuplet of fields in $\As_I\, dz^I$ we have a matter quartet $\SF{X}^\mu$ which extends $X^\mu$ in an appropriate manner.
In particular the superfield $\SF{X}^\mu(z)$ transforms in the fundamental representation of $\UT$, viz.,  $(\SF{\Lambda},\SF{X}^\mu)_\Kref = \SF{\Lambda} \,\delKMS \SF{X}^\mu$.\footnote{ This behaviour is different from the transformation of the coordinate field $X^\mu$ in standard equivariant cohomology which 
 transforms as a vector ${\cal L}_k X^\mu = \xi^\mu_k$, see \eqref{eq:lieX}. The distinction lies in the fact that the usual discussion gauges the group action in  target space, whereas in hydrodynamics we gauge the worldvolume action of the KMS $\UT$ gauge symmetry.
}
 We will see a detailed example of this in \S\ref{sec:langevin}.

 In \cite{Haehl:2015uoc} we have sketched a construction of the topological sigma model necessary to describe dissipative hydrodynamics.\footnote{ For an alternate construction which eschews the idea of thermal equivariance and treats the KMS condition as a discrete ${\mathbb Z}_2$ transformation see \cite{Crossley:2015evo}. As we discuss elsewhere in the high temperature limit relevant for hydrodynamics, the two formalisms agree on the supersymmetry algebra of import.} We will give a detailed discussion of the full equivariant construction in a separate paper, opting for now to describe a much simpler system, a Brownian particle, demonstrating in this classic textbook example the general principles of the Schwinger-Keldysh theory.

%~~~~~~~~~~~~~~~~~~~~~~~~~~~~~~~~~~~~~~~~~~~~~~~
\section{Textbook example: Langevin dynamics}
\label{sec:langevin}
%~~~~~~~~~~~~~~~~~~~~~~~~~~~~~~~~~~~~~~~~~~~~~~

We now turn to studying Brownian motion in a thermal background as a paradigmatic example, which illustrates the Schwinger-Keldysh structures in a clean and simple fashion. Such dynamics is described by the Langevin equation for stochastic motion  in a medium with friction. It is well-known that Langevin theory naturally can be written in a form manifesting the topological symmetries. This was understood in the early analysis of Martin, Siggia, and Rose (henceforth MSR) \cite{Martin:1973zz}. Similar techniques used in this analysis were also employed in the discussion of disordered systems by Parisi and Sourlas \cite{Parisi:1979ka} who then went on to demonstrate that these topological symmetries appear in stochastic systems \cite{Parisi:1982ud}; their names are usually associated with the class of structures we describe. Readers unfamiliar with the subject can find a comprehensive discussion in the textbook \cite{ZinnJustin:2002ru}.\footnote{ These statements are also well known in the topological field theory literature, see \cite{Das:1988vd,Birmingham:1991ty}.} While these discussions appreciate the topological BRST charges inherent in the system and make the connection to the supersymmetric quantum mechanics discussion of \cite{Witten:1982im} manifest, they do not attempt to utilize the language of the extended equivariant cohomology algebra.

Our rationale for employing the extended ${\cal N}_T =2 $ algebra is that this system provides a simple toy model for the more complicated setting of hydrodynamics. As explained in \cite{Haehl:2015foa} we can imagine immersing into the fluid probe branes of various codimension. These probes are buffeted by the thermal and quantum fluctuations undergoing as a consequence a generalized Brownian motion in the fluid. We therefore have chosen to call them Brownian branes in recognition of this fact. The worldvolume theory of Brownian branes  is an ${\cal N}_T=2$ topological sigma model. The simplest of these is the theory of a Brownian particle (a 0-brane) which undergoes Langevin motion. We will focus on this special case elaborating on the discussion presented in Appendix A of \cite{Haehl:2015foa} to illustrate our point. The discussion of the space filling Brownian brane which is the hydrodynamic effective theory is described in \cite{Haehl:2015uoc}.

To keep the discussion simple and not get bogged down in technicalities of the target space in which the Brownian particle navigates we are going to consider motion in one spacetime direction. This allows us to write down the effective action without worrying about various curvature couplings which are present  for higher dimensional target space.\footnote{  Readers familiar with discussions of supersymmetric interpretations of Morse theory \cite{Witten:1982im} will immediately realize that the theory we are after is given by Witten's construction.}

With these motivations in mind, consider a Brownian particle which is characterized by a single degree of freedom, viz., its position $X(t)$. The parameter $t$ should be viewed as the worldline proper time coordinate, and serves to define an intrinsic clock for the particle.  We will assume that this particle is subject to a time independent (conservative) force, arising from a potential $U(X)$, in addition to the drag it encounters from the fluid medium it is immersed in. The stochastic Brownian motion is then described by the Langevin equation:
\begin{equation}
m\frac{d^2 X}{dt^2} +\frac{\partial U}{\partial X}  +  \nu\ \delKMS X
= \mathbb{N} \,,
\label{eq:langevin}
\end{equation}
where $m$ is the mass, and $\nu$ is the viscous friction coefficient responsible for the dissipation of the particle subject to the forces of the fluid. The right hand side of \eqref{eq:langevin} describes a particular realization of ($X$-independent) stochastic white noise, drawn from a Gaussian ensemble, viz.,
\begin{equation}
{\cal P}(\mathbb{N}) = \exp\left(- \int dt \, \frac{1}{4\,\nu} \, \mathbb{N}^2\right)\,.
\label{eq:PNdef}
\end{equation}	

We have also written the friction term with the benefit of some hindsight in terms of the Lie derivative operator $\delKMS$. For the most part we will restrict to the high temperature limit, where 
\begin{equation}
\delKMS\equiv - i \left( 1 - e^{-i\delta_\Kbeta}\right) \approx  \delta_\Kbeta \equiv  \Kbeta \frac{d}{dt} \,.
\label{eq:DelLang}
\end{equation}
We can think of $\delKMS$ and $\Kbeta \, \frac{d}{dt}$ as being interchangeable, but for the most part work with the former since it allows a cleaner connection with the algebraic structures we wish to unearth.

 An explicit solution of the model would involve solving \eqref{eq:langevin} for various different $\mathbb{N}$ and finally taking a statistical average over all the noise realizations in the given ensemble. Through the fluctuation-dissipation theorem, the width of the Gaussian ensemble will be linked to the viscosity $\nu$, which has already been incorporated in \eqref{eq:PNdef}. We will now try to construct an effective action for this dissipative system and show how the fluctuation-dissipation relation arises as a consequence of the Schwinger-Keldysh supersymmetry.

%~~~~~~~~~~~~~~~~~~~~~~~~~~~~~~~~~~~~~~~~~~~~~~~
\subsection{The MSR effective action for Langevin dynamics}  
\label{sec:LangevinMSR}
%~~~~~~~~~~~~~~~~~~~~~~~~~~~~~~~~~~~~~~~~~~~~~~

We first turn to arguing from an alternative perspective that a natural description of the Brownian particle should have ${\cal N}_T =2$ symmetry. The logic we employ is to give a simple argument that dates back many decades to the work of Martin-Siggia-Rose \cite{Martin:1973zz} (see also \cite{Janssen:1976fk,DeDominicis:1977fw} and the Parisi-Sourlas construction \cite{Parisi:1982ud}) who construct a Schwinger-Keldysh effective action whose dynamics is captured by \eqref{eq:langevin}.

We will first argue using a familiar trick due to Faddeev-Popov that the canonical effective action for the Brownian particle involves a quadruplet of fields,   $\{X, \xpsi, \xpsib, \tx\}$ from an effective action perspective. As may be envisioned based on our superspace discussions, we aim to assemble these fields into a single ${\cal N}_T=2$ supermultiplet in due course.

 We will first construct the effective action following the aforementioned references (see also \cite{ZinnJustin:2002ru}) which will allow us to infer the action of the SK and KMS supercharges. To motivate this, imagine computing  correlation functions for the position at different time instances. This can be done, for example, by solving the equation \eqref{eq:langevin} for a particular noise realization $\mathbb{N}$, taking the on-shell values of $X_{\mathbb{N}}(t)$  and multiplying them out to obtain the correlator. We then use the given noise probability distribution to evaluate the ensemble average.  In other words we would write:
\begin{equation}
\begin{split}
G(t_1,t_2,,\cdots t_m) & \equiv \vev{X(t_1)\, X(t_2) \, \cdots X(t_m)}_{\mathbb{N}} 
 \\
&=
\int [d\mathbb{N]}\, {\cal P}(\mathbb{N}) \; \left(\prod_{i=1}^m X_{\mathbb{N}}(t_i) \right)
 \\
&= 
\int[d\mathbb{N]}\, {\cal P}(\mathbb{N}) \;  \int [dX] \, 
\left(\prod_{i=1}^m X(t_i) \right)\,  \delta(X-X_{\mathbb{N}})\,.
\end{split}
\end{equation}	
Using the standard identity for delta functionals we can we write delta functional implementing the on-shell condition in terms of the equation of motion \eqref{eq:langevin}.  Writing the Langevin equation of motion by separating out the non-noise part
\begin{equation}
\begin{split}
{\cal E}_X \equiv  -m\frac{d^2X}{dt^2}-\frac{\partial U}{\partial X} - \nu\,  \delKMS X \,,
\end{split}
\end{equation}
we may re-express our  observable as the functional integral:
\begin{equation}
\begin{split}
G(t_1,t_2,,\cdots t_m)
= 
\int[d\mathbb{N]}\, {\cal P}(\mathbb{N}) \;  \int [dX] \, 
\left(\prod_{i=1}^m X(t_i) \right)\,   \delta \left({\cal E}_X + \mathbb{N} \right) \;  
\det\prn{\frac{\delta {\cal E}_X}{\delta X} } .
\end{split}
\label{eq:measureLangevin}
\end{equation}

We are now in a position to introduce the quadruplet. Following the classic Faddeev-Popov trick we can exponentiate the delta-function with a Lagrange multiplier or Nakanishi-Lautrup field  $\tx$, and rewrite the determinant in terms of a pair of Grassmann valued scalar ghosts $\{ \xpsi, \xpsib \}$ to functionally represent the determinant:
\begin{equation}
\begin{split}
G(t_1,t_2,,\cdots & t_m)  = 
\int[d\mathbb{N]}\, {\cal P}(\mathbb{N}) \;  \int [dX][d\tx] [d\xpsi] [d\xpsib]\;
\\
& \qquad \qquad   \left(\prod_{i=1}^m X(t_i) \right)\; 
\exp\left( i \, \int dt\, \left[ \tx\, {\cal E}_X + \tx \, \mathbb{N} + \xpsib\, 
\frac{\delta {\cal E}_X}{\delta X} \, \xpsi \right]\right)
\\
& \equiv
 \int \, [d\SF{X}] \; \left(\prod_{i=1}^m X(t_i) \right)\;
\exp\left( i \, \int dt\, \left[ \tx\, {\cal E}_X + i\, \nu \, \tx^2 + \xpsib\, 
\frac{\delta {\cal E}_X}{\delta X} \, \xpsi \right]\right) .
\end{split}
\end{equation}
To obtain the second line we have performed the integral over the noise field using the Gaussian distribution  and also abbreviated the functional integral measure 
$[d\SF{X}] = [dX \, d\tx\, d\xpsi \, d\xpsib]$.

We can now read off the Lagrangian density for the Langevin dynamics (as obtained by the MSR construction), which as noted earlier involves the quadruplet of fields: 
\begin{equation}
\begin{split}
\Lagref_{_{\sf{B0}}}^{_{\sf{(MSR)}}} &=
 \tx\ {\cal E}_X+ i \, \nu \,  \tx^2
+ \xpsib \prn{\frac{\delta E_X}{\delta X} }  \xpsi\\
&= -\brk{ \tx\ \frac{\partial U}{\partial X} +  \xpsib \frac{\partial^2 U}{\partial X^2} \xpsi }
- m \brk{  \tx\ \frac{d^2X}{dt^2} +  \xpsib\ \frac{d^2 \xpsi}{dt^2} }\\
&\qquad \qquad  -\nu \brk{ \tx\  \delKMS X - \xpsib\  \delKMS  \xpsi} + i \, \nu \,  \tx^2\,.
\end{split}
\label{eq:LSKLan}
\end{equation}
We have used the subscript $\sf{B0}$ to alert the reader that the Langevin particle should be thought of as a Brownian $0$-brane. As presaged we have the quadruplet of fields. What remains to be seen is that they transform in an appropriate fashion under the ${\cal N}_T = 2$ algebra, which we now proceed to explain.\footnote{ We note in passing that \cite{Kovtun:2014hpa} attempt to construct hydrodynamic effective actions by similarly exponentiating the conservation equations. While this is fraught with some subtleties (cf., \cite{Haehl:2015pja}), it does capture the general structure, as our analysis will make clear below.}

By Taylor expanding \eqref{eq:LSKLan} in the difference field $\tx$ and expressing the result in terms of the left-right basis
\begin{equation}\label{eq:langevinRL}
X_\skR \equiv X -  \frac{1}{e^{i\delta_\beta}-1} \, \tx \,,\qquad
X_\skL \equiv X - \left(1+  \frac{1}{e^{i\delta_\beta}-1} \right) \, \tx \,.
\end{equation}
we can rewrite $\Lagref_{_{\sf{B0}}}^{_{\sf{(MSR)}}}$ in terms of the left-right basis and infer the Feynman-Vernon influence functional (i.e., the part of the Lagrangian which is not of a left-right factorized form). While it is in principle straightforward to do this for an arbitrary potential $U(X)$, for simplicity we restrict to the case $U(X) = \lambda \, X^2$. In this case we find:  
\begin{equation}
\begin{split}
\Lagref_{_{\sf{B0}}}^{_{\sf{(MSR)}}} 
&= \left[\frac{m}{2} \,\prn{\frac{dX_\skR}{dt}}^2- \lambda \, X_\skR^2   \right] - \left[ \frac{m}{2}\, \prn{\frac{dX_\skL}{dt}}^2 - \lambda \, X_\skL^2\right] \\ 
&\qquad
- \lambda \, \left( X_\skR-X_\skL \right) \, \frac{1-e^{-i\delta_\beta}}{1+e^{-i\delta_\beta}} \,  \left( X_\skR-X_\skL \right) 
\\
&\qquad +\nu \bigbr{- \half(X_\skR-X_\skL) \brk{ \prn{ \delKMS -2i} X_\skR + 
\prn{ \delKMS+2i} X_\skL} }\\
&\qquad  
 - m \xpsib \ \frac{d^2\xpsi}{dt^2} -  2\lambda\, \xpsib\,\xpsi+ \nu \xpsib\ \delKMS\xpsi   \,,
\end{split}
\end{equation}
which can be further simplified by using $(1-e^{-i\delta_\beta})(1+e^{-i\delta_\beta})^{-1} \approx \frac{i\Kref}{2} \frac{d}{dt}$ in the second line, and $\delKMS \approx \Kref \frac{d}{dt}$ in the third and fourth lines. 
This Lagrangian consists of the following parts:
\begin{enumerate}
\item[($i$)] the first line denotes the standard Schwinger-Keldysh action of the form $\Lagref[X_\skR]-\Lagref[X_\skL]$, 
\item[($ii$)] the second and third lines contain the influence functional $\Lagref_\text{IF}[X_\skR,X_\skL]$ which couples left and right fields in a non-trivial way and is crucial for dissipation (note that we discarded a total derivative term which derives from rewriting the kinetic terms for $X_{\skR,\skL}$), 
\item[($iii$)] the last line gives the ghost couplings. As we will see below, these are precisely such that they render the Lagrangian supersymmetric.  
\end{enumerate}

%~~~~~~~~~~~~~~~~~~~~~~~~~~~~~~~~~~~~~~~~~~~~~~~
\subsection{Equivariance 1: the SK and KMS charges}
\label{sec:SKlangevin}
%~~~~~~~~~~~~~~~~~~~~~~~~~~~~~~~~~~~~~~~~~~~~~~

In this subsection we set up the basic notation for naive transformation rules under the SK and KMS supercharges. We will defer the issue of gauge covariance to the next subsection. All expressions in the present subsection should be understood as gauge non-covariant in the sense that we implicitly set the gauge superfield $\As=0$. In \S\ref{sec:equiv2} we will properly define the gauge transformation in question and covariantize all formulae correspondingly. 

Our first task is to introduce the position supermultiplet, which we write as: 
\begin{equation}
\SF{X}(t,\theta,\bar\theta) \equiv X(t) + \bar\theta\, \xpsi(t) + \theta \, \xpsib(t) + \bar\theta \theta \,  \tx(t) \,.
\label{eq:xsf}
\end{equation}
The fields $\{X, \tx\}$ are  interpreted as the retarded and advanced fields respectively which  turns out to be the natural basis to describe Langevin dynamics. We can of course relate back to the Schwinger-Keldysh right/left basis by \eqref{eq:langevinRL}, or equivalently by the usual set of (differential) equations:
\begin{equation}
\delKMS\, X \equiv -i  \, (X_\skR - e^{-i\,\delta_\Kbeta} \, X_\skL ) 
 \,, \qquad 
  \tx \equiv X_\skR- X_\skL\,.
\label{eq:arL}
\end{equation}	

With this structure in mind, we are now in a position to analyze the symmetries of the Langevin effective action \eqref{eq:LSKLan}. From a superspace point of view, all the symmetries are encapsulated in the simple actions of $\{\QSK\simeq\partial_{\bar\theta},\,\QSKb\simeq\partial_\theta\}$ on the superfield \eqref{eq:xsf} and on the top interior super-contraction
\begin{equation}
 \IKMSzero = \Qzero + \thb \, \QKMS - \theta \, \QKMSb + \thb \theta \, \Qbeta \,.
\end{equation}
For completeness we will now nevertheless give the action of the individual Hilbert space operators.

Firstly, we see using the MSR construction that one can immediately motivate the action of the SK charges as in Eq.\ \eqref{eq:qskkmsaction}:
\begin{equation}\label{eq:QSKLangevin}
\begin{split}
\gradcomm{\QSK}{X}&=  \xpsi,\quad
\gradcomm{\QSK}{\xpsi } = 0\, ,\quad
\gradcomm{\QSK}{\xpsib} = -  \tx\, ,\quad
\gradcomm{\QSK}{ \tx} = 0\,, \\
\gradcomm{\QSKb}{X} &= \xpsib \,,\quad
\gradcomm{\QSKb}{\xpsib } = 0 \,,\quad
\gradcomm{\QSKb}{\xpsi} =   \tx \,,\quad
\gradcomm{\QSKb}{\tx} = 0 \,.
\end{split}
\end{equation}
Likewise one can argue that the KMS charges ought to act on the fields consistent with \eqref{eq:qskkmsaction}, viz.,
\begin{equation}\label{eq:QKMSLangevin}
\begin{split}
\gradcomm{\QKMS}{X}&= 0,\quad
\gradcomm{\QKMS}{\xpsi } = 0\, ,\quad
\gradcomm{\QKMS}{\xpsib}=\delKMS X\, ,\quad
\gradcomm{\QKMS}{ \tx} = \delKMS \xpsi\,, \\
\gradcomm{\QKMSb}{X} &= 0\,,\quad
\gradcomm{\QKMSb}{\xpsib } = 0 \,,\quad
\gradcomm{\QKMSb}{\xpsi} = \delKMS X \,,\quad
\gradcomm{\QKMSb}{ \tx} = -\delKMS \xpsib \,.
\end{split}
\end{equation}

While this takes care of the two interior contractions and the Weil charges, we also need the third interior contraction. Following the logic outlined in \S\ref{sec:sknt2} we define the Grassmann-even supercharge 
$\Qzero$ to have the following action:
\begin{equation}\label{eq:Q0Langevin}
\begin{split}
\gradcomm{\Qzero}{X}&= 0,\quad
\gradcomm{\Qzero}{\xpsi } = 0\, ,\quad
\gradcomm{\Qzero}{\xpsib} = 0\, ,\quad
\gradcomm{\Qzero}{ \tx} = \delKMS X\,.
\end{split}
\end{equation}
Finally, $\Qbeta$ simply realizes the democratic action of thermal translations:
\begin{equation}
\gradcomm{\Qbeta}{{\cal X}} = \delKMS {\cal X} \quad \text{for} \quad {\cal X}\in \{X,\xpsi,\xpsib,\tx\} \,.
\end{equation}
This then completes the quartet of thermal super-translations $\{ \QKMS, \QKMSb,\Qbeta,\Qzero\}$. These fully characterize all translations in the Grassmann-odd and thermal directions, respectively.

%~~~~~~~~~~~~~~~~~~~~~~~~~~~~~~~~~~~~~~~~~~~~~~~
\subsection{Equivariance 2: the gauge multiplet}
\label{sec:equiv2}
%~~~~~~~~~~~~~~~~~~~~~~~~~~~~~~~~~~~~~~~~~~~~~~

What we have uncovered in the previous subsection is the canonical action of the ${\cal N}_T =2$ generators  on the position multiplet. While this is interesting, the construction there is not yet fully covariant, as it appears to eschew the gauge multiplet that is canonically present in the algebra. After all, we have not yet quite spelled out what symmetry we are equivariant with respect to. The fact that the Lie derivative operator $\Qbeta$ generates thermal super-translations is the key to unraveling the structure of the gauge group and writing down a fully ${\cal N}_T =2$ covariant effective action.

We use the observation that in the present $(0+1)$-dimensional example the Lie derivation simply acts as $\delKMS = \Kref\, \frac{d}{dt}$ and ought to generate the gauge transformations. We can thence understand the gauge symmetry as being induced by the thermal diffeomorphisms. Following our previous discussion we will refer to this as the $\UT$ KMS gauge invariance. To describe the algebra we find it convenient to make use of the thermal bracket, which acts to gauge transform all the matter fields, c.f., \S\ref{sec:sknt2}. Given a gauge parameter $\LamS$ (which we take to be an adjoint superfield here, for the sake of generality) we denote a gauge transformation of a matter multiplet $\SF{X}$ via
\begin{equation}
(\LamS,\SF{X})_\Kref=\LamS  \,\lieD_\Kref \SF{X}  =\LamS  \,\delKMS \SF{X}
= \LamS\, \Kref\frac{d}{dt} \SF{X}\,.
\label{eq:betabrkX}
\end{equation}	
One may view this as the action of the gauge symmetry on the fundamental representation. We can work out the analog of the adjoint representation by examining the Wess-Zumino commutator of the successive gauge transformations as in \S\ref{sec:sknt2}. The Jacobi identity, for instance, fixes the action of the thermal bracket on another adjoint superfield $\SF{\Lambda}'$, so that under an infinitesimal $\UT$ transformation $\LamS' \mapsto \LamS' + (\LamS,\LamS')_\Kref $  with
\begin{align}
(\LamS,\LamS')_\Kref &=\LamS\lieD_\Kref \LamS'-\LamS' \lieD_\Kref \LamS \,.
\label{eq:adbetabrk}
\end{align}

We are now in a position to remedy the lack of explicit action of the $\UT$ symmetry in 
\S\ref{sec:SKlangevin} by incorporating the gauge dodecuplet into the construction. We introduce the gauge superfield one-form as in \eqref{eq:A1fsf} and the similarly associated field strength and covariant derivative. The  main change from Eqs.~\eqref{eq:covDI} and \eqref{eq:fdef} is that the gauge algebra is generated by the thermal bracket \eqref{eq:adbetabrk} as appropriate for adjoint-valued gauge fields.

While it is possible to write down the full gauge field one-form, to keep the present discussion under control we are going to exploit the following fact. The Brownian particle has a one-dimensional worldline which means that any gauge field associated with it can be trivially gauge fixed to zero. This implies that we can w.l.o.g.\ set the temporal component $\As_t =0$. Then we are only required to deal with the octet of fields in the $\As_\theta$ and $\As_{\thb}$ superfields. The structure of the Weil charge action discussed in \S\ref{sec:sknt2} leads to the construction 
\begin{equation}
\begin{split} 
\As_{\thb} \equiv&\
	 \GT +  \thb  \bigbr{ \phiT-  \half (\GT,\GT)_\Kref  } 
	+ \theta \bigbr{ \BT-  \half (\GbT,\GT)_\Kref } \\
&
	\ - \thb   \theta \bigbr{ \etaT +  (\GbT,\phiT)_\Kref - (\GT,\BT )_\Kref +
	 \half (\GT,(\GbT,\GT)_\Kref )_\Kref  \ }\\
\As_\theta \equiv&\ 
	\GbT + \theta  \bigbr{ \phibT-  \half (\GbT,\GbT)_\Kref  } +  
	\thb  \bigbr{\phiT^0- \BT-  \half (\GbT,\GT)_\Kref  } \\
&\ 
	+ \thb   \theta \bigbr{\etabT +  (\GT,\phiT^0)_\Kref -  
	(\GbT, \phiT^0- \BT)_\Kref+ \half  (\GbT, (\GbT,\GT)_\Kref )_\Kref  } 
\end{split}
\label{eq:Asuper}
\end{equation}
In writing the above we have given a subscript $\sf{T}$ to the gauge multiplets to denote their origin from the thermal KMS invariance. The thermal bracket acts on the Grassmann-odd parameters as in \eqref{eq:adbetabrk}, but with the usual extra sign; to wit,
\begin{equation}
 (\hat{\Op{A}},\hat{\Op{B}})_\Kbeta \equiv \hat{\Op{A}}\delKMS \hat{\Op{B}} - (-)^{\Op{A}\Op{B}} \,\hat{\Op{B}} \delKMS \hat{\Op{A}}
\end{equation}
for the FP ghosts $\{\GT,\BT,\GbT\}$ as well as the adjoint operators such as the Vafa-Witten quintet $\{\phiT,\etaT,\phiT^0,\phibT,\etabT\}$.

With the gauge superfield $\As$ at hand, we can immediately define a gauge invariant field strength in complete analogy with the field strength in generic extended equivariant cohomology, cf., Eq.~\eqref{eq:fdef}:
\begin{equation}
\SF{\mathscr{F}}_{IJ} \equiv  (1-\frac{1}{2}\,\delta_{IJ}) \left( \partial_I \As_J - (-)^{IJ}  \partial_J \As_I + (\As_I,\As_J)_\Kref\right) \,.
\end{equation}
The identities of \S\ref{sec:nt2} (in particular Eq.\ \eqref{eq:quintet}) of course hold for this field strength with the commutator action given by the thermal bracket.

To construct covariant objects with nice $\UT$ transformation properties, we should also rewrite the position multiplet approrpiately to account for the gauge transformation \eqref{eq:betabrkX}. 
As explained in \S\ref{sec:sknt2}, the components of the position superfield $\SF{X}$ as written in Eq.\ \eqref{eq:xsf} should be decomposed into covariant and curvature pieces. This is necessitated by the idea that $\{\QKMS,\QKMSb,\Qzero\}$ should act as interior contraction operators and hence annihilate the covariant components of the superfield. Explicitly, we write this split as follows:
\begin{equation} 
\begin{split}
\SF{X} &= X + \thb \, \xpsi + \theta \xpsib + \thb\theta \, \tx \\
	&\equiv X + \thb  \bigbr{\xpsi^\cov -  \GT \, \Kref   \frac{dX}{dt} } 
	+ \theta \bigbr{\xpsib^\cov -  \GbT \, \Kref \frac{dX}{dt} } \\
&\quad 
	+ \thb   \theta \Bigl\{\tx^\cov 
	+ \GT \, \Kref \frac{d}{dt} \left(\xpsib^\cov-  \GbT \, \Kref \frac{dX}{dt} \right)
	- \GbT\,  \Kref \frac{d}{dt} \left(\xpsi^\cov- \GT\, \Kref \, \frac{dX}{dt}\right)  \Bigr.\\
&\qquad \Bigl. \qquad
-\BT\, \Kref\frac{dX}{dt} - 
\GbT \GT \,\Kref \frac{d}{dt} \left(\Kref \frac{dX}{dt}\right)  \Bigr\}  .
\end{split}
\label{eq:Xsuper}
\end{equation}
The interior contractions $\{\QKMS,\QKMSb,\Qzero\}$ annihilate all the covariant parts, and only act on the Faddeev-Popov ghost triplet $\{\GT,\BT\GbT\}$ as usual (see \eqref{eq:KMSG}). 

It is now clear that there are some Cartan charges $\Q$ and $\Qb$, which act as covariant derivatives:
\begin{equation}\label{eq:dCdCbdef}
\begin{split}
 \Q\, \mathfrak{F}
 &\equiv \Dsf_{\bar\theta}\,\SF{\mathfrak{F}}|  \equiv \left(\partial_{\bar\theta}\,\SF{\mathfrak{F}} + (\As_{\bar\theta},\SF{\mathfrak{F}} )_\Kbeta\right) |\,,\\
 \Qb\, \mathfrak{F}
 &\equiv \Dsf_{\theta}\,\SF{\mathfrak{F}} |\equiv \left(\partial_{\theta}\,\SF{\mathfrak{F}} + (\As_{\theta},\SF{\mathfrak{F}} )_\Kbeta\right) |\,,
\end{split}
\end{equation}
for any superfield $\SF{\mathfrak{F}}$ with bottom component $\mathfrak{F}$. Here, $\{\Q,\Qb\}$ are the SK-KMS Cartan charges, defined via the above equation. Note that this definition of $\{\Q,\Qb\}$ on any given field $\mathfrak{F}$ requires the construction of a superfield $\SF{\mathfrak{F}}$ whose bottom component is $\SF{\mathfrak{F}}| = \mathfrak{F}$. 
Their corresponding ordinary space expansion in terms of Weil differentials and interior contractions can be read off from Eq.\ \eqref{eq:QCQCbfull}; we will, however, not need the latter. We will give a gauge fixed version of them below.

\paragraph{Partial gauge fixing to Wess-Zumino gauge:} We can now continue to work directly with the multiplets \eqref{eq:Asuper} and \eqref{eq:Xsuper}, but a partial gauge fixing allows for a cleaner presentation. 
We realize that in the construction of gauge invariant data we do not need the Faddeev-Popov ghost triplet. By a procedure completely isomorphic to the discussion in \S\ref{sec:WZ2}, we can perform a Wess-Zumino gauge fixing, where $\{\GT,\BT,\GbT\}$ are set to zero. With this choice, the field content of our theory simplifies enormously: 
\begin{equation}
\label{eq:AWZ}
\begin{split}
 (\SF{X})_{_{WZ}} &= X + \thb \, X_\psi^\cov + \theta \, X_{\psib}^\cov + \thb \theta \, \tilde{X}^\cov \,,\\
 (\As)_{_{WZ}} &= \SF{\Ascr}_t \, dt + \SF{\mathscr{A}}_\theta \, d\theta + \SF{\Ascr}_{\bar\theta} \, d\bar\theta \\
 &\equiv \left(\theta \, \phibT + \bar\theta \, \phiT^0 + \bar\theta\theta \, \etabT\right) \,d\theta 
 + \left(\bar\theta \,\phiT -\bar\theta\theta \, \etaT\right) \, d\bar\theta  \,.
\end{split}
\end{equation}
In the following, we will always assume WZ gauge and hence drop the explicit subscript ``$WZ$'' in expressions such as the above. This also means, we can drop the superscript ``$cov.$'', for the full components agree in WZ gauge with their covariant parts: $(X_\psi)_{_{WZ}} = X_\psi^\cov$ etc.. 

In this Wess-Zumino gauge we can easily evaluate the Cartan charge action. We would write as before 
\begin{equation}
\begin{split}
\Q &= \QSK + \phiT \, \QKMSb + \phiT^0 \, \QKMS +  \etaT\, \Qzero \,,\\
\Qb &= \QSKb + \phibT \, \QKMS \,,
\end{split}
\label{eq:QClangevin}
\end{equation}
so that on the position multiplet the action by SK-KMS Cartan charges $\{\Q,\Qb\}$ is given by 
\begin{equation}\label{eq:PositionCartan}
\begin{split}
  \gradcomm{\Q}{X} \equiv \Dsf_{\bar\theta} \SF{X} | = \xpsi  \,,\qquad
  & \gradcomm{\Q}{\xpsib} \equiv \Dsf_{\bar\theta} \Dsf_{\theta} \SF{X} | =  -  \tx + \phiT^0 \delKMS X\,,\\
  \gradcomm{\Q}{\xpsi} \equiv \Dsf_{\bar\theta}\Dsf_{\bar\theta} \SF{X} | =  \phiT \delKMS X \,,\qquad
  & \gradcomm{\Q}{ \tx} \equiv \Dsf_{\bar\theta}\Dsf_\theta \Dsf_{\bar\theta} \SF{X} | =  \phiT^0 \, \delKMS\xpsi - \phiT \delKMS\xpsib +\etaT \delKMS X \,,\\
  \gradcomm{\Qb}{X}  \equiv \Dsf_\theta \SF{X}| = \xpsib \,,\qquad
 & \gradcomm{\Qb}{\xpsib} \equiv \Dsf_\theta \Dsf_\theta \SF{X} | =  \phibT \delKMS X  \,,\\
  \gradcomm{\Qb}{\xpsi} \equiv \Dsf_\theta\Dsf_{\bar\theta}\SF{X}|= \tx  \,,\qquad &
  \gradcomm{\Qb}{ \tx} \equiv \Dsf_\theta\Dsf_\theta \Dsf_{\bar\theta} \SF{X} | =  \phibT \delKMS \xpsi \,.\\
\end{split}
\end{equation}
This set of equations was previously written down in \cite{Haehl:2015foa}.
Similarly, \eqref{eq:dCdCbdef} defines also the action on the ghost of ghost quintet. Using the superfield expansions \eqref{eq:Fvectors} for gauge invariant field strength components and acting on them with the covariant derivative maps \eqref{eq:DamapAdj}, one can readily verify the following relations:
\begin{equation}\label{eq:GaugeCartan}
\begin{split}
  \gradcomm{\Q}{\phiT^0} \equiv \Dsf_{\bar\theta} \SF{\mathscr{F}}_{\theta\bar\theta}|  = \etaT   \,,\qquad& \gradcomm{\Qb}{\phiT^0} \equiv \Dsf_{\theta} \SF{\mathscr{F}}_{\theta\bar\theta}| = \etabT \,,\\
  \gradcomm{\Q}{\phiT} \equiv \Dsf_{\bar\theta} \SF{\mathscr{F}}_{\bar\theta\bar\theta}|  = 0\,,\qquad& \gradcomm{\Qb}{\phiT} \equiv \Dsf_{\theta} \SF{\mathscr{F}}_{\bar\theta\bar\theta}| = -\etaT \,,\\
  \gradcomm{\Q}{\phibT} \equiv \Dsf_{\bar\theta}  \SF{\mathscr{F}}_{\theta\theta}| = -\etabT \,,\qquad& \gradcomm{\Qb}{\phibT}\equiv \Dsf_{\theta}  \SF{\mathscr{F}}_{\theta\theta}| =0 \,,\\
  \gradcomm{\Q}{\etaT} \equiv  \Dsf_{\bar\theta}^2 \SF{\mathscr{F}}_{\theta\bar\theta}|= (\phi,\phiT^0)_\Kbeta \,,\qquad&
  \gradcomm{\Qb}{\etaT} \equiv \Dsf_{\theta} \Dsf_{\bar\theta} \SF{\mathscr{F}}_{\theta\bar\theta}| = (\phiT,\phibT)_\Kbeta\,,\\
  \gradcomm{\Q}{\etabT}  \equiv\Dsf_{\bar\theta} \Dsf_\theta \SF{\mathscr{F}}_{\theta\bar\theta}|   = (\phibT,\phiT)_\Kbeta\,,\qquad&
  \gradcomm{\Qb}{\etabT} \equiv \Dsf_{\theta}^2 \SF{\mathscr{F}}_{\theta\bar\theta}| = (\phibT,\phiT^0)_\Kbeta  \,.
\end{split}
\end{equation}
Note that all these transformations ensure that $\Q^2$, $\Qb^2$, $\gradcomm{\Q}{\Qb}$ are pure gauge, i.e., they generate time translations with gauge parameters $\phiT$, $\phibT$, $\phiT^0$ respectively. We have now all the ingredients to formulate the Schwinger-Keldysh effective theory of Langevin dynamics, using equivariant language.

%~~~~~~~~~~~~~~~~~~~~~~~~~~~~~~~~~~~~~~~~~~~~
\subsection{Effective worldline action with $\UT$ gauge invariance}
\label{sec:LangActions}
%~~~~~~~~~~~~~~~~~~~~~~~~~~~~~~~~~~~~~~~~~~~~

Our philosophy is to write down a worldvolume theory of the Brownian particle that explicitly makes manifest the full ${\cal N}_T =2 $ symmetry. To this end we have already identified the various fields and their transformation properties in the preceding subsections. We now should simply enumerate the allowed terms in the worldvolume action that respect the symmetries and confirm that the resulting Lagrangian describes Langevin dynamics. More precisely, the resulting Lagrangian should coincide in a suitable sense with the Lagrangian obtained in the MSR construction, Eq.\ \eqref{eq:LSKLan}. 

Following the usual discussion of cohomological field theories we realize that all we need are terms that are $\Q$ and $\Qb$ exact to appear in the Lagrangian.  Since we have already identified the action of the Cartan supercharges with the superspace covariant derivatives in \eqref{eq:dCdCbdef}, we can equivalently write down an action that is a superspace density with vanishing ghost number. The integrals over the supercoordinates $\{\theta, \thb\}$ would then allow us to reconstruct the $\Q$ and $\Qb$ action. To wit, the general form of the superspace action should take the form
\begin{equation}
S_{_{\sf{B0}}} = \int dt\, d\theta\, d\thb \, \SF{\Lagref}_{_{\sf{B0}}} = \int \, dt \gradcomm{\Qb}{\gradcomm{\Q}{\SF{{\Lagref}}_{_{\sf{B0}}}|}}  \,, 
\end{equation}	
with total ghost number $\gh{\SF{\Lagref}} =0$.

Having identified that we need to construct a superspace density function $\SF{\Lagref}_{_{\sf{B0}}}$, we can immediately intuit the set of allowed terms. We have 
\begin{itemize}
\item{\bf Inertial term:} These are simply the kinetic terms for the superfield $\SF{X}$, which for instance can be $\frac{m}{2}\, \left(\frac{d\SF{X}}{dt}\right)^2 $. More generally, this term can be written in as a functional  of the pullback of the target space metric onto the worldvolume as one expects for sigma model kinetic terms.
\item {\bf Potential term:} To account for the external potential felt by the particle in target space we can introduce a superpotential term $U(\SF{X})$. Such a term breaks target space diffeomorphisms, but preserves all the worldline symmetries.
\item{\bf Friction term:} In addition to the superfield $\SF{X}$ we also have its gauge covariant derivatives 
$\Dsf_\theta \SF{X}$ and $\Dsf_{\thb}\SF{X}$. While neither of these can appear in isolation owing to the fact that they have non-trivial ghost number, their product is indeed an allowed operator of ghost number zero. This means that we can include contributions of the form $\Dsf_\theta \SF{X} \Dsf_{\thb}\SF{X}$ in the action. 
\end{itemize}

These three classes of terms exhaust the possibilities for cohomologically trivial terms as can be inferred from \cite{Dijkgraaf:1996tz}, cf., their Theorem 2.1 (and also the discussion of section 4.3 there). One can also understand these in terms of the supersymmetric quantum mechanics theory as in \cite{Witten:1982im}.
For the Brownian particle to reproduce the dynamics of the Langevin equation \eqref{eq:langevin} we should simply make some specific choices for the three classes of allowed terms, leading to
\begin{equation}\label{eq:langevinS}
\begin{split}
  S_{_{\sf{B0}}} &= \int dt\, d\theta\, d\thb \, \left\{\frac{m}{2} \left(\frac{d\SF{X}}{dt}\right)^2 - U(\SF{X}) -i\nu \, \Dsf_\theta \SF{X} \,\Dsf_{\thb} \SF{X} \right\} \\
  &= \int dt \, \gradcomm{\Qb}{\gradcomm{\Q}{\frac{m}{2} \left(\frac{dX}{dt}\right)^2 - U(X) -i\nu \, \xpsib\, \xpsi }} \,.
  \end{split}
\end{equation}
We now turn to a discussion of various features that can be derived from the action \eqref{eq:langevinS}.

\paragraph{Relating back to MSR construction:}
We start by relating the action \eqref{eq:langevinS} back to the MSR construction, which led to the Lagrangian \eqref{eq:LSKLan}. 
It is straightforward to either perform the superspace integral, or implement the action of $\Qb$, $\Q$ in order to obtain the real space Schwinger-Keldysh action from \eqref{eq:langevinS}. Doing either of these exercises, we find 
\begin{equation}
\begin{split}
  S_{_{\sf{B0}}}=  \int dt \, \Bigg\{ &-m\left(  \tx \frac{d^2X}{dt^2} + \xpsib \,\frac{d^2\xpsi}{dt^2} \right)  - \left( \tx \, \frac{\partial U}{\partial X} + \xpsib \, \frac{\partial^2 U}{\partial X^2} \, \xpsi  \right) \\
 & -i\,\nu\,\bigg[- \tx^2 +\phibT \,\xpsi \delKMS\xpsi +\phiT^0 \,  \tx \delKMS X - \phiT\, \phibT\, (\delKMS X)^2 \\
  & \qquad\quad\;\;\; + (\etabT \,\xpsi + \etaT \,\xpsib) \delKMS X - (\phiT^0 \,\xpsi - \phiT, \xpsib) \delKMS \xpsib\bigg] \Bigg\}\,,
\end{split}
\end{equation}
where we discarded some total derivative terms. Structurally, this is similar to the MSR Lagrangian constructed earlier in \eqref{eq:LSKLan} but there are various additional contributions from the ghost of ghost quintet as expected. To truncate to the MSR action, we can consider gauge fixing the above Lagrangian with the choice:
\begin{equation}
\phiT = \etaT = \etabT = \phibT = 0 \,, \qquad  \vev{\phiT^0} \neq 0 \,,
\label{eq:ggfix}
\end{equation}	
where  we are eliminating all the ghost of ghost fields carrying non-vanishing ghost number and are only allowing the possibility for the single field $\phiT^0$ to acquire a non-vanishing expectation value.  With this choice we find 
\begin{equation}\label{eq:SKlangevinFixed}
\begin{split}
  S_{_{\sf{B0}}}\bigg{|}_\text{fixed} =  \int dt \, \Big\{ &-m\left(  \tx \frac{d^2 X}{dt^2} + \xpsib \,\frac{d^2\xpsi}{dt^2} \right)  - \left( \tx \, \frac{\partial U}{\partial X} + \xpsib \, \frac{\partial^2 U}{\partial X^2} \, \xpsi  \right) \\
 &\qquad
  -i\,\nu\,\bigg(- \tx^2  +\vev{\phiT^0} \,  \tx \delKMS X - \vev{\phiT^0} \xpsi \delKMS \xpsib 
 \bigg) \Big\}\,,
\end{split}
\end{equation}
where `$\sf{fixed}$' refers to the gauge fixing \eqref{eq:ggfix}. 
At this point we can decouple the interactions between the difference field $\tx$ and the rest of the multiplet by using a Hubbard-Stratonovich noise field $\mathbb{N}$. This allows us to absorb the terms quadratic in the fluctuation field $ \tx$ and write \eqref{eq:SKlangevinFixed} in the form
\begin{equation}
\begin{split}
  S_{_{\sf{B0}}}\bigg{ |}_\text{fixed} = \int dt \, \bigg\{ &\left( - m \,\frac{d^2X}{dt^2}- \frac{\partial U}{\partial X} - i\, \nu \, \vev{\phiT^0}\, \delKMS X + \mathbb{N} \right) \tx  \\
 & -  \xpsib \left( m\,\frac{d^2}{dt^2} + \frac{\partial^2U}{\partial X^2} \right) \xpsi + i\,\nu\, \vev{\phiT^0} \, \xpsi \delKMS \xpsib  + \frac{i}{4\nu} \, \mathbb{N}^2  \bigg\}\,.
\end{split}
\label{eq:langevinFinal}
\end{equation}

From this form of the Lagrangian, we can immediately see that we have achieved our goal: variation with respect to the difference field $ \tx$ gives exactly the dissipative Langevin equation of motion \eqref{eq:langevin}, if we choose $\phiT^0$'s vev to be: 
\begin{equation}
\vev{\phiT^0}= -i \,.
\label{eq:phiT0fix}
\end{equation}
With this choice, we recognize the gauge fixed action \eqref{eq:SKlangevinFixed} as being precisely the MSR effective action \eqref{eq:LSKLan}.
In effect the origin of dissipative term in the Langevin dynamics can be associated with a {\em ghost condensate}. Note that of the various fields in the gauge multiplet, $\phiT^0$ has a distinguished role; it is a covariant field strength of the underlying gauge symmetry  \eqref{eq:quintet} and has vanishing ghost number. We will argue momentarily that the vev \eqref{eq:phiT0fix} should be viewed as arising from spontaneous breaking of microscopic {\sf CPT} invariance in the system.

For completeness, we can write down the equations of motion for the remaining fields by varying, for example, Eq.\ \eqref{eq:langevinFinal}: 
\begin{equation}\label{eq:ghostEOM}
\begin{split}
m\frac{d^2 X}{dt^2} +\frac{\partial U}{\partial X}  +  \nu\ \delKMS X
&= \mathbb{N} \,,\\
  m \, \frac{d^2\xpsi}{dt^2} + \frac{\partial^2U}{\partial X^2} \, \xpsi  - \nu \, \delKMS \xpsi &= 0 \,,\\
  m \, \frac{d^2\xpsib}{dt^2} + \frac{\partial^2U}{\partial X^2} \, \xpsib  + \nu \, \delKMS \xpsib &= 0 \,,\\
  m \, \frac{d^2 \tx}{dt^2} + \frac{\partial^2 U}{\partial X^2} \,  \tx +  \frac{\partial^3U}{\partial X^3} \, (\xpsib\,\xpsi) - \nu \, \delKMS  \tx &= 0 \,,
\end{split}
\end{equation}
where we used \eqref{eq:phiT0fix}. One can immediately make various simple consistency checks of the BRST supersymmetry structure; e.g. all fields of the position multiplet have the same mass. 

We note that the convergence of the path integral with action \eqref{eq:langevinFinal} forces a sign on the viscous coefficient $\nu$; indeed, inspection of the ghost kinetic term requires that $\nu \geq 0$ for the action to be well-defined. This is the first sign for a second law statement. As a consequence, we realize the following intriguing feature of \eqref{eq:ghostEOM}: while the fields $X$ and $ \xpsib$ are damped (i.e., they dissipate with a friction $+\nu$), the ghost $\xpsi$ and the fluctuation field $\tx$ are anti-damped, i.e., they {\em anti-dissipate} (they come with a coefficient $-\nu$).

\paragraph{Superspace dynamics:}
Instead of performing the superspace integral in \eqref{eq:langevinS} and working in real space, we could also just work in superspace to check that \eqref{eq:langevinS} is indeed the correct Lagrangian for Langevin dynamics. To wit, we can compute the Lagrangian equations of motion for $\SF{X}$. Varying the action with respect to $\SF{X}$, we find 
\begin{equation}\label{eq:deltalangevinS}
\begin{split}
  \delta_{\SF{X}}\,   S_{_{\sf{B0}}} &= \int dt\, d\theta\, d\thb \, \left\{ \left(-m \, \frac{d^2\SF{X}}{dt^2} - \frac{\partial U}{\partial \SF{X}} + i\, \nu \, (\Dsf_\theta\Dsf_{\bar\theta}-\Dsf_{\bar\theta}\Dsf_{\theta}) \, \SF{X} \right) \, \delta \SF{X} + \cdots \right\} \,,
  \end{split}
\end{equation}
where the ellipses indicate total derivatives terms of the form $\Dsf_\theta(\ldots)$, $\Dsf_{\bar\theta}(\ldots)$,  or $\frac{d}{dt}(\ldots)$. Since we will eventually set $\phibT=\etabT=\etaT=\phiT=0$ and $\phiT^0 = -i$, it is clear that none of these total derivatives will contribute to equations of motion. If we use 
\begin{equation}
(\Dsf_\theta\Dsf_{\bar\theta} + \Dsf_{\bar\theta}\Dsf_{\theta})\, \SF{X} = (\Fs_{\theta\thb},\SF{X})_\Kref = (\phiT^0+\ldots) \, \delKMS \SF{X} \,,
\end{equation}
where $\ldots$ denotes terms vanishing in the limit \eqref{eq:ggfix},
we see that the equation of motion in \eqref{eq:deltalangevinS} gives:
\begin{equation}
-m \, \frac{d^2\SF{X}}{dt^2} - \frac{\partial U}{\partial \SF{X}} - i \, \nu \, \vev{ \phiT^0} \, \delKMS \SF{X} 
 = -2 i \, \nu \, \Dsf_{\theta}\Dsf_{\bar\theta} \SF{X} \,.
\end{equation}
If we now project down to real space, using the identity $\Dsf_{\theta}\Dsf_{\bar\theta} \SF{X} | = \tx$, this gives a real space equation of motion:
\begin{equation}
\begin{split}
 m \, \frac{d^2X}{dt^2} + \frac{\partial U}{\partial X} +  i\,\nu  \,\vev{\phiT^0} \delKMS X 
 &= 2 i \, \nu   \tx \,.
\end{split}
\end{equation}
With the identification \eqref{eq:phiT0fix}, this is precisely the Langevin equation \eqref{eq:langevin} where the noise is realized by the difference field. This relation can be confirmed by comparing the real space action with Hubbard-Stratonovich noise \eqref{eq:langevinFinal} with the previous \eqref{eq:SKlangevinFixed}, where we can indeed identify $\mathbb{N} \simeq 2i \, \nu\,  \tx$ up to a constant renormalization of the path integral measure. This again establishes the basic idea of the fluctuation dissipation theorem and the fact that $ \tx$ should be thought of as the fluctuation field.

%~~~~~~~~~~~~~~~~~~~~~~~~~~~~~~~~~~~~~~~~~~~~
\subsection{Fluctuation-dissipation theorem and Jarzynski work relation}
\label{sec:Jarzynski}
%~~~~~~~~~~~~~~~~~~~~~~~~~~~~~~~~~~~~~~~~~~~~

The fluctuation-dissipation theorem gives a relation between the friction coefficient responsible for the latter and the stochastic or thermal fluctuations. It is immediate to see this from the effective action; indeed, the dissipative coefficient $\nu$ determines the width of the Gaussian of the statistical noise fluctuations in \eqref{eq:langevinFinal}.
 More precisely, the action tells us that the noise correlations are Gaussian with $\vev{ \mathbb{N} \, \mathbb{N} } \sim \nu$ as presaged in \eqref{eq:PNdef}. It is  rather remarkable that all these features are encoded in the simple and elegant superspace Lagrangian \eqref{eq:langevinS}. In fact, below we will derive further consequences and a precise version of the fluctuation-dissipation theorem from the symmetries of the Lagrangian.

We also should highlight the  fact that $\phiT^0$ takes a non-zero expectation value as in \eqref{eq:phiT0fix} is
 a tantalizing feature of the formalism. As indicated this expectation value can be seen as the order parameter for dissipation through a spontaneous breaking of {\sf CPT} invariance. The rich phenomenology of such a symmetry breaking pattern is briefly mentioned in \cite{Haehl:2015uoc}, following earlier observations in the statistical physics literature; we sketch some of its consequences below.

As an application and as a demonstration of the power of the supersymmetric formulation of SK Langevin dynamics, we will now derive a second law statement in the form of Jarzynski's non-equilibrium work relations \cite{Jarzynski:1997ab,Jarzynski:1997aa}. We remind the reader that these relations encode a general form of out-of-equilibrium fluctuation-dissipation results. They can equivalently be stated in terms of the ratio of probability distributions of work done under a given dynamical protocol and its time-reversed counterpart, which go by the name of Crooks relations \cite{Crooks:1999fk}.

To understand these statements let us allow for an explicitly time-dependent potential $U(X(t),t)$. We assume that at initial time $t_i$, the system is in thermal equilibrium. During $t_i < t < t_f$ the external potential drives the system away from equilibrium and does work on our system. At $t=t_f$ the sources are switched off and we let the system relax back to some new equilibrium state (characterized by a different free energy) which is attained asymptotically. What we are interested in is to characterize the work done by the source in the time it is turned on, averaging over various protocols to go from the chosen initial state to the instantaneous state at the end of the period of the sourcing. As such we expect to generate some free energy, which we can take to be the difference between the instantaneous value at $t=t_f$ (which we note differs from the asymptotic late time free energy after re-equilibration) and the initial free energy.  This difference only depends on the instantaneous data at the two times corresponding to the start and end of the disturbance. The key insight of Jarzynski was to relate this quantity to the ensemble average of the work done, cf., \eqref{eq:Jarzynski}, which then effectively tells us about the entropy generated in the process of driving the system, cf., \eqref{eq:Jarzynski2ndLaw}. One can essentially view the relation as a formalization of the second law of thermodynamics with a suitable 
notion of averaging over protocols.

We will now derive Jarzynski's theorem as the supersymmetric Ward identity associated with {\sf CPT} transformations following the discussion of  \cite{Mallick:2010su}. A very nice discussion of {\sf CPT} transformations an their implications for driven systems is given in \cite{Gaspard:2012la,Gaspard:2013vl}. We will adapt their language to our Schwinger-Keldysh and superspace discussion.\footnote{ The discussion below has been modified in version 3 to make clear the role of {\sf CPT} transformations and $\UT$ thermal symmetries. We also refer the reader to discussions of KMS transformations and its inter-relation with the Schwinger-Keldysh formalisms in the discussion of \cite{Mallick:2010su} and \cite{Sieberer:2015hba} quite useful in the following.} 

Let us start by defining time reversal in our formalism. Let us first recall the definition of SK time-ordering in generic correlation functions: 
\begin{equation}
\begin{split}
& \langle {\cal T}_{SK} \, \SKR{O}^{(1)} \cdots \SKR{O}^{(n-k)}\SKL{O}^{(n-k+1)} \cdots \SKL{O}^{(n)} \rangle  = \text{Tr} \left\{ {\cal T} \left( \OpH{O}^{(1)} \cdots  \OpH{O}^{(n-k)} \right) \rho_{_T} \bar{\cal T} \left( \OpH{O}^{(n-k+1)} \cdots \OpH{O}^{(n)} \right) \right\} \,,
 \end{split}
\end{equation}
where ${\cal T}$ and ${\bar{\cal T}}$ denote standard time ordering and anti-time ordering, respectively, and we abbreviate $\Op{O}^{(q)} \equiv \Op{O}^{(q)}(t_q)$. 
Effectively, we will be considering a {\sf CPT} transformation. As explained in \S{8} of \cite{Haehl:2016pec} we can implement {\sf CPT} on the SK path integral via an exchange of sources:
\begin{equation} \label{eq:CPTdef}
{\sf CPT}: \qquad \mathcal{Z}_{SK}[\mathcal{J}_\skR,\mathcal{J}_\skL] \;\; \mapsto \;\; \mathcal{Z}_{SK}[\mathcal{J}_\skL,\mathcal{J}_\skR]^* \,.\qquad
\end{equation}
This means that we can define a reversed SK time ordering, $\overline{\cal T}_{SK}$ defined as\footnote{ For simplicity, we assume that operators are real here.}
\begin{equation}
\begin{split}
& \langle \overline{\cal T}_{SK} \, \SKR{O}^{(1)} \cdots \SKR{O}^{(n-k)}\SKL{O}^{(n-k+1)} \cdots \SKL{O}^{(n)} \rangle   = \text{Tr} \left\{ {\cal T} \left( \OpH{O}^{(n-k+1)} \cdots \OpH{O}^{(n)} \right) \rho_{_T} \bar{\cal T} \left( \OpH{O}^{(1)} \cdots  \OpH{O}^{(n-k)} \right)\right\} \,,
 \end{split}
\end{equation}
In this implementation of {\sf CPT} time does not actually get reversed. Instead we effectively traverse the SK contour in the opposite direction. To make contact with the supersymmetric formulation of the theory, we now focus on the generating functional with explicit sources from which correlators such as the above can be computed by taking functional derivatives. To write this down, let us first introduce some notation for the coupling of the Langevin theory to external sources:
\begin{equation}
\begin{split}
 \mathcal{Z}_{SK}[\SF{{\cal J}}] 
 &\equiv \text{Tr} \left\{ \, \rho_{_T} \, {\cal T}_{SK} \; \text{exp} \left( i S_{_{\sf B0}}[\SF{X},\Dsf_\theta \SF{X}, \Dsf_\thb \SF{X}] -i \int dt \,d\theta\, d\thb \, \SF{\cal J} \SF{X} \right) \right\} \,.
\end{split}
\end{equation}
At the level of the generating functional, {\sf CPT} then acts as follows: 
\begin{equation}\label{eq:SKCPT}
\begin{split}
 {\sf CPT}: \qquad \mathcal{Z}_{SK}[\SF{\cal J}] 
 \; \mapsto\; \text{Tr} \left\{ \, \rho_{_T} \, \overline{\cal T}_{SK} \; \text{exp} \left( i S_{_{\sf B0}}[\SF{X},\Dsf_\thb \SF{X}, \Dsf_\theta \SF{X}] - i\int dt\, d\theta\, d\thb \, \SF{\cal J} \SF{X} \right) \right\} .
\end{split}
\end{equation}
Note that the sign in front of the action does not change as the sign flip of $i$ is compensated for by a sign flip in the measure of the superspace integral. 

To probe the spontaneous breaking of {\sf CPT} in our dissipative system, we now consider a particular $\UT$ transformation in the {\sf CPT} transformed theory. The $\UT$ transformation we will be considering corresponds to choosing a $\UT$ gauge parameter $\SF{\Lambda} = - \thb \theta\, \mathscr{F}_{\theta\thb} \equiv - \thb \theta\, \phiT^0$,\footnote{ We recall that we made a choice in the presentation of the $\mathcal{N}_\smallT =2$ equivariant cohomology algebra where we chose representatives that were not manifestly covariant with respect to the $\mathfrak{sl}(2)$ automorphism, c.f., footnote \ref{fn:sl2}. The reason was to ensure a simple action of {\sf CPT} symmetry along one of the generators  $\phiT^0$. \label{fn:cptphi0}} which has the effect of shifting top components of superfields by a time derivative. This is a useful symmetry transformation to consider because it allows us to probe the spontaneous breaking of \eqref{eq:SKCPT} by giving a {\sf CPT} breaking expectation value to $\phiT^0$. Further, it is simple enough to lead us to Jarzynski's equality. If we combine this $\UT$ transformation with the action of {\sf CPT} as described above, we end up with the following $\mathbb{Z}_2$ transformation: 
\begin{equation}\label{eq:CPTUT}
\begin{split}
 \UT\circ {\sf CPT}:\qquad \left\{\phibT,\etabT,\phiT^0,\etaT,\phiT\right\} &\mapsto \left\{ \phiT,\etaT,\phiT^0,\etabT,\phibT\right\} \,,\\
  \left\{ i ; \theta, \thb ; {\cal T}_{SK} , \overline{\cal T}_{SK} \right\} & \mapsto \left\{ -i ; \thb, \theta;  \overline{\cal T}_{SK} , {\cal T}_{SK} \right\} \,, \qquad\\ 
   \SF{X} &\mapsto 
   	\SF{X} - \thb\theta \, \phiT^0 \,\delKMS X \,,\\
   \Dsf_\theta \SF{X} & \mapsto
   	 \Dsf_{\bar \theta} \SF{X} - \thb\theta \, \phiT^0\,\delKMS \xpsi\,,\\
   \Dsf_{\thb} \SF{X} &\mapsto 
   	\Dsf_\theta \SF{X}- \thb \theta\, \phiT^0\,\delKMS \xpsib \,.
\end{split}
\end{equation}
We have chosen to not indicate the superfields in the last terms of  the r.h.s.\ since the explicit $\thb\theta$ ensures that we end up only with the bottom component of the corresponding superfield.\footnote{ We note that we could combine $\UT$ with a different discrete symmetry to derive similar results. We will elaborate on this briefly at the end of the present section.} 

Given the symmetry transformation, we can compute the change in the action \eqref{eq:langevinS}. The fact that we are shifting the entries by a term proportional to $\thb\theta$ is now helpful. We find that the change can be expressed quite simply as
\begin{equation}\label{eq:SlangFlip}
 \UT\circ {\sf CPT}:\qquad  i   S_{_{\sf{B0}}}\;\mapsto\; i \,  S_{_{\sf{B0}}} -i \vev{\phiT^0} \, \Kbeta\, \bigg( G_f-G_i + W \bigg) \,.
\end{equation}
We have expressed the change in terms of quantities appearing in the Jarzynski relation. The  difference in free energy $G$ between initial time (at $t=t_i$) and the instantaneous final state (at $t=t_f$), and the total work done $W$, are respectively given by
\begin{equation}
\begin{split}
 G_f-G_i &= \left[ \frac{m}{2} (\delKMS X)^2 - U(X(t),t) - i\,\nu \, \xpsib\,\xpsi \right]_{t_i}^{t_f} \,,\\
 W &= \int_{t_i}^{t_f} dt \; \frac{\partial U(X(t),t)}{\partial t} \,.
\end{split}
\end{equation}
By giving explicit time dependence to the potential, we essentially mean that the system is coupled to an external source, for instance: 
\begin{equation}
  \int_{t_i}^{t_f} dt \; \frac{\partial U(X(t),t)}{\partial t} \equiv \int_{t_i}^{t_f} dt \; \frac{d {\cal J}(t)}{d t} \, X(t) = - \int_{t_i}^{t_f} dt \;  {\cal J}(t) \, \frac{d X(t)}{d t} \,,
\end{equation}
where ${\cal J} = \SF{\cal J}|$ is the ``retarded'' source.
Note that the free energy difference takes a very intuitive form: it is simply the free energy density defined by the Lagrangian \eqref{eq:langevinS} evaluated at the initial and final times of the evolution protocol. Similarly, the work done is just the non-conservative part of the potential integrated along the evolution protocol.

From \eqref{eq:SlangFlip} we see that if $\vev{ \phiT^0} = 0$, then the system $  S_{_{\sf{B0}}}$ is just invariant. However, as described before, this choice is not consistent with the usual picture of thermal fluctuations. Following \eqref{eq:phiT0fix}, we therefore set $\vev{ \phiT^0} = -i$. We now see that this choice is associated with a spontaneous breaking of {\sf CPT} symmetry.

As a Ward identity for the $ \UT+{\sf CPT}$ symmetry, we conclude the following from \eqref{eq:SlangFlip}:
\begin{equation}\label{eq:UTward}
\begin{split}
 \text{Tr} \left\{ \, \rho_{_T} \, {\cal T}_{SK} \; e^{i (S_{_{\sf B0}} - \int \SF{\cal J} \SF{X})} e^{-i \,\phiT^0 \Kbeta \,W} \right\} =
 e^{-\Kbeta (G_f-G_i)}\, \text{Tr} \left\{ \, \rho_{_T} \, \overline{\cal T}_{SK} \; e^{i (S_{_{\sf B0}} - \int \SF{\cal J} \SF{X})} \right\} .
\end{split}
\end{equation}
By choosing not to take any functional derivatives, we find the Jarzynski equation: 
\begin{equation}\label{eq:Jarzynski}
\big{\langle} e^{-\Kbeta\, W} \big{\rangle} = e^{-\Kbeta\, (G_f-G_i)} \,. 
\end{equation}
The right hand side of this equation describes the difference in free energy between the two equilibrium states at times $t=t_i$ and $t=t_f$. The left hand side then corresponds to the work done as a statistical average over all protocols going between these equilibrium configurations. We therefore have a remarkable relation between the work done (averaged over generic non-equilibrium protocols) on one hand, and difference in free energies of equilibrium configurations on the other hand. This relation should be thought of as a Ward identity in the {\sf CPT} broken phase, associated to the supersymmetry of the system. 

The equality \eqref{eq:Jarzynski} is known to be the appropriate version of the second law in statistical systems out of equilibrium (even far from equilibrium): while both entropy producing and entropy destroying processes are in principle allowed, the former are exponentially more likely to occur. Indeed, we can apply the standard Jensen inequality $\langle  e^x \rangle \geq e^{\langle x \rangle}$ to \eqref{eq:Jarzynski} and find
\begin{equation}\label{eq:Jarzynski2ndLaw}
 \langle W \rangle \geq G_f-G_i \,,
\end{equation}
which is, of course, the familiar statement that the second law of thermodynamics is respected in a statistical sense. We hence see that in a modern path integral formulation, the second law of thermodynamics is fundamentally not an inequality, but should be thought of as an equality relating statistical quantities.

Finally, let us derive a set of off-equilibrium thermal sum rules. By this we mean certain correlator identities which are enforced by the symmetries of the system. By taking $n$ functional derivatives of \eqref{eq:UTward} with respect to ${\cal J}$, we find:
\begin{equation}\label{eq:ThermalSum}
 \Big{\langle} \mathcal{T}_{SK} \,   \prod_{k=1}^n \left( \tilde{X}(t_k) + i \, \Kref \, \frac{d X(t_k)}{d t}\right) e^{-W_J}\Big{\rangle} = 0 \,,
\end{equation}
where we used the fact that $\langle \overline{\mathcal{T}}_{SK} \,   \prod_{k=1}^n \tilde{X}(t_k) \rangle = 0$ due to unitarity. The vanishing of these correlators encodes non-equilibrium Ward identities of the equivariant cohomology underlying the thermal physics. 
We give an alternative derivation of this identity in Appendix \ref{sec:susyWard}, using only the action of the supersymmetry generators $\Q$, $\Qb$.

\paragraph{Addendum:} In the previous version of this review, we used a slightly different symmetry to derive the Jarzynski relation. Instead of the implementation of {\sf CPT} as defined by \eqref{eq:CPTdef}, we chose to consider the linear time reversal transformation\footnote{ Note further that the previous versions of this paper had a typo: time reversal as used here is linear, in particular it does {\it not} involve mapping $i \mapsto -i$.}
\begin{equation}
 {\sf T}:\qquad t \mapsto -t \,, \qquad \theta \mapsto \thb \,, \qquad \thb \mapsto \theta \,.
 \label{eq:cptsk}
\end{equation}	
Combined with the $\UT$ transformation advocated above, this leads to the transformation
\begin{equation}\label{eq:CPTUT2}
\begin{split}
\UT \circ {\sf T}: \qquad  \left\{\phibT,\etabT,\phiT^0,\etaT,\phiT\right\} &\mapsto \left\{ \phiT,\etaT,\phiT^0,\etabT,\phibT\right\} \,,\qquad \\
   \SF{X}(t) &\mapsto 
   	\SF{X}(-t) - \thb\theta \, \phiT^0 \,\delKMS X(-t) \,,\\
   \Dsf_\theta \SF{X}(t) & \mapsto
   	 \Dsf_{\bar \theta} \SF{X} (-t)- \thb\theta \, \phiT^0\,\delKMS \xpsi(-t)\,,\\
   \Dsf_{\thb} \SF{X}(t) &\mapsto 
   	\Dsf_\theta \SF{X} (-t)- \thb \theta\, \phiT^0\,\delKMS \xpsib(-t) \,.
\end{split}
\end{equation}
This transformation can also be used to derive the Jarzynski relation, by following the same arguments as in \cite{Mallick:2010su}.\footnote{ The symmetry \eqref{eq:CPTUT2} is also relevant in other recent discussions of dissipative effective field theories. It corresponds to the `dynamical KMS symmetry' discussed in \cite{Crossley:2015evo}.} This ambiguity can be traced back to the fact that implementation of {\sf CPT} in Schwinger-Keldysh formalism is not unambiguous (see \S8 of \cite{Haehl:2016pec}). With the benefit of hindsight, we find it most natural in the superspace formalism to work with the {\sf CPT} symmetry as in \eqref{eq:CPTUT}.

%~~~~~~~~~~~~~~~~~~~~~~~~~~~~~~~~~~~~~~~~~~~~
\section{Discussion}
\label{sec:discussion}
%~~~~~~~~~~~~~~~~~~~~~~~~~~~~~~~~~~~~~~~~~~~~

In this paper we have reviewed both the supersymmetric algebraic structure underlying any Schwinger-Keldysh theory with a KMS condition, and the language of equivariant cohomology. We have then shown that the SK-KMS symmetries generate precisely the algebra relevant for an $\mathcal{N}_T=2$ equivariant construction. We used these insights to formulate a worldline theory of Brownian motion as a gauged sigma-model. The main physical insight is that the symmetry which is being gauged is a $\UT$ invariance under thermal translations. This structure is then tight enough to constrain the set of possible worldline theories consistent with the symmetries to correspond precisely to the Brownian particle theory. 

We believe that the scope of this formalism extends far beyond the simple example of Langevin dynamics. As explained in \cite{Haehl:2016pec} there is a wide range of problems, which one can address using the SK formalism. The reformulation of standard SK rules in terms of BRST supercharges and the associated algebraic structures might be extremely useful in addressing certain problems such as renormalization group flow in SK theories. The most direct generalization of ideas presented in the present paper, was initiated in \cite{Haehl:2015uoc}, where we upgraded the construction of the dissipative 0-brane (i.e., the Brownian particle) to the dissipative sector of the sigma-model of a space-filling brane (i.e., hydrodynamics).\footnote{ A related discussion for construction of dissipative hydrodynamic effective actions appears in \cite{Crossley:2015evo}. We also note that \cite{Kovtun:2014hpa} directly attempt to use the MSR construction on the hydrodynamic equations of motion. We will explain the connections between these  approaches elsewhere.} In case of the 0-brane we saw explicitly in the present paper that the number of all possible terms in the supersymmetric Lagrangian is highly constrained. In a similar spirit, the dissipative sector of hydrodynamics was exhaustively explored in \cite{Haehl:2015uoc}, with the main new ingredient being the non-linearity introduced by generic dissipative transport coefficients. 
A similarly exhaustive analysis of the adiabatic transport in hydrodynamics will be an important future problem. Eventually, the most general higher dimensional non-linear generalization of our B0-brane action \eqref{eq:langevinS} should reproduce all of hydrodynamic transport as classified exhaustively in \cite{Haehl:2014zda,Haehl:2015pja}.   

Finally, our discussion was restricted to thermal correlation functions computed in the Schwinger-Keldysh formalism, which leads to $\mathcal{N}_\smallT =2$ equivariant cohomological structures. In the classification of balanced topological field theories \cite{Dijkgraaf:1996tz} argue for natural generalizations to higher numbers of topological charges. We argue in \cite{Haehl:2016pec} that these the extended algebraic structures are relevant for the computation of out-of-time-order correlation functions in thermal and near-thermal systems. It is interesting to speculate that the $\mathcal{N}_\smallT = 2k$ thermal equivariant cohomology theory could shed light on $k$-fold out-of-time-order correlation functions, which appear to be useful probes to understand chaos and scrambling in quantum systems. In particular, perhaps constraints arising from $\mathcal{N}_\smallT =4$ thermal equivariant cohomology directly give the chaos bound \cite{Maldacena:2015waa}. We hope to report on these issues in the near future.

%~~~~~~~~~~~~~~~~~~~~~~~~~~~~~~~~~~~~~~~~~~~~~~
\acknowledgments

%~~~~~~~~~~~~~~~~~~~~~~~~~~~~~~~~~~~~~~~~~~~~~~

 FH and RL would like to thank the QMAP at UC Davis for hospitality during the course of this project.
 FH \& MR would like to thank the Yukawa Institute for Theoretical Physics, Kyoto  and   Perimeter Institute for Theoretical Physics  (supported by the Government of Canada through the Department of Innovation, Science and Economic Development and by the Province of Ontario through the Ministry of Research and Innovation) for hospitality during the course of this project. MR would also like to thank Nordita and Arnold Sommerfeld Center, Munich for hospitality during the concluding states of this project.  FH gratefully acknowledges support through a fellowship by the Simons Foundation. 
RL gratefully acknowledges support from International Centre for
 Theoretical Sciences (ICTS), Tata institute of fundamental research, Bengaluru and Ramanujan fellowship from Govt. of India. RL would
also like to acknowledge his debt to all those who have generously supported and encouraged the pursuit of science in India.

\appendix

%~~~~~~~~~~~~~~~~~~~~~~~~~~~~~~~~~~~~~~~~~~~~
\section{Algebra with explicit gauge parameters}
\label{app:gauge}
%~~~~~~~~~~~~~~~~~~~~~~~~~~~~~~~~~~~~~~~~~~~~

We can extend the algebraic relations of \S\ref{sec:equivariance} to involve a gauge transformation parameter $\alpha^j$ (which may depend on the coordinates of ${\cal M}$) and consider the Lie derivation and contraction along $\alpha = \alpha^j t_j \in\mathfrak{g}$. 

\paragraph{Graded algebra discussion:} 
Let us define first the contraction operator $ \IbarEG{\alpha}\equiv \alpha^j \otimes \IbarEG{j}$, to act as
\begin{equation}
\begin{split}
\IbarEG{\alpha} G^i &= -(-)^\alpha\, \alpha^i \,,\qquad
\IbarEG{\alpha} \phi^i = 0 \,.
\end{split}
\label{eq:Ibaral}
\end{equation}	
where we have allowed for the possibility of $\alpha$ being Grassmann odd and incorporated its intrinsic Grassmann parity into the definition. The Lie derivation operation then follows by taking the definition in terms of the graded commutator between the Weil differential and the interior contraction, viz.,
we get
\begin{equation}
\begin{split} 
\lieEG{\alpha} G^i &= \gradcomm{\QWEG}{\IbarEG{\alpha}} G^i  %= (\alpha^k \otimes \lieEG{k}   )G^i 
=  f^i_{jk} \alpha^j \,G^k \,,\\  
\lieEG{\alpha} \phi^i &=  \gradcomm{\QWEG}{\IbarEG{\alpha}}   \phi^i  =   f^i_{jk}  \alpha^j   \phi^k\,.
\end{split}
\label{eq:Lal}
\end{equation}
We recognize these as the correct gauge transformation for the connection and curvature. 
Note that the algebra \eqref{eq:LieSAlg} is equivalent to 
\begin{equation}
\begin{aligned}
\gradcomm{\IbarEG{\alpha}}{\IbarEG{\beta}} &= 0 \,, &   \qquad
\gradcomm{\QWEG}{\IbarEG{\alpha}} &= \lieEG{\alpha} \,,
 \\
\gradcomm{\lieEG{\alpha}}{\IbarEG{\beta}} &=-  \IbarEG{\comm{\alpha}{\beta}} \,,  &  \qquad
\gradcomm{\QWEG}{\lieEG{\alpha} } &= 0 \,,
 \\
\gradcomm{\lieEG{\alpha}}{\lieEG{\beta}} &= - \lieEG{\comm{\alpha}{\beta}} \,,  &  \qquad
\gradcomm{\QWEG}{\QWEG} &= 0 \,,
\end{aligned}
\label{eq:LieSAlgParam}
\end{equation}
where $\comm{\alpha}{\beta} \equiv (f^k_{ij} \alpha^i \beta^j) t_k$.  

\paragraph{Superspace discussion:} Similar arguments lead to a generalization of the superspace operations to involve an explicit gauge parameter. For instance, the super-contraction operator of \S\ref{sec:superLie} reads
\begin{equation}
\begin{split}
  \SF{\Ibar}_\alpha & \equiv \Ibar_\alpha + \thb {\cal L}_\alpha \equiv \alpha^k \left( \Ibar_k + (-)^\alpha \thb \, {\cal L}_k \right) \,,
\end{split}
\end{equation}
This super-operator has the expected action our fundamental field content:
\begin{equation}
\begin{split}
   \SF{\Ibar}_\alpha \SF{\mathfrak{F}} = 0 \,, \qquad
   \SF{\Ibar}_\alpha \As_J = - (-)^\alpha \, \delta_{J\thb} \, \alpha\,,\qquad
   \SF{\Ibar}_\alpha \SF{X}^\mu = 0 \,.
\end{split}
\end{equation}
% 

%~~~~~~~~~~~~~~~~~~~~~~~~~~~~~~~~~~~~~~~~~~~~
\section{Mathai-Quillen conjugation and the BRST model}
\label{sec:MQ}
%~~~~~~~~~~~~~~~~~~~~~~~~~~~~~~~~~~~~~~~~~~~~

In this appendix we establish a third model of equivariant cohomology, viz., the {\it BRST model} \cite{Dijkgraaf:1996tz}, which is useful in some instances. 

\paragraph{Mathai-Quillen conjugation in $\mathcal{N}_\smallT=1$ equivariant cohomology:}
To obtain the BRST model, one defines the BRST differential via the following `Mathai-Quillen conjugation':
\begin{equation}\label{eq:dBRST}
 \QBRST \equiv e^{-G^i \otimes (\Ibar_i)_{\cal M}} \, \QW \, e^{G^i \otimes (\Ibar_i)_{\cal M}}  =  \QW - G^k \otimes ({\cal L}_k)_{_{\cal M}} + \phi^k \otimes (\Ibar_k)_{_{\cal M}} \,.
\end{equation}
The interior contraction, Lie derivative and Cartan differential can be conjugated in a similar fashion:
\begin{equation}\label{eq:ILBRST}
\begin{split}
 (\Ibar_k)_{_{\text{\tiny BRST}}} &\equiv e^{-G^i \otimes (\Ibar_i)_{\cal M}} \, \Ibar_k \, e^{G^j \otimes (\Ibar_j)_{\cal M}} = \IbarEG{k} \otimes \mathbb{1}_{_{\cal M}}  \,,\\
 ({\cal L}_k)_{_{\text{\tiny BRST}}} &\equiv e^{-G^i \otimes (\Ibar_i)_{\cal M}} \, {\cal L}_k \, e^{G^j \otimes (\Ibar_j)_{\cal M}} ={\cal L}_k \,,\\
 (\QC)\BRST &\equiv e^{-G^i \otimes (\Ibar_i)_{\cal M}}  \, \QC \, e^{G^i \otimes (\Ibar_i)_{\cal M}}  = \QBRST + G^i \, ({\cal L}_i)\BRST + \frac{1}{2} \, f^k_{ij} G^i G^j \, (\Ibar_k)\BRST \,. 
 \end{split}
\end{equation} 
On various fields, the differentials $\QBRST$ and $(\QC)\BRST$ act the same way as $\QW$ and $\QC$ with the following exceptions:
\begin{equation}
\begin{split}
 \QBRST X^\mu = \psiW^\mu - G^k \xi^\mu_k \,, \qquad\qquad\;\;\; & 
 (\QC)\BRST X^\mu = \psiW^\mu \,,\\
 \QBRST \psiW^\mu = \phi^k \xi_k^\mu - G^k \, \psiW^\sigma \partial_\sigma \xi_k^\mu \,,\qquad &
 (\QC)\BRST \psiW^\mu = \phi^k \xi_k^\mu \,.
\end{split}
\end{equation}
 Finally, we can easily infer that the conjugation map acts on $p$-forms as a basis change between Weil and Cartan bases:
\begin{equation}
e^{-G^i \otimes (\Ibar_i)_{\cal M}} \, V_{\psiC} \, e^{G^i \otimes (\Ibar_i)_{\cal M}} = V_{\psiW} \,,
\end{equation}
using the notation of \eqref{eq:Vpdef}. 

A similar construction can be made for $\mathcal{N}_\smallT =2$ algebras.

% \paragraph{Mathai-Quillen conjugation in $\mathcal{N}_\smallT=2$ equivariant cohomology:}

%~~~~~~~~~~~~~~~~~~~~~~~~~~~~~~~~~~~~~~~~~~~~
\section{Off-equilibrium sum rules as supersymmetry Ward identities}
\label{sec:susyWard}
%~~~~~~~~~~~~~~~~~~~~~~~~~~~~~~~~~~~~~~~~~~~~

In this section we elaborate on the Ward identities associated with the Langevin action $S_{_{\sf B0}}$ and its invariance under $\Q$ and $\Qb$. These can be stated as $n$-point correlation function identities which include the off-equilibrium version of the fluctuation-dissipation theorem as a special case for $n=2$. We will also give an alternative derivation of the correlator identities \eqref{eq:ThermalSum}. The content of this appendix is essentially a review of the discussion of \cite{Mallick:2010su} (see their section III) to make the connection with our superspace formalism precise.

Working to linear order in an infinitesimal parameter $\varepsilon$ (with Grassmann parity 1), invariance of $S_{_{\sf B0}}$ under the supersymmetry generator $\Qb$ implies: 
\begin{equation}
\begin{split}
 \mathcal{Z}_{SK}[\SF{{\cal J}}] 
 &= \text{Tr} \left\{ \rho_{_T} \, {\cal T}_{SK} \; \text{exp} \left[ i\, S_{_{\sf{B0}}}[\SF{X}+{\varepsilon}\,\Dsf_\theta \SF{X}] -i \int dt \, d\theta \, d\bar{\theta} \, \SF{{\cal J}} \SF{X} \right] \right\}\\
 &= \text{Tr} \left\{ \rho_{_T} \, {\cal T}_{SK} \; \text{exp} \left[ i\, S_{_{\sf{B0}}}[\SF{X}] -i \int dt \, d\theta \, d\bar{\theta} \, \SF{{\cal J}} (\SF{X} -{\varepsilon}\,\Dsf_\theta \SF{X}) \right] \right\}  \\
 &= \text{Tr} \left\{ \rho_{_T} \, {\cal T}_{SK}  \, \left( 1-{\varepsilon}\int dt \, d\theta \, d\bar{\theta} \, \SF{{\cal J}} \,\Dsf_\theta \SF{X} \right)  \, \text{exp} \left[ i\, S_{_{\sf{B0}}}[\SF{X}] -i \int dt \, d\theta \, d\bar{\theta} \, \SF{{\cal J}} \SF{X} \right]\right\} \,.
\end{split}
\end{equation}
Using that $\varepsilon$ was an arbitrary parameter, we infer: 
\begin{equation}\label{eq:Ward1}
\begin{split}
 0 &= \text{Tr} \left\{ \rho_{_T} \, {\cal T}_{SK} \;\left( \int dt \, d\theta \, d\bar{\theta} \, \SF{{\cal J}} \,\Dsf_\theta \SF{X} \right)  \, \text{exp} \left[ i\, S_{_{\sf{B0}}}[\SF{X}] -i \int dt \, d\theta \, d\bar{\theta} \, \SF{{\cal J}} \SF{X} \right]\right\} \\
 &= \int dt \; \left\{ \tilde{{\cal J}} \, \frac{\delta \mathcal{Z}_{SK}}{\delta {\cal J}_\psi} - {\cal J}_{\psib}\, \left( \frac{\delta \mathcal{Z}_{SK}}{\delta {\cal J} } - \phiT^0 \delKMS \left(  \frac{\delta \mathcal{Z}_{SK}}{\delta \tilde{{\cal J}}}  \right)  \right) + {\cal J} \, \phiT^0 \delKMS \left(  \frac{\delta \mathcal{Z}_{SK}}{\delta {\cal J}_\psi}  \right)\right\} \,.
\end{split}
\end{equation}
where we only keep track of $\phiT^0$, but not of the other fields of the Vafa-Witten quintet (which we will eventually set to zero).

Repeating the same argument with $\Q$ instead of $\Qb$, we conclude analogously: 
\begin{equation}\label{eq:Ward2}
\begin{split}
 0 &= \int [\mathcal{D}\SF{X}] \, \left( \int dt \, d\theta \, d\bar{\theta} \, \SF{{\cal J}} \,\Dsf_{\thb} \SF{X} \right)  \, \text{exp} \left\{ i\, S_{_{\sf{B0}}}[\SF{X}] -i \int dt \, d\theta \, d\bar{\theta} \, \SF{{\cal J}} \SF{X}\right\} \\
 &= \int dt \; \left( \tilde{{\cal J}} \, \frac{\delta \mathcal{Z}_{SK}}{\delta {\cal J}_{\psib}} - {\cal J}_\psi \,  \frac{\delta \mathcal{Z}_{SK}}{\delta {\cal J}}   \right) \,.
\end{split}
\end{equation}

To relate the Ward identities \eqref{eq:Ward1} and \eqref{eq:Ward2}, we now act with $\frac{\delta^2}{\delta {\cal J}_{\psib}(t) \delta \tilde{{\cal J}}(t')}$ on \eqref{eq:Ward1} and subsequently set the sources to zero. This gives:
\begin{equation}\label{eq:Ward1mod}
\begin{split}
 0  &=  \frac{\delta^2 \mathcal{Z}_{SK}}{\delta {\cal J}_{\psib}(t)\delta {\cal J}_\psi(t')} -  \frac{\delta^2 \mathcal{Z}_{SK}}{\delta  {\cal J}(t) \tilde{{\cal J}}(t')} + \phiT^0 \,\frac{\delta}{ \delta \tilde{{\cal J}}(t')} \delKMS \left(  \frac{\delta\mathcal{Z}_{SK}}{\delta \tilde{{\cal J}}(t)}  \right) \,.
\end{split}
\end{equation}
Similarly, we can act with $\frac{\delta^2}{\delta \tilde{{\cal J}}(t) \delta {\cal J}_\psi(t')}$ on \eqref{eq:Ward2} and find
\begin{equation}\label{eq:Ward2mod}
\begin{split}
 0  &=  \frac{\delta^2 \mathcal{Z}_{SK}}{\delta {\cal J}_{\psi}(t')\delta {\cal J}_{\psib}(t)} -  \frac{\delta^2 \mathcal{Z}_{SK}}{\delta {\cal J}(t') \delta \tilde{{\cal J}}(t)}    \,.
\end{split}
\end{equation}

Equating the ghost correlators that appear in both \eqref{eq:Ward1mod} and \eqref{eq:Ward2mod}, we find 
\begin{equation}
 \phiT^0 \,\frac{\delta}{ \delta \tilde{{\cal J}}(t')} \delKMS \left(  \frac{\delta\mathcal{Z}_{SK}}{\delta \tilde{{\cal J}}(t)}  \right) =  \frac{\delta^2 \mathcal{Z}_{SK}}{\delta  {\cal J}(t) \tilde{{\cal J}}(t')} - \frac{\delta^2 \mathcal{Z}_{SK}}{\delta {\cal J}(t') \delta \tilde{{\cal J}}(t)} \,.
\end{equation}
Or, as a correlator statement, and after setting $\langle \phiT^0 \rangle = -i$: 
\begin{equation} 
\begin{split}
i\, \Kref\, \langle \mathcal{T}_{SK} \;  \frac{dX(t)}{dt} X(t') \rangle &= \langle \mathcal{T}_{SK} \; X(t) \tilde{X}(t') \rangle - \langle \mathcal{T}_{SK} \; \tilde{X}(t) X(t') \rangle \\
&= \langle \comm{\hat{X}(t)}{\hat{X}(t')} \rangle \,,
\end{split}
\end{equation}
where $\hat{X}$ denotes the physical single copy operator. We recognize this as the equilibrium fluctuation-dissipation relation. 
%Let us now w.l.o.g.\ take $t>0$ and $t'=0$.
%%
%\begin{equation} 
%\begin{split}
%i\delKMS \left( \langle \left\{\hat{X}(t),\hat{X}(0) \right\}\rangle - \frac{1}{2}\text{coth}\left(\frac{i \delta_\Kref}{2}\right) \langle \comm{\hat{X}(t)}{\hat{X}(0)} \rangle \right)   &= \langle \comm{\hat{X}(t)}{\hat{X}(t')} \rangle \,,
%\end{split}
%\end{equation}
%%

We can also derive a more general off-equilibrium sum rule from the supersymmetry Ward identities. To this end, we act on \eqref{eq:Ward1} with the following operator: 
\begin{equation}
\frac{\delta}{\delta {\cal J}_{\psib}(t)} \prod_{k=1}^n \left( \frac{\delta}{\delta {{\cal J}}(t_k)} - \phiT^0 \delKMS \frac{\delta}{\delta \tilde{{\cal J}}(t_k)} \right) \,.
\end{equation}
Acting with this on \eqref{eq:Ward1} and setting sources to zero afterwards, we find: 
\begin{equation}\label{eq:Wardcorr}
 \Big{\langle} \mathcal{T}_{SK} \,   \prod_{k=1}^n \left( \tilde{X}(t_k) +i \, \Kref \, \delKMS X(t_k)\right) \Big{\rangle} = 0 \,.
\end{equation}
This relation holds for time independent potentials.

Let us now generalize \eqref{eq:Wardcorr} away from equilibrium. Since \eqref{eq:Wardcorr} follows from the $\Qb$ Ward identity, we need to address the fact that the action $S_{_{\sf B0}}$ coupled to a time-dependent potential is not entirely invariant under shifting fields by $\Qb$. Schematically, we can see this as follows: 
%%
%\begin{equation}
%\begin{split}
%S_{_{\sf B0}}[\SF{X}+ \varepsilon \Dsf_\theta \SF{X}] &= \int dt \, d\theta \, d\thb \; \SF{\Lag}_{_{\sf B0}}[\SF{X}+ \varepsilon \Dsf_\theta \SF{X}] \\
%&= \varepsilon\int dt \, \Dsf_\theta \, \Dsf_\thb \,\left( \Dsf_\theta \SF{X} \, \frac{\delta \SF{\Lag}_{_{\sf B0}} [ \SF{X}] }{\delta \SF{X}} \right)\bigg{|}  \\
%&= \varepsilon\int dt \, \left\{ \Dsf_\theta \left( ( \Dsf_\thb \, \Dsf_\theta \SF{X} )  \frac{\delta \SF{\Lag}_{_{\sf B0}}  }{\delta \SF{X}} \right) \bigg{|} + X_{\psib} \, \Dsf_\theta \, \Dsf_\thb \, \frac{\delta \SF{\Lag}_{_{\sf B0}}  }{\delta \SF{X}} \bigg{|} +\ldots \right\} \\
%&= \varepsilon\,\phiT^0\, \Kref\int dt \,  \left( \frac{d (\Dsf_\theta \SF{\Lag}_{_{\sf B0}}|)}{dt} + \frac{\partial^2 U(X(t),t)}{\partial X \, \partial t}  X_{\psib} \right) + \ldots 
%\end{split}
%\end{equation}
%%
%
\begin{equation}
\begin{split}
S_{_{\sf B0}}[\SF{X}+ \varepsilon \Dsf_\theta \SF{X}] &= \int dt \, d\theta \, d\thb \; \SF{\Lag}_{_{\sf B0}}[\SF{X}+ \varepsilon \Dsf_\theta \SF{X}] \\
&= \varepsilon\int dt \, \Dsf_\theta \, \Dsf_\thb \, \Dsf_\theta \, \SF{\Lag}_{_{\sf B0}} [ \SF{X}] \Big{|}  \\
&= \varepsilon\,\phiT^0\, \int dt \,  \Dsf_\theta \, \delKMS \, \SF{\Lag}_{_{\sf B0}} \Big{|}+ \ldots \\
&= \varepsilon\,\phiT^0\, \Kref\int dt \,  \left( \frac{d (\Dsf_\theta \SF{\Lag}_{_{\sf B0}}|)}{dt} + \frac{\partial^2 U(X(t),t)}{\partial X \, \partial t}  X_{\psib} \right) + \ldots 
\end{split}
\end{equation}
where we discard terms that are proportional to Vafa-Witten quintet ghosts other than $\phiT^0$. To go from the second to the third line, we used $\gradcomm{\Dsf_\thb}{\Dsf_\theta} = \SF{\mathscr{F}}_{\thb\theta} \delKMS$. In the final line, the first term is a total derivative and hence does not pose a problem for the current argument. However, the second term corresponds to work done by the time dependent potential.
We remedy this non-invariance by adding a Jarzynski work term to the action such that its invariance under $\Qb$ is restored:
\begin{equation}\label{eq:actionmod}
iS_{_{\sf B0}}[\SF{X}] - i\, \phiT^0\,\Kref \, W_J \qquad \text{with} \qquad W_J \equiv \int dt  \, \frac{\partial U(X(t),t)}{\partial t} 
\end{equation}
is clearly invariant under $\SF{X} \rightarrow \SF{X} + \varepsilon \Dsf_\theta \SF{X}$. Therefore, the modified action \eqref{eq:actionmod} satisfies the Ward identity \eqref{eq:Ward1} and we can repeat the derivation of the sum rule for this case. This leads to the following off-equilibrium generalization of \eqref{eq:Wardcorr}: 
\begin{equation}\label{eq:Wardcorr2}
 \Big{\langle} \mathcal{T}_{SK} \,   \prod_{k=1}^n \left( \tilde{X}(t_k) +i  \, \delKMS X(t_k)\right) e^{-\Kref \, W_J}\Big{\rangle} = 0 \,.
\end{equation}
The same identity was derived in \S\ref{sec:Jarzynski} using slightly different techniques.

%%%%%%%%%%%%%%%%%%%%%%%%%%%%%%%%%%%%%%%%%%%%%%%
% \bibliographystyle{JHEP}
% \bibliography{sk-kms}

\providecommand{\href}[2]{#2}\begingroup\raggedright\endgroup

\end{document}